\documentclass[aps,pra,reprint,superscriptaddress,nofootinbib,longbibliography, showkeys]{revtex4-2}

\usepackage[utf8]{inputenc}
\usepackage[T1]{fontenc}
\usepackage{amsmath,amssymb,amsthm,mathtools,bm}
\usepackage{booktabs,array,graphicx}
\usepackage{xcolor}
\usepackage{braket}
\usepackage[colorlinks=true,linkcolor=blue!55!black,
  citecolor=blue!55!black,urlcolor=blue!55!black]{hyperref}
\usepackage{microtype}
\usepackage{comment}
\usepackage{tikz}
\usetikzlibrary{arrows.meta,positioning,calc,decorations.pathreplacing}

\allowdisplaybreaks
\newtheorem{theorem}{Theorem}
\newtheorem{proposition}[theorem]{Proposition}
\newtheorem{lemma}[theorem]{Lemma}
\newtheorem{corollary}[theorem]{Corollary}
\theoremstyle{definition}
\newtheorem{definition}[theorem]{Definition}
\theoremstyle{remark}

\newcommand{\Sp}{\operatorname{Sp}}
\newcommand{\R}{\mathbb{R}}
\newcommand{\I}{\mathbb{I}}
\newcommand{\Tr}{\operatorname{Tr}}
\newcommand{\Sym}{\operatorname{Sym}}
\newcommand{\diag}{\operatorname{diag}}
\newcommand{\IC}{I_{\mathrm C}}
\newcommand{\IQ}{I_{\mathrm Q}}
\newcommand{\FG}{F_{\mathrm G}}
\newcommand{\dd}{\mathrm d}

\tikzset{
  optical line/.style={-{Latex[length=2mm]},thick},
  optical component/.style={draw,rounded corners=1.5pt,fill=blue!6,
    minimum height=7mm,minimum width=13mm,align=center,font=\small},
  beam splitter/.style={draw,fill=gray!15,minimum height=17mm,
    minimum width=7mm,align=center,font=\small},
  detector/.style={draw,circle,fill=green!8,minimum size=8mm,
    inner sep=1pt,font=\small},
  annotation/.style={font=\footnotesize,align=center}
}

\begin{document}
\title{Fisher-Efficiency Optimization of Multimode Gaussian Measurements}
\author{Leonardo A. M. Souza}
\email{leonardoamsouza@ufv.br}
\affiliation{Dipartimento di Ingegneria, Universit\`a degli Studi di Palermo, Viale delle Scienze, 90128 Palermo, Italy}
\affiliation{Universidade Federal de Vi\c{c}osa - Campus Florestal, LMG 818 Km6 S/N, Florestal, 35690-000, Minas Gerais, Brazil}
\author{Enrico Russo}
\affiliation{Dipartimento di Ingegneria, Universit\`a degli Studi di Palermo, Viale delle Scienze, 90128 Palermo, Italy}
\affiliation{University San Pablo-CEU, CEU Universities, Department of Applied Mathematics and Data Science, Campus de Moncloa, C/Juli\'{a}n Romea 23 28003, Madrid, Spain}
\author{Gerardo Adesso}
\affiliation{School of Mathematical Sciences and Centre for the Mathematics and Theoretical Physics of Quantum Non-Equilibrium Systems, University of Nottingham, University Park, Nottingham NG7 2RD, UK}
\affiliation{Algorithmiq S.r.L., Via della Chiusa 15, 20123 Milano, Italy}
\author{Rosario Lo Franco}
\affiliation{Dipartimento di Ingegneria, Universit\`a degli Studi di Palermo, Viale delle Scienze, 90128 Palermo, Italy}
\date{\today}
\begin{abstract}
We develop a spectral framework for designing Gaussian measurements that retain quantum Fisher information uniformly over local one-parameter Gaussian perturbations of bosonic states. The resulting Fisher-efficiency spectrum compares the information extracted by a covariant general-dyne measurement with the corresponding directional quantum Fisher information, while its smallest eigenvalue provides a robust measurement-design criterion. For arbitrary faithful multimode Gaussian states, we derive the complete spectrum for seeds aligned with the Williamson structure and solve the optimization over all finite physical Gaussian seeds within this covariant measurement class. A pure seed matched to the
symplectic geometry of the reference state is globally optimal, and the optimal value depends only on the smallest symplectic eigenvalue. For
two-mode squeezed thermal states with arbitrary symplectic spectrum, product heterodyne is optimal over all separable Gaussian seeds, including correlated ones, whereas the matched collective measurement is strictly better for every
nonzero two-mode squeezing. This establishes a strict collective-separable efficiency gap within the covariant general-dyne class:
classical correlations in the seed cannot attain the collective optimum, and crossing the separable optimum necessarily requires an entangled seed. We also provide Gaussian-optical implementations and a proof-of-principle protocol for testing these predictions.
\end{abstract}
\maketitle

\section{Introduction}

Quantum metrology investigates how accurately physical parameters can be inferred from quantum systems and which measurements attain the corresponding precision limits \cite{Helstrom1976,Holevo2011,Paris2009,
Giovannetti2006,Giovannetti2011}. For a prescribed measurement, the attainable
local precision is governed by the classical Fisher information (CFI), whereas the quantum Fisher information (QFI) gives the optimal single-parameter value over all quantum measurements \cite{BraunsteinCaves1994,Paris2009}. A probe may therefore possess a large QFI while an experimentally accessible measurement extracts only a small part of it. This makes measurement design, together with probe optimization, a central problem in quantum sensing.

Bosonic Gaussian systems offer a natural setting in which to address this
problem. Gaussian states, transformations, and measurements are central to continuous-variable quantum optics and can be implemented using squeezing, linear optics, ancillary Gaussian modes, and homodyne or heterodyne
detection \cite{Caves1981,BraunsteinVanLoock2005,Weedbrook2012,
AdessoRagyLee2014,Serafini2017}. Although their Hilbert space is
infinite-dimensional, Gaussian states are completely characterized by their
first and second moments. Consequently, faithful Gaussian states form a
finite-dimensional statistical manifold with a well-developed symplectic and
information-geometric structure \cite{Weedbrook2012,LinkStrunz2015,
Miller2025}. Explicit QFI formulas are available for general single- and
multimode Gaussian models \cite{Monras2013,Pinel2013,Jiang2014,
SafranekLeeFuentes2015,Safranek2019}, and have been applied to phase,
temperature, loss, displacement, squeezing, and
gravitational-decoherence estimation in optomechanical
platforms \cite{Monras2006,MonrasIlluminati2011,Oh2019,Cenni2022,
Morelli2021,Rigovacca2017,Souza2024,Sorelli2024,Souza2026Gravitational}.

Most of these studies begin with a specified parameter or generator and then optimize the probe, the measurement, or both, for that particular encoding. Recent works have examined optimal Gaussian measurements for phase estimation
and thermometry \cite{Oh2019,Cenni2022}, collective Gaussian measurements on repeated copies \cite{Navarro2025}, joint Gaussian readout of independent thermal modes \cite{Liu2026GaussianSuperadditivity}, and Gaussian measurement optimization for state discrimination \cite{Turner2026}. Gaussian-state tomography has also been compared under homodyne and heterodyne detection \cite{Rehacek2015}, while optimal moment-based reconstruction has been studied for general Gaussian states \cite{TripierMondancin2026}. Related developments include asymptotic discrimination under collective Gaussian measurements \cite{Girardi2026} and a symplectic decomposition of the Gaussian QFI \cite{Chatterjee2026}. These works share important ingredients with the present analysis but address different operational tasks. Related and complementary results on Fisher-information recovery by Gaussian measurements were reported very recently in Ref. \cite{Grzywa2026}, whose Hermite-Gauss formulation intersects with the displacement and covariance sectors studied here. The relation between the two independently developed approaches is summarized in the Note added at the end of this work.

In particular, the asymptotic discrimination of repeated copies and the joint readout of constituent modes in a single realization are distinct measurement scenarios. The absence of a collective Gaussian advantage for centered states
in the repeated-copy discrimination problem \cite{Girardi2026} therefore does not conflict with the modewise collective advantage established below. Here we address the complementary problem of designing a single Gaussian measurement that remains informative for every local one-parameter Gaussian encoding through a calibrated reference state. Since each nonzero Gaussian tangent direction can be realized as the derivative of such a local submodel, we can compare, direction by direction, the CFI extracted by the fixed measurement with the corresponding QFI. The smallest of these ratios quantifies the measurement efficiency along its least efficiently resolved Gaussian direction, and optimizing it selects a measurement that performs uniformly well without assigning prior weights to displacement, squeezing, correlation, or thermal perturbations. This directional criterion should not be confused with the simultaneous estimation of all Gaussian parameters, which constitutes a genuinely multiparameter problem and generally requires compatibility conditions or the Holevo bound \cite{Holevo2011,Nichols2018,Liu2020,ChangGenoniAlbarelli2025}.

In this work, we focus on covariant Gaussian general-dyne POVMs, which are characterized by centered physical Gaussian seeds whose covariance matrices determine the noise added to the phase-space outcomes. This class encompasses heterodyne detection and general-dyne measurements with finite squeezing, with ideal homodyne detection recovered only as a singular limit. The theory of covariant phase-space observables, including their informational completeness and optical realization, is well established \cite{KiukasLahti2008,Lahti2010,Kiukas2012,KiukasSchultz2013,Genoni2014,Genoni2016}.

The relation between Fisher information and measurement frames provides the starting point for our construction, together with semiparametric formulations in which efficiency is expressed on score tangent spaces and information transmission through statistical channels is represented by a positive information map \cite{TsangAlbarelliDatta2020, Tsang2026Semiparametric}. In finite dimensions, measurement
efficiency can be represented through the spectrum of a frame operator associated with an informationally complete measurement \cite{Saini2026}. Frame-theoretic stability can also be formulated as a resource monotone for
informational completeness \cite{Saini2026c}. In continuous variables scenario, a direct extension to the full space of bosonic observables is delicate because informational completeness does not by itself imply a positive lower frame
bound. A reference-state-regularized frame geometry provides a suitable setting for stable observable reconstruction in this case \cite{Innocenti2026}.

Here we work on the finite-dimensional sector generated by Gaussian tangent directions. We do not construct a new frame on the full operator space. Instead, the Gaussian Fisher-efficiency operator represents the classical Fisher form relative to the quantum Fisher form on this restricted sector. Its spectrum identifies Gaussian directions resolved with different efficiencies, while its smallest eigenvalue defines the worst-direction objective optimized below. The specific contributions of the present work are the analytic Gaussian efficiency spectrum and its optimization over physical Gaussian seeds.

Within this measurement class, we obtain three main results. First, for state and seed covariances aligned in a common Williamson basis, we derive the complete Fisher-efficiency spectrum for arbitrary faithful Gaussian states,
including nondegenerate symplectic spectra. Second, we solve the optimization over all finite physical Gaussian seeds of a covariant general-dyne POVM and prove that a pure seed matched to the symplectic geometry of the reference
state is globally optimal. Third, for two-mode squeezed thermal states, we derive the exact product and separable benchmark and establish a strict gap between this benchmark and the collective optimum. Gaussian separability
 \cite{WernerWolf2001,Simon2000,Duan2000} and classical data processing \cite{Ferrie2014} imply that correlated separable seeds cannot outperform
the optimal product seed. The matched collective seed is nevertheless strictly better for every nonzero two-mode squeezing. Thus, within the
covariant general-dyne class, seed entanglement is necessary, but not sufficient, to surpass the separable optimum.

The matched measurement also admits a direct optical implementation: one first applies the inverse of the Gaussian transformation that brings the reference state to its Williamson form, and then performs local heterodyne detection on the resulting modes. In the single-mode case, this amounts to inverse squeezing followed by standard double-homodyne detection, whereas for two modes the inverse two-mode squeezer can be decomposed into balanced beam splitters and single-mode squeezers acting on opposite  quadratures \cite{Fedorov2017, CariolaroPierobon2016, CariolaroPierobon2018}. 

This work is organized as follows. Section \ref{sec:gaussianSystems} reviews Gaussian states and covariant general-dyne measurements. Section \ref{sec:gaussianTangentEstimation} introduces local quantum estimation on the Gaussian tangent space, and Sec. \ref{sec:fisherEfficiencyOperator} defines the Fisher-efficiency operator and develops its operational interpretation. The multimode spectrum and the global, product, separable, and collective optimization results are presented in Sec. \ref{sec:mainResults}. Section \ref{sec:opticalImplementation} describes Gaussian-optical implementations. Section \ref{sec:discussion} discusses the scope and limitations of the results, and Sec. \ref{sec:conclusion} concludes the paper. Technical proofs, operator-domain issues, and the analysis of the pure-state boundary are collected in the appendices.

\section{Gaussian systems and measurements}
\label{sec:gaussianSystems}
\subsection{Gaussian states}

We consider an $N$-mode bosonic system with quadrature-operator vector \cite{Weedbrook2012, serafini2005quantifying, cerf2007quantum, adesso2007entanglementThesis, AdessoIlluminati2007, AdessoRagyLee2014, Souza2024, Souza2015GaussianInterferometricPower}
\begin{equation}
  \hat{\bm R}
  =
  (\hat x_1,\hat p_1,\ldots,\hat x_N,\hat p_N)^T,
  \qquad
  [\hat R_j,\hat R_k]
  =
  i(\Omega_N)_{jk}, \nonumber
\end{equation}
where
\begin{equation}
  \Omega_N
  =
  \bigoplus_{j=1}^{N}J,
  \qquad
  J=
  \begin{pmatrix}
    0 & 1\\
   -1 & 0
  \end{pmatrix}.\nonumber
\end{equation}
The first-moment vector and covariance matrix of a state are defined as
\begin{align}
  \bm\varepsilon
  &:=
  \langle\hat{\bm R}\rangle,\nonumber\\
  \sigma_{jk}
  &:=
  \frac12
  \left\langle
    \left\{
      \Delta\hat R_j,\Delta\hat R_k
    \right\}
  \right\rangle,
  \qquad
  \Delta\hat{\bm R}
  :=
  \hat{\bm R}-\bm\varepsilon.\nonumber
\end{align}
With these conventions, the covariance matrix of the $N$-mode vacuum is
$\I_{2N}/2$. A real symmetric matrix $\sigma$ is a physical covariance matrix
if and only if it satisfies the Robertson-Schrödinger uncertainty relation
\begin{equation}
  \sigma+\frac{i}{2}\Omega_N\succeq0,
  \label{eq:RS}
\end{equation}

A Gaussian state is faithful, equivalently, has full rank, if and only if all
its symplectic eigenvalues satisfy
\begin{equation}
  \nu_j>\frac12,
  \qquad j=1,\ldots,N.\nonumber
\end{equation}
Throughout this work, the reference state is assumed to be faithful. Pure
reference states are therefore excluded from the statements of the main
results. Whenever a boundary value is considered, it is understood as a limit
of faithful states, provided that the corresponding limit exists.

\subsection{Covariant general-dyne measurements}

Let $\rho_m$ be a centered $N$-mode Gaussian state that serves as the
\emph{measurement seed} of the covariant general-dyne POVM defined below \cite{Turner2026}. Its
first moments vanish,
\begin{equation}
  \Tr(\rho_m\hat{\bm R})=0,\nonumber
\end{equation}
and its covariance matrix $\sigma_m$, hereafter referred to as the
\emph{seed covariance matrix}, satisfies
\begin{equation}
  \sigma_m+\frac{i}{2}\Omega_N\succeq0.
  \label{eq:seedPhysicality}
\end{equation}
The seed state $\rho_m$ need not be faithful. In particular,
$\sigma_m=\I_{2N}/2$ is the covariance matrix of the pure multimode vacuum
seed, which generates the standard multimode heterodyne POVM.

We define the Weyl displacement operators by
\begin{equation}
  D(\bm r)
  :=
  \exp\!\left(-i\bm r^T\Omega_N\hat{\bm R}\right),
  \qquad
  \bm r\in\R^{2N},\nonumber
\end{equation}
so that
\begin{equation}
  D(\bm r)^\dagger\hat{\bm R}D(\bm r)
  =
  \hat{\bm R}+\bm r.\nonumber
\end{equation}
The covariant Gaussian phase-space POVM generated by $\rho_m$ is
\begin{equation}
  \dd G(\bm r)
  =
  \frac{\dd^{2N}\bm r}{(2\pi)^N}
  D(\bm r)\rho_mD(\bm r)^\dagger,
  \qquad
  \int_{\R^{2N}}\dd G(\bm r)=\I,
  \label{eq:POVM}
\end{equation}
where $\dd^{2N}\bm r$ denotes Lebesgue measure on the outcome space
$\R^{2N}$.

For an input Gaussian state with first-moment vector $\bm\varepsilon$ and
covariance matrix $\sigma$, the probability density of the measurement
outcomes is
\begin{equation}
  p(\bm r)
  =
  \frac{\Tr[\rho\,\dd G(\bm r)]}{\dd^{2N}\bm r}
  =
  \frac{
    \exp\!\left[
      -\frac12
      (\bm r-\bm\varepsilon)^T
      \tau^{-1}
      (\bm r-\bm\varepsilon)
    \right]
  }{
    (2\pi)^N\sqrt{\det\tau}
  },
  \label{eq:outcomeDistribution}
\end{equation}
with effective outcome covariance
\begin{equation}
  \tau=\sigma+\sigma_m.
  \label{eq:tau}
\end{equation}
Thus, $\sigma_m$ is the covariance of the Gaussian noise added by the
measurement. Equation \eqref{eq:tau} follows from the displacement convention
specified above. Alternative phase-space conventions may represent the seed by
a reflected or transposed covariance matrix, but they lead to the same physical
measurement when used consistently. Throughout this work, a measurement design means a single covariant general-dyne POVM of the form \eqref{eq:POVM}, specified by one fixed finite-covariance Gaussian seed at the reference point. The optimization over $\sigma_m$ therefore does not include randomized or labelled schedules of
different measurement settings, even when the setting label is retained in the recorded data. Such schedules define a larger measurement class and a different optimization problem.

A covariant phase-space measurement is informationally complete if and only if
the zero set of the Fourier-Weyl transform of its seed has dense complement
in phase space \cite{Kiukas2012}. For every Gaussian seed with finite
covariance matrix, this transform is a nonvanishing Gaussian function.
Consequently, all measurements defined by Eq. \eqref{eq:POVM} are
informationally complete, including those generated by pure Gaussian seeds such
as the vacuum seed used in heterodyne detection. More general criteria for the
informational completeness of Gaussian observables were established
in Ref. \cite{KiukasSchultz2013}.

Ideal homodyne detection corresponds instead to the infinite-squeezing limit
of general-dyne measurements \cite{Genoni2016}; equivalently, its
POVM elements are projectors onto quadrature eigenstates, formally regarded as
infinitely squeezed states \cite{Weedbrook2012}.
It is therefore not represented within the covariant phase-space class by a
trace-class Gaussian seed with a finite covariance matrix and is included here
only as a singular limit \cite{Kiukas2012,Genoni2016}.

\section{Local quantum estimation on the Gaussian tangent space}
\label{sec:gaussianTangentEstimation}

\subsection{Local quantum estimation theory}
Let $\theta\mapsto\rho_\theta$ be a differentiable one-parameter quantum
statistical model with faithful reference state $\rho_0=\rho$ \cite{Paris2009, Helstrom1976, Souza2024, Souza2015GaussianInterferometricPower, holevo2003statistical, ballester2005estimation}. Its symmetric
logarithmic derivative (SLD) $L$ is defined by
\begin{equation}
  \partial_\theta\rho_\theta
  =\frac12\{\rho,L\}.\nonumber
\end{equation}
The quantum Fisher information is
\begin{equation}
  I_{\mathrm Q}=\Tr(\rho L^2).\nonumber
\end{equation}
For a fixed POVM with outcome density $p_\theta(y)$, the corresponding
classical Fisher information is
\begin{equation}
  I_{\mathrm C}
  =\int \frac{[\partial_\theta p_\theta(y)]^2}{p_\theta(y)}\,\dd y.\nonumber
\end{equation}
For every measurement, $I_{\mathrm C}\le I_{\mathrm Q}$, and the two
quantities determine the classical and quantum Cram\'er-Rao bounds,
respectively \cite{BraunsteinCaves1994}. The comparison made below is
directional: each Gaussian tangent vector defines a local one-parameter
submodel. It does not describe the simultaneous attainability of a genuinely
multiparameter bound, for which compatibility conditions or the Holevo
bound must be considered \cite{Liu2020,Nichols2018}.

\subsection{Gaussian tangent space}

Let $\theta\mapsto\rho_\theta$ be a differentiable one-parameter family of
Gaussian states defined in a neighborhood of $\theta=0$, with
$\rho_0=\rho$, where $\rho$ is faithful. Since a Gaussian state is completely
determined by its first-moment vector and covariance matrix
 \cite{Miller2025, Weedbrook2012,LinkStrunz2015}, the tangent vector to this family at
$\theta=0$ is specified by
\begin{align}
  \bm u
  &:=
  \left.
    \frac{\dd\bm\varepsilon_\theta}{\dd\theta}
  \right|_{\theta=0}
  \in\R^{2N},\nonumber\\
  X
  &:=
  \left.
    \frac{\dd\sigma_\theta}{\dd\theta}
  \right|_{\theta=0}
  \in\Sym(2N,\R).\nonumber
\end{align}
where 
\begin{equation}
  \Sym(2N,\R)
  :=
  \left\{
    X\in\R^{2N\times2N}:X^T=X
  \right\}.\nonumber
\end{equation}

The Gaussian tangent space at a faithful state can therefore be identified
with
\begin{equation}
  \mathcal T_G
  \cong
  \R^{2N}\oplus\Sym(2N,\R).\nonumber
\end{equation}
Since
\begin{equation}
  \dim\Sym(2N,\R)
  =
  \frac{(2N)(2N+1)}{2}
  =
  2N^2+N,\nonumber
\end{equation}
its dimension is
\begin{equation}
  \dim\mathcal T_G
  =
  2N+(2N^2+N)
  =
  2N^2+3N.
  \nonumber
\end{equation}

The identification above relies on the faithfulness assumption. Indeed,
$\nu_j>1/2$ for every $j$ is equivalent to
\begin{equation}
  \sigma+\frac{i}{2}\Omega_N\succ0.\nonumber
\end{equation}
Because positive definiteness is an open condition, for every
$X\in\Sym(2N,\R)$ there exists $\delta>0$ such that
\begin{equation}
  \sigma+\theta X+\frac{i}{2}\Omega_N\succeq0
  \qquad
  \text{for all }|\theta|<\delta.\nonumber
\end{equation}
Thus, every pair $(\bm u,X)\in
\R^{2N}\oplus\Sym(2N,\R)$ is realized by the local Gaussian family
\begin{equation}
  \bm\varepsilon_\theta
  =
  \bm\varepsilon+\theta\bm u,
  \qquad
  \sigma_\theta
  =
  \sigma+\theta X,\nonumber
\end{equation}
for sufficiently small $|\theta|$. At the boundary of the physical covariance
cone, including pure Gaussian states, this identification requires additional
qualifications and is not assumed in the present work.

\subsection{Fisher information on the Gaussian tangent space}

Following Monras \cite{Monras2013}, we introduce the linear superoperator
\begin{align}
  \mathcal D_\sigma&:\Sym(2N,\R)\longrightarrow\Sym(2N,\R),\nonumber\\
  \mathcal D_\sigma(Y)&:=2\sigma Y\sigma+\frac12\Omega_NY\Omega_N.
  \label{eq:Dsigma}
\end{align}
To relate Eq. \eqref{eq:Dsigma} to the conventions of
Ref. \cite{Monras2013}, recall that Monras uses the covariance matrix
$\Gamma=2\sigma$ and defines
\begin{equation}
  \mathcal D_\Gamma(Y)
  :=
  \Gamma Y\Gamma^T-\Omega_NY\Omega_N^T.\nonumber
\end{equation}
Since $\sigma^T=\sigma$ and $\Omega_N^T=-\Omega_N$, it follows that
\begin{align}
  \mathcal D_\Gamma(Y)
  &=
  4\sigma Y\sigma+\Omega_NY\Omega_N \nonumber \\
  &=
  2\mathcal D_\sigma(Y).
  \nonumber
\end{align}
Thus,
\begin{equation}
  \mathcal D_\Gamma^{-1}
  =
  \frac12\mathcal D_\sigma^{-1}
  \nonumber
\end{equation}
whenever these inverses exist.

For a faithful Gaussian state, $\mathcal D_\sigma$ is a
self-adjoint, positive-definite, and therefore invertible linear operator on
$\Sym(2N,\R)$ as proved in Lemma \ref{lem:positiveExtension}
of Appendix \ref{app:productSeedAuxiliaries},
with respect to the Frobenius inner product
\begin{equation}
  \langle X,Y\rangle_{\mathrm F}:=\Tr(XY).\nonumber
\end{equation}
Using $\Gamma=2\sigma$, $\dot\Gamma=2X$, and
$\dot{\bm\varepsilon}=\bm u$ in the QFI formula of
Ref. \cite{Monras2013}, one obtains
\begin{align}
  \IQ(\bm u,X)
  &=
  \frac12
  \Tr\!\left[
    \dot\Gamma\,
    \mathcal D_\Gamma^{-1}(\dot\Gamma)
  \right]
  +
  2\bm u^T\Gamma^{-1}\bm u\nonumber\\
  &=
  \Tr\!\left[
    X\mathcal D_\sigma^{-1}(X)
  \right]
  +
  \bm u^T\sigma^{-1}\bm u.
\end{align}
Hence, in the conventions adopted in this work,
\begin{equation}
  {
  \IQ(\bm u,X)
  =
  \bm u^T\sigma^{-1}\bm u
  +
  \Tr\!\left[
    X\mathcal D_\sigma^{-1}(X)
  \right].}
  \label{eq:QFI}
\end{equation}

For a fixed measurement seed, the outcome distribution
\eqref{eq:outcomeDistribution} has mean derivative $\bm u$ and covariance
derivative $X$, since
\begin{equation}
  \dot{\bm\varepsilon}=\bm u,
  \qquad
  \dot\tau=\dot\sigma=X.\nonumber
\end{equation}
The standard Fisher-information formula for a multivariate normal
distribution therefore gives
\begin{equation}
  {
  \IC(\bm u,X)
  =
  \bm u^T\tau^{-1}\bm u
  +
  \frac12
  \Tr\!\left[
    \tau^{-1}X\tau^{-1}X
  \right].}
  \label{eq:CFI}
\end{equation}
Equivalently, the covariance contribution can be written as
$\frac12\Tr[(\tau^{-1}X)^2]$. The factor $1/2$ is the standard factor in
the covariance contribution to the Fisher information of a multivariate
normal distribution. This Gaussian-output Fisher-information formula is also
used in the optimization of Gaussian measurements for thermometry; see, for
example, Ref. \cite{Cenni2022}.

Equations \eqref{eq:QFI} and \eqref{eq:CFI} contain no mixed term between
$\bm u$ and $X$. Consequently, the displacement and covariance sectors are
orthogonal with respect to both Fisher bilinear forms. In the classical case,
the displacement score is linear in the centered outcome variables, whereas
the covariance score is a centered quadratic polynomial. Their product has
vanishing expectation because all odd-order centered moments of a Gaussian
distribution vanish. In the quantum case, the same orthogonal decomposition
follows directly from Eq. \eqref{eq:QFI}.

\section{Gaussian Fisher-efficiency operator}
\label{sec:fisherEfficiencyOperator}

\subsection{Definition from the Fisher bilinear forms}

Let
\begin{equation}
  \xi=(\bm u,X),
  \qquad
  \zeta=(\bm v,Y)
  \qquad
  \text{belong to }\mathcal T_G.\nonumber
\end{equation}
The quantum and classical Fisher quadratic forms
\eqref{eq:QFI} and \eqref{eq:CFI} determine the real symmetric bilinear forms
\begin{align}
  G_Q(\xi,\zeta)
  &:=
  \bm u^T\sigma^{-1}\bm v
  +
  \Tr\!\left[
    X\mathcal D_\sigma^{-1}(Y)
  \right],
  \label{eq:GQbilinear}\\
  G_C(\xi,\zeta)
  &:=
  \bm u^T\tau^{-1}\bm v
  +
  \frac12
  \Tr\!\left[
    \tau^{-1}X\tau^{-1}Y
  \right],
  \label{eq:GCbilinear}
\end{align}
where $\tau=\sigma+\sigma_m$ and $\mathcal D_\sigma$ is defined in
Eq. \eqref{eq:Dsigma}. These expressions are the bilinear extensions of the
Gaussian QFI and of the Fisher information of the Gaussian outcome
distribution, respectively \cite{Monras2013,Cenni2022}. In particular,
\begin{equation}
  \IQ(\xi)=G_Q(\xi,\xi),
  \qquad
  \IC(\xi)=G_C(\xi,\xi).
  \label{eq:FisherQuadraticForms}
\end{equation}

\begin{proposition}[Positive definiteness of the Gaussian Fisher forms]
\label{prop:strictPositivity}
Let $\rho$ be a faithful Gaussian state, and let $\sigma_m$ be the covariance
matrix of a physical Gaussian measurement seed with finite covariance. Then
$G_Q$ and $G_C$ are positive definite on $\mathcal T_G$. In particular,
\begin{equation}
  \IQ(\xi)>0,
  \qquad
  \IC(\xi)>0
  \qquad
  \text{for every }\xi\neq0.\nonumber
\end{equation}
\end{proposition}

\begin{proof}
The manifold structure of faithful Gaussian states and its tangent-space
description in terms of infinitesimal variations of the first and second
moments are discussed in Refs. \cite{Miller2025, LinkStrunz2015}. For a faithful Gaussian state, the QFI formula \eqref{eq:QFI}, derived in
Ref. \cite{Monras2013}, is strictly positive for every nonzero
$\xi\in\mathcal T_G$ and therefore defines a positive-definite quadratic form on $\mathcal T_G$. Hence,
\begin{equation}
  G_Q(\xi,\xi)>0
  \qquad
  \text{for every }\xi\neq0.\nonumber
\end{equation}

For the classical form, the physical covariance matrices $\sigma$ and
$\sigma_m$ are real, symmetric, and positive definite. Therefore,
\begin{equation}
  \tau=\sigma+\sigma_m\succ0.\nonumber
\end{equation}
For $\xi=(\bm u,X)$, Eq. \eqref{eq:CFI} can then be written as
\begin{equation}
  G_C(\xi,\xi)
  =
  \left\|
    \tau^{-1/2}\bm u
  \right\|_2^2
  +
  \frac12
  \left\|
    \tau^{-1/2}X\tau^{-1/2}
  \right\|_{\mathrm F}^2.
  \label{eq:GCpositive}
\end{equation}
Both terms are nonnegative. Since $\tau^{-1/2}$ is invertible, the first term
vanishes if and only if $\bm u=0$, while the second vanishes if and only if
$X=0$. Consequently,
\begin{equation}
  G_C(\xi,\xi)=0
  \quad\Longleftrightarrow\quad
  \xi=0.\nonumber
\end{equation}
Thus, $G_C$ is positive definite on $\mathcal T_G$.
\end{proof}

For each fixed $\zeta\in\mathcal T_G$, the map
\begin{equation}
  \xi\longmapsto G_C(\xi,\zeta)\nonumber
\end{equation}
is a linear functional on the finite-dimensional real inner-product space
$(\mathcal T_G,G_Q)$. By the finite-dimensional Riesz representation
theorem \cite{Conway1990}, there exists a unique vector
$\FG\zeta\in\mathcal T_G$ such that
\begin{equation}
  G_C(\xi,\zeta)
  =
  G_Q(\xi,\FG\zeta)\label{eq:RieszFG}
\end{equation}
for every $\xi\in\mathcal T_G$.
The bilinearity of $G_C$ and the uniqueness of the Riesz representation imply that the map $\zeta\mapsto\FG\zeta$ is linear.

\begin{definition}[Gaussian Fisher-efficiency operator]
The unique linear operator
\begin{equation}
  \FG:\mathcal T_G\longrightarrow\mathcal T_G\nonumber
\end{equation}
satisfying
\begin{equation}
  {
  G_C(\xi,\zeta)
  =
  G_Q(\xi,\FG\zeta)
  \qquad
  \text{for all }\xi,\zeta\in\mathcal T_G}
  \label{eq:FGdefinition}
\end{equation}
is called the \emph{Gaussian Fisher-efficiency operator}.
\end{definition}

At the abstract level, our Fisher-efficiency operator is the Gaussian-model realization of the information map defined by Ref. \cite{Tsang2026Semiparametric}. While the Gaussian example considered there varies only the first moments at fixed covariance, the present construction includes both displacement and covariance tangent directions and optimizes over physical general-dyne seeds.

The symmetry of $G_C$ implies that $\FG$ is self-adjoint with respect to $G_Q$. Indeed,
\begin{align}
  G_Q(\xi,\FG\zeta)
  &=
  G_C(\xi,\zeta)\nonumber\\
  &=
  G_C(\zeta,\xi)\nonumber\\
  &=
  G_Q(\zeta,\FG\xi)\nonumber\\
  &=
  G_Q(\FG\xi,\zeta).\nonumber
\end{align}

To express the operator in coordinates, choose any basis of $\mathcal T_G$
and let $\mathbf G_Q$ and $\mathbf G_C$ denote the Gram matrices of $G_Q$
and $G_C$ in that basis. If $\mathbf F_G$ is the matrix representing $\FG$,
then Eq. \eqref{eq:FGdefinition} gives
\begin{equation}
  \mathbf G_C
  =
  \mathbf G_Q\mathbf F_G,\nonumber
\end{equation}
and hence
\begin{equation}
  {
  \mathbf F_G
  =
  \mathbf G_Q^{-1}\mathbf G_C.}
  \label{eq:FGmatrix}
\end{equation}

Although $\mathbf F_G$ need not be symmetric with respect to the Euclidean
inner product, it is self-adjoint with respect to the QFI inner product and is similar to the real symmetric matrix
\begin{equation}
  \widehat{\mathbf F}_G
  :=
  \mathbf G_Q^{-1/2}
  \mathbf G_C
  \mathbf G_Q^{-1/2}.
  \label{eq:FGsymmetric}
\end{equation}
More precisely,
\begin{equation}
  \mathbf F_G
  =
  \mathbf G_Q^{-1/2}
  \widehat{\mathbf F}_G
  \mathbf G_Q^{1/2}.
  \label{eq:FGsimilarity}
\end{equation}
Consequently, $\FG$ is diagonalizable and has a real spectrum by the
finite-dimensional spectral theorem \cite{HornJohnson2013}.

For every nonzero Gaussian tangent direction $\xi$, the defining identity
\eqref{eq:FGdefinition} gives
\begin{equation}
  \frac{\IC(\xi)}{\IQ(\xi)}
  =
  \frac{G_C(\xi,\xi)}{G_Q(\xi,\xi)}
  =
  \frac{G_Q(\xi,\FG\xi)}{G_Q(\xi,\xi)}.
  \label{eq:FGRayleigh}
\end{equation}
Thus, the ratio $\IC(\xi)/\IQ(\xi)$ is the Rayleigh quotient of $\FG$ with
respect to the QFI-induced inner product. Its minimum and maximum over nonzero Gaussian tangent directions are the smallest and largest eigenvalues of $\FG$, respectively, by the
Rayleigh-Ritz principle \cite{HornJohnson2013}.

\subsection{Representation in terms of first and second moments}

For later use, we denote the restrictions of the QFI inner product to the
displacement and covariance sectors by
\begin{align}
  \langle\bm u,\bm v\rangle_{Q,1}
  &:=
  \bm u^T\sigma^{-1}\bm v,
  \label{eq:QinnerDisplacement}\\
  \langle X,Y\rangle_{Q,2}
  &:=
  \Tr\!\left[
    X\mathcal D_\sigma^{-1}(Y)
  \right].
  \label{eq:QinnerCovariance}
\end{align}
These expressions follow directly from the Gaussian QFI formula in the
conventions adopted here \cite{Monras2013}.

\begin{proposition}[Explicit first- and second-moment representation]
\label{prop:explicitFG}
Let $\rho$ be a faithful $N$-mode Gaussian state with covariance matrix
$\sigma$, let $\sigma_m$ be a physical Gaussian measurement-seed covariance,
and set
\begin{equation}
  \tau:=\sigma+\sigma_m.\nonumber
\end{equation}
Define
\begin{align}
  \mathcal N_\tau&:
  \Sym(2N,\R)\longrightarrow\Sym(2N,\R),\nonumber\\
  \mathcal N_\tau(Y)&:=\tau^{-1}Y\tau^{-1}.
  \label{eq:Ntau}
\end{align}
With respect to the decomposition
\begin{equation}
  \mathcal T_G
  =
  \R^{2N}\oplus\Sym(2N,\R)\nonumber
\end{equation}
into displacement and covariance directions, the Gaussian Fisher-efficiency
operator is block diagonal:
\begin{equation}
  \FG
  =
  F_G^{(1)}\oplus F_G^{(2)}.
  \label{eq:FGdirectSum}
\end{equation}
The displacement and covariance blocks are, respectively,
\begin{align}
  F_G^{(1)}
  &=
  \sigma\tau^{-1}
  :
  \R^{2N}\longrightarrow\R^{2N},
  \label{eq:FG1}\\
  F_G^{(2)}
  &=
  \frac12
  \mathcal D_\sigma\circ\mathcal N_\tau
  :
  \Sym(2N,\R)\longrightarrow\Sym(2N,\R).
  \label{eq:FG2}
\end{align}
Each block is self-adjoint with respect to the corresponding QFI-induced
inner product. Moreover,
\begin{equation}
  \IC(\bm u,X)
  =
  \left\langle
    \bm u,F_G^{(1)}\bm u
  \right\rangle_{Q,1}
  +
  \left\langle
    X,F_G^{(2)}X
  \right\rangle_{Q,2}.
  \label{eq:ICoperator}
\end{equation}
\end{proposition}

\begin{proof}
The quantum and classical Fisher forms contain no mixed terms between
displacement and covariance perturbations. Hence, the displacement and
covariance sectors are mutually orthogonal with respect to both $G_Q$ and
$G_C$.

We first show that $\FG$ preserves the displacement sector. Let
$\zeta=(\bm v,0)$ be a displacement direction and let $\xi=(0,X)$ be an
arbitrary covariance direction. From the defining relation
\eqref{eq:FGdefinition}
\begin{equation}
  G_Q\!\left(
    (0,X),\FG(\bm v,0)
  \right)
  =
  G_C\!\left(
    (0,X),(\bm v,0)
  \right)
  =
  0.\nonumber
\end{equation}
Write
\begin{equation}
  \FG(\bm v,0)=(\bm w,Y).\nonumber
\end{equation}
The orthogonality of the two sectors then gives
\begin{align}
  G_Q\bigl((0,X),\FG(\bm v,0)\bigr)
  &=G_Q^{(2)}(X,Y)=0,\nonumber\\[-1mm]
  &\hspace{-7mm}\text{for every }X\in\Sym(2N,\R).\nonumber
\end{align}
Since $G_Q$ is positive definite on $\mathcal T_G$, its restriction
$G_Q^{(2)}$ to the covariance sector is also positive definite and hence
nondegenerate. Therefore, $Y=0$, and the covariance component of
$\FG(\bm v,0)$ vanishes.

Similarly, let $\zeta=(0,Y)$ be a covariance direction and let
$\xi=(\bm u,0)$ be an arbitrary displacement direction. Write
\begin{equation}
  \FG(0,Y)=(\bm w,Z).\nonumber
\end{equation}
Equation \eqref{eq:FGdefinition} gives
\begin{align}
  0
  &=G_C\bigl((\bm u,0),(0,Y)\bigr)\nonumber\\
  &=G_Q\bigl((\bm u,0),\FG(0,Y)\bigr)\nonumber\\
  &=G_Q^{(1)}(\bm u,\bm w)\nonumber
\end{align}
for every $\bm u\in\R^{2N}$. Since $G_Q^{(1)}$ is nondegenerate, it follows
that $\bm w=0$. Thus, $\FG$ also preserves the covariance sector and is
therefore block diagonal.

For displacement directions $\bm u,\bm v\in\R^{2N}$,
Eq. \eqref{eq:FGdefinition} becomes
\begin{equation}
  \bm u^T\tau^{-1}\bm v
  =
  \bm u^T\sigma^{-1}
  F_G^{(1)}\bm v
  \qquad
  \text{for all }\bm u,\bm v\in\R^{2N}.\nonumber
\end{equation}
Because this identity holds for every $\bm u$ and $\bm v$,
\begin{equation}
  \sigma^{-1}F_G^{(1)}
  =
  \tau^{-1}.\nonumber
\end{equation}
Multiplying from the left by $\sigma$ yields
\begin{equation}
  F_G^{(1)}
  =
  \sigma\tau^{-1}.\nonumber
\end{equation}

For covariance directions $X,Y\in\Sym(2N,\R)$, cyclicity of the trace gives
\begin{align}
  G_C^{(2)}(X,Y)
  &=
  \frac12
  \Tr\!\left[
    \tau^{-1}X\tau^{-1}Y
  \right]\nonumber\\
  &=
  \frac12
  \Tr\!\left[
    X\tau^{-1}Y\tau^{-1}
  \right]\nonumber\\
  &=
  \frac12
  \Tr\!\left[
    X\mathcal N_\tau(Y)
  \right].\nonumber
\end{align}
The defining relation for the covariance block is therefore
\begin{equation}
  \Tr\!\left[
    X\mathcal D_\sigma^{-1}
    \!\left(
      F_G^{(2)}Y
    \right)
  \right]
  =
  \frac12
  \Tr\!\left[
    X\mathcal N_\tau(Y)
  \right]\nonumber
\end{equation}
for every $X,Y\in\Sym(2N,\R)$. Equivalently,
\begin{equation}
  \Tr\!\left[
    X
    \left(
      \mathcal D_\sigma^{-1}
      \!\left(
        F_G^{(2)}Y
      \right)
      -
      \frac12\mathcal N_\tau(Y)
    \right)
  \right]
  =
  0\nonumber
\end{equation}
for every $X\in\Sym(2N,\R)$.

Both $\mathcal D_\sigma^{-1}(F_G^{(2)}Y)$ and
$\mathcal N_\tau(Y)=\tau^{-1}Y\tau^{-1}$ are symmetric matrices. Hence, the
matrix in parentheses belongs to $\Sym(2N,\R)$. Since the trace pairing
\begin{equation}
  (X,Z)\longmapsto\Tr(XZ)\nonumber
\end{equation}
is nondegenerate on $\Sym(2N,\R)$, it follows that
\begin{equation}
  \mathcal D_\sigma^{-1}
  \!\left(
    F_G^{(2)}Y
  \right)
  =
  \frac12\mathcal N_\tau(Y).\nonumber
\end{equation}
Because $\rho$ is faithful, $\mathcal D_\sigma$ is invertible on
$\Sym(2N,\R)$. Applying it to both sides gives
\begin{equation}
  F_G^{(2)}Y
  =
  \frac12
  \mathcal D_\sigma
  \!\left(
    \mathcal N_\tau(Y)
  \right),\nonumber
\end{equation}
and therefore
\begin{equation}
  F_G^{(2)}
  =
  \frac12
  \mathcal D_\sigma\circ\mathcal N_\tau.\nonumber
\end{equation}

The symmetry of $G_C$ and Eq. \eqref{eq:FGdefinition} imply that $\FG$ is
self-adjoint with respect to $G_Q$. Since $\FG$ preserves the displacement
and covariance sectors, its two restrictions are self-adjoint with respect
to the corresponding restricted QFI-induced inner products.

Finally, setting
\begin{equation}
  \xi=\zeta=(\bm u,X)\nonumber
\end{equation}
in Eq. \eqref{eq:FGdefinition} gives
\begin{align}
  \IC(\bm u,X)
  &=
  G_C\bigl((\bm u,X),(\bm u,X)\bigr)\nonumber\\
  &=
  G_Q\!\left(
    (\bm u,X),
    \FG(\bm u,X)
  \right)\nonumber\\
  &=
  \left\langle
    \bm u,F_G^{(1)}\bm u
  \right\rangle_{Q,1}
  +
  \left\langle
    X,F_G^{(2)}X
  \right\rangle_{Q,2},\nonumber
\end{align}
which proves Eq. \eqref{eq:ICoperator}.
\end{proof}

The matrix $F_G^{(1)}=\sigma\tau^{-1}$ need not be symmetric with respect to
the Euclidean inner product. This does not conflict with
Proposition \ref{prop:explicitFG}: the relevant notion of self-adjointness is
defined by the QFI-induced inner product
$\langle\cdot,\cdot\rangle_{Q,1}$, not by the Euclidean one.

\subsection{Spectral bounds and operational interpretation}

\begin{theorem}[Spectral bounds for Gaussian Fisher efficiency]
\label{thm:GaussianSpectralBounds}
Let $\rho$ be a faithful Gaussian state, and let the measurement be a
covariant general-dyne POVM generated by a physical Gaussian seed with finite
covariance matrix. Then, for every nonzero Gaussian tangent direction
$\xi\in\mathcal T_G$,
\begin{equation}
  {
  0
  <
  \lambda_{\min}(\FG)
  \leq
  \frac{\IC(\xi)}{\IQ(\xi)}
  \leq
  \lambda_{\max}(\FG)
  \leq
  1.}
  \label{eq:RayleighBounds}
\end{equation}
The lower and upper Rayleigh bounds are attained precisely on the eigenspaces
of $\FG$ associated with $\lambda_{\min}(\FG)$ and
$\lambda_{\max}(\FG)$, respectively. More generally, if
\begin{equation}
  \FG\xi=\lambda\xi,
  \qquad
  \xi\neq0,\nonumber
\end{equation}
then
\begin{equation}
  \frac{\IC(\xi)}{\IQ(\xi)}
  =
  \lambda.
  \label{eq:eigenvalueEfficiency}
\end{equation}
\end{theorem}

\begin{proof}
Choose a basis of $\mathcal T_G$, and let $\bm x$ be the coordinate vector
representing $\xi$. Equations \eqref{eq:FisherQuadraticForms} and
\eqref{eq:FGmatrix} give
\begin{equation}
  \frac{\IC(\xi)}{\IQ(\xi)}
  =
  \frac{
    \bm x^T\mathbf G_C\bm x
  }{
    \bm x^T\mathbf G_Q\bm x
  }.
  \label{eq:generalizedRayleighQuotient}
\end{equation}
Since both Fisher forms are positive definite,
\begin{equation}
  \mathbf G_Q\succ0,
  \qquad
  \mathbf G_C\succ0.\nonumber
\end{equation}

Introduce
\begin{equation}
  \bm y
  :=
  \mathbf G_Q^{1/2}\bm x.\nonumber
\end{equation}
Using Eq. \eqref{eq:FGsymmetric},
Eq. \eqref{eq:generalizedRayleighQuotient} becomes
\begin{equation}
  \frac{\IC(\xi)}{\IQ(\xi)}
  =
  \frac{
    \bm y^T\widehat{\mathbf F}_G\bm y
  }{
    \bm y^T\bm y
  }.
  \label{eq:ordinaryRayleighQuotient}
\end{equation}
The matrix $\widehat{\mathbf F}_G$ is real, symmetric, and positive definite.
The Rayleigh-Ritz theorem therefore yields
\begin{equation}
  \lambda_{\min}(\widehat{\mathbf F}_G)
  \leq
  \frac{\IC(\xi)}{\IQ(\xi)}
  \leq
  \lambda_{\max}(\widehat{\mathbf F}_G),\nonumber
\end{equation}
with equality precisely on the corresponding extremal
eigenspaces \cite{HornJohnson2013}.

By Eq. \eqref{eq:FGsimilarity}, $\mathbf F_G$ and
$\widehat{\mathbf F}_G$ have the same eigenvalues. Moreover, the change of
variables
\begin{equation}
  \bm x
  =
  \mathbf G_Q^{-1/2}\bm y\nonumber
\end{equation}
maps an eigenvector of $\widehat{\mathbf F}_G$ to an eigenvector of
$\mathbf F_G$ with the same eigenvalue. This proves the two Rayleigh bounds
and their equality conditions.

For any fixed POVM, the classical Fisher information cannot exceed the SLD
quantum Fisher information in any tangent direction \cite{Saini2026, BraunsteinCaves1994}:
\begin{equation}
  \IC(\xi)\leq\IQ(\xi)
  \qquad
  \text{for every }\xi\in\mathcal T_G.
  \label{eq:CFIleqQFI}
\end{equation}
Equivalently,
\begin{equation}
  \mathbf G_C\preceq\mathbf G_Q\nonumber
\end{equation}
in the Löwner order. Applying the congruence transformation defined by
$\mathbf G_Q^{-1/2}$ gives
\begin{equation}
  \widehat{\mathbf F}_G
  =
  \mathbf G_Q^{-1/2}
  \mathbf G_C
  \mathbf G_Q^{-1/2}
  \preceq
  \I.\nonumber
\end{equation}
Therefore,
\begin{equation}
  \lambda_{\max}(\FG)
  =
  \lambda_{\max}(\widehat{\mathbf F}_G)
  \leq1.\nonumber
\end{equation}
Likewise, $\mathbf G_C\succ0$ implies
$\widehat{\mathbf F}_G\succ0$, and hence
\begin{equation}
  \lambda_{\min}(\FG)>0.\nonumber
\end{equation}

Finally, if $\FG\xi=\lambda\xi$, then
Eq. \eqref{eq:FGRayleigh} gives
\begin{equation}
  \frac{\IC(\xi)}{\IQ(\xi)}
  =
  \frac{
    G_Q(\xi,\lambda\xi)
  }{
    G_Q(\xi,\xi)
  }
  =
  \lambda,\nonumber
\end{equation}
which completes the proof.
\end{proof}

\subsubsection{Operational meaning of the worst-direction efficiency}

The spectrum of $\FG$ has a direct directional interpretation. Each
eigenvalue is the fraction of the QFI retained by the chosen measurement along
the corresponding Gaussian eigendirection. In particular,
\begin{equation}
  \lambda_{\min}(\FG)
  =
  \min_{\xi\neq0}
  \frac{\IC(\xi)}{\IQ(\xi)}
  \label{eq:FGminimum}
\end{equation}
is the smallest Fisher efficiency over all nonzero Gaussian tangent
directions.

This minimum defines a natural robust measurement-design criterion. We seek a
single measurement that performs as uniformly as possible over all local
one-parameter Gaussian encodings through the reference state, without
assigning prior weights to particular directions. Equivalently,
$\lambda_{\min}(\FG)$ is the largest constant $\eta$ for which
\begin{equation}
  \IC(\xi)\geq\eta\,\IQ(\xi)
  \qquad
  \text{for every }\xi\in\mathcal T_G.
  \label{eq:uniformEfficiencyBound}
\end{equation}
Because both Fisher informations scale quadratically under
$\xi\mapsto a\xi$, their ratio is dimensionless and invariant under a local
reparametrization of the corresponding one-parameter model. Maximizing
$\lambda_{\min}(\FG)$ over the allowed measurement seeds therefore selects the
measurement with the highest worst-direction efficiency, rather than one that
performs exceptionally well only for selected encodings. If some directions
are known in advance to be more relevant than others, a weighted or averaged
criterion may be more appropriate, but it defines a different
measurement-design problem. 

\subsubsection{Directional criterion and multiparameter scope}

The Fisher-efficiency ratio has a directional, locally one-parameter
interpretation. Every nonzero tangent vector $\xi\in\mathcal T_G$ can be realized as the
derivative at the reference state of a local one-parameter Gaussian submodel.
For such a direction, the ratio$ \frac{I_C(\xi)}{I_Q(\xi)}
$ compares the Fisher information extracted by the fixed measurement with the quantum Fisher information available for the corresponding one-parameter
submodel.

This criterion does not describe the simultaneous estimation of all
coordinates of $\mathcal T_G$. In a genuinely multiparameter problem, incompatibility among the SLDs may
prevent the simultaneous saturation of the SLD quantum Cramér-Rao matrix
bound. Attainable precision must then be assessed through the relevant
compatibility conditions or, more generally, the Holevo bound
\cite{Liu2020,Nichols2018}. Thus, the eigenvalues of $\FG$
should not be interpreted as additive information fractions for the
simultaneous estimation of all Gaussian parameters.

\section{Main results}
\label{sec:mainResults}

\subsection{Williamson covariance and aligned spectra}

Every faithful $N$-mode Gaussian covariance matrix admits a Williamson
decomposition
\begin{equation}
  \sigma
  =
  SD_\nu S^T,
  \qquad
  D_\nu
  :=
  \bigoplus_{j=1}^{N}\nu_j\I_2,
  \qquad
  \nu_j>\frac12,
  \label{eq:generalWilliamson}
\end{equation}
where $S\in \Sp(2N,\R)$ is a real symplectic matrix satisfying
\begin{equation}
  S\Omega_NS^T=\Omega_N,\nonumber
\end{equation}
and $\{\nu_j\}_{j=1}^N$ are the symplectic eigenvalues of
$\sigma$ \cite{Weedbrook2012}. No degeneracy of the symplectic spectrum is
assumed.

We consider a physical Gaussian measurement seed aligned with the same
Williamson basis:
\begin{equation}
  \sigma_m
  =
  SM_\mu S^T,
  \qquad
  M_\mu
  :=
  \bigoplus_{j=1}^{N}\mu_j\I_2,
  \qquad
  \mu_j\geq\frac12.
  \label{eq:alignedSeedGeneral}
\end{equation}
For later convenience, define
\begin{equation}
  t_j:=\nu_j+\mu_j,
  \qquad
  j=1,\ldots,N.
  \label{eq:tjDefinition}
\end{equation}

\begin{lemma}[Covariance under a common symplectic transformation]
\label{lem:commonSymplecticCovariance}
Let $\sigma$ be the covariance matrix of a faithful Gaussian state, let
$\sigma_m$ be an arbitrary physical Gaussian seed covariance matrix, and let
$S\in\Sp(2N,\R)$. Define
\begin{equation}
  \sigma_0
  :=
  S^{-1}\sigma S^{-T},
  \qquad
  \sigma_{m,0}
  :=
  S^{-1}\sigma_mS^{-T}.
  \label{eq:commonSymplecticCovariances}
\end{equation}
Then the Fisher-efficiency operators
\begin{equation}
  \FG(\sigma,\sigma_m)
  \qquad\text{and}\qquad
  \FG(\sigma_0,\sigma_{m,0})\nonumber
\end{equation}
are related by a similarity transformation
induced by an invertible map of the Gaussian tangent space. Therefore, they
have identical spectra.

In particular, if
\begin{equation}
  \sigma=SD_\nu S^T,
  \qquad
  \sigma_m=SM_\mu S^T,\nonumber
\end{equation}
then
\begin{equation}
  \sigma_0=D_\nu,
  \qquad
  \sigma_{m,0}=M_\mu.\nonumber
\end{equation}
\end{lemma}

A detailed proof of Lemma \ref{lem:commonSymplecticCovariance} is given in
Appendix \ref{app:commonSymplecticCovariance}.

\subsection{Decomposition into modal blocks}

In the common Williamson basis \cite{Weedbrook2012},
\begin{equation}
  \sigma_0=D_\nu,
  \qquad
  \sigma_{m,0}=M_\mu,
  \qquad
  \tau_0
  =
  D_\nu+M_\mu
  =
  \bigoplus_{j=1}^{N}t_j\I_2.
  \label{eq:WilliamsonBasisCovariances}
\end{equation}
The displacement block of the Fisher-efficiency operator is therefore
\begin{equation}
  F_G^{(1)}
  =
  D_\nu\tau_0^{-1}
  =
  \bigoplus_{j=1}^{N}
  \frac{\nu_j}{t_j}\I_2.
  \label{eq:FG1Williamson}
\end{equation}
For each mode, define the two-dimensional displacement subspace
\begin{equation}
  V_j
  :=
  \operatorname{span}
  \left\{
    \bm e_{x_j},\bm e_{p_j}
  \right\}.\nonumber
\end{equation}
Equation \eqref{eq:FG1Williamson} shows directly that every $V_j$ is invariant
under $F_G^{(1)}$ and that
\begin{equation}
  \lambda_{\mathrm d,j}
  =
  \frac{\nu_j}{t_j}\nonumber
\end{equation}
is an eigenvalue of multiplicity two.

To analyze the covariance sector, write
$X\in\Sym(2N,\R)$ as an $N\times N$ array of real $2\times2$ blocks:
\begin{equation}
  X=(X_{jk})_{j,k=1}^{N},
  \qquad
  X_{kj}=X_{jk}^T.\nonumber
\end{equation}
Since
\begin{equation}
  [\mathcal N_{\tau_0}(X)]_{jk}
  =
  \frac{X_{jk}}{t_jt_k},\nonumber
\end{equation}
Eq. \eqref{eq:FG2} gives
\begin{equation}
  \left[
    F_G^{(2)}(X)
  \right]_{jk}
  =
  \frac{1}{4t_jt_k}
  \left(
    4\nu_j\nu_kX_{jk}
    +
    JX_{jk}J
  \right).
  \label{eq:FGblockGeneral}
\end{equation}

Introduce the linear map
\begin{equation}
  \mathcal R:
  M_2(\R)\longrightarrow M_2(\R),
  \qquad
  \mathcal R(A):=JAJ.
  \label{eq:Rmap}
\end{equation}
Since $J^2=-\I_2$,
\begin{equation}
  \mathcal R^2(A)
  =
  J(JAJ)J
  =
  J^2AJ^2
  =
  A.\nonumber
\end{equation}
Thus, $\mathcal R$ is an involution and its only possible eigenvalues are
$r=\pm1$.

A convenient basis of $M_2(\R)$ is
\begin{align}
  \I_2&=\begin{pmatrix}1&0\\0&1\end{pmatrix},&
  J&=\begin{pmatrix}0&1\\-1&0\end{pmatrix},\nonumber\\
  Z&=\begin{pmatrix}1&0\\0&-1\end{pmatrix},&
  K&=\begin{pmatrix}0&1\\1&0\end{pmatrix}.\nonumber
\end{align}
The action of $\mathcal R$ on this basis is
\begin{align}
  \mathcal R(\I_2)&=-\I_2,&
  \mathcal R(J)&=-J,\nonumber\\
  \mathcal R(Z)&=Z,&
  \mathcal R(K)&=K.
  \label{eq:Raction}
\end{align}

\begin{theorem}[Spectrum for arbitrary symplectic eigenvalues]
\label{thm:Nmode}
Under the hypotheses of
Eqs. \eqref{eq:generalWilliamson}-\eqref{eq:alignedSeedGeneral}, the complete
spectrum of the Gaussian Fisher-efficiency operator consists of the following
eigenvalues.

For each mode $j=1,\ldots,N$, the displacement sector contributes
\begin{equation}
  \lambda_{\mathrm d,j}
  =
  \frac{\nu_j}{t_j},
  \qquad
  \operatorname{mult}(\lambda_{\mathrm d,j})=2.
  \label{eq:lambdaDj}
\end{equation}
The diagonal covariance block of mode $j$ contributes
\begin{align}
  \lambda_{jj,-}
  &=
  \frac{4\nu_j^2-1}{4t_j^2},
  &
  \operatorname{mult}(\lambda_{jj,-})
  &=1,
  \label{eq:lambdaJJminus}\\
  \lambda_{jj,+}
  &=
  \frac{4\nu_j^2+1}{4t_j^2},
  &
  \operatorname{mult}(\lambda_{jj,+})
  &=2.
  \label{eq:lambdaJJplus}
\end{align}
For each unordered pair of distinct modes $j<k$, the off-diagonal covariance
sector contributes
\begin{align}
  \lambda_{jk,-}
  &=
  \frac{4\nu_j\nu_k-1}{4t_jt_k},
  &
  \operatorname{mult}(\lambda_{jk,-})
  &=2,
  \label{eq:lambdaJKminus}\\
  \lambda_{jk,+}
  &=
  \frac{4\nu_j\nu_k+1}{4t_jt_k},
  &
  \operatorname{mult}(\lambda_{jk,+})
  &=2.
  \label{eq:lambdaJKplus}
\end{align}
The spectrum depends only on the symplectic eigenvalues
$\{\nu_j\}$ and $\{\mu_j\}$ and not on the common symplectic matrix $S$.
\end{theorem}

\begin{proof}
The displacement eigenvalues and their multiplicities follow immediately from
Eq. \eqref{eq:FG1Williamson}. Consider first a diagonal covariance block $j=k$. Since the global covariance
perturbation $X$ is symmetric, its diagonal block
\begin{equation}
  A:=X_{jj}\nonumber
\end{equation}
belongs to $\Sym(2,\R)$, which has basis
$\{\I_2,Z,K\}$. By Eq. \eqref{eq:Raction}, the restriction of
$\mathcal R$ to $\Sym(2,\R)$ has the eigenspace decomposition
\begin{align}
  E_-
  &=
  \operatorname{span}\{\I_2\},
  &
  \dim E_-&=1,\nonumber\\
  E_+
  &=
  \operatorname{span}\{Z,K\},
  &
  \dim E_+&=2.\nonumber
\end{align}
If $\mathcal R(A)=rA$, where $r=\pm1$, then
Eq. \eqref{eq:FGblockGeneral} gives
\begin{equation}
  \left[
    F_G^{(2)}(X)
  \right]_{jj}
  =
  \frac{4\nu_j^2+r}{4t_j^2}A.\nonumber
\end{equation}
The choices $r=-1$ and $r=+1$ therefore yield
$\lambda_{jj,-}$ and $\lambda_{jj,+}$ with multiplicities one and two,
respectively.

Now fix an unordered pair of distinct modes $j<k$. The associated covariance
subspace consists of matrices whose only potentially nonzero blocks are
\begin{equation}
  X_{jk}=A,
  \qquad
  X_{kj}=A^T,
  \qquad
  A\in M_2(\R).
  \label{eq:offDiagonalPair}
\end{equation}
This subspace has dimension four. On $M_2(\R)$, the eigenspaces of
$\mathcal R$ are
\begin{align}
  E_-
  &=
  \operatorname{span}\{\I_2,J\},
  &
  \dim E_-&=2,\nonumber\\
  E_+
  &=
  \operatorname{span}\{Z,K\},
  &
  \dim E_+&=2.\nonumber
\end{align}
Moreover, if $JAJ=rA$, then
\begin{equation}
  JA^TJ
  =
  (JAJ)^T
  =
  rA^T,\nonumber
\end{equation}
so the blocks $A$ and $A^T$ transform with the same eigenvalue. Therefore,
each eigenmatrix of $\mathcal R$ defines a symmetric global covariance
direction in the pair sector \eqref{eq:offDiagonalPair}. From
Eq. \eqref{eq:FGblockGeneral},
\begin{equation}
  \left[
    F_G^{(2)}(X)
  \right]_{jk}
  =
  \frac{4\nu_j\nu_k+r}{4t_jt_k}A.\nonumber
\end{equation}
The eigenspaces with $r=-1$ and $r=+1$ thus yield
$\lambda_{jk,-}$ and $\lambda_{jk,+}$, each with multiplicity two.

Finally, the total multiplicity is
\begin{align}
  2N
  +3N
  +4\binom{N}{2}
  &=
  2N^2+3N\nonumber\\
  &=
  \dim\mathcal T_G.\nonumber
\end{align}
Hence, the listed eigenvalues exhaust the complete spectrum.

The independence from $S$ follows from
Lemma \ref{lem:commonSymplecticCovariance}, which shows that the original and
Williamson-basis Fisher-efficiency operators are similar.
\end{proof}

The factors $4\nu_j\nu_k\pm1$ appearing in the covariance sectors are also present in the Williamson-basis even-odd decomposition of the Gaussian QFI \cite{Chatterjee2026}. A closely related sectorwise CFI-to-QFI construction was recently developed in Ref. \cite{Grzywa2026} using a decomposition into Hermite-Gauss orders. Restricted to local Gaussian perturbations, its first- and second-order sectors correspond, respectively, to the displacement and covariance sectors diagonalized in Theorem \ref{thm:Nmode}. The present theorem provides the explicit Williamson-aligned multimode eigenvalues and multiplicities for arbitrary symplectic spectra and finite aligned general-dyne seeds.

\begin{corollary}[Fully degenerate symplectic spectra]
\label{cor:isossymplecticSpectrum}
Suppose that
\begin{equation}
  \nu_j=\nu>\frac12,
  \qquad
  \mu_j=\nu_m\geq\frac12
  \qquad
  \text{for every }j.\nonumber
\end{equation}
Equivalently,
\begin{align}
  \sigma
  &=
  \nu SS^T,
  \label{eq:isothermalState}\\
  \sigma_m
  &=
  \nu_mSS^T.
  \label{eq:matchedSeed}
\end{align}
Then the eigenvalues in Theorem \ref{thm:Nmode} collapse into three families:
\begin{align}
  \lambda_-
  &=
  \frac{4\nu^2-1}{4(\nu+\nu_m)^2},
  &
  \operatorname{mult}(\lambda_-)
  &=
  N^2,
  \label{eq:lambdamin}\\
  \lambda_+
  &=
  \frac{4\nu^2+1}{4(\nu+\nu_m)^2},
  &
  \operatorname{mult}(\lambda_+)
  &=
  N(N+1),
  \label{eq:lambdaplus}\\
  \lambda_{\mathrm d}
  &=
  \frac{\nu}{\nu+\nu_m},
  &
  \operatorname{mult}(\lambda_{\mathrm d})
  &=
  2N.
  \label{eq:lambdad}
\end{align}
They satisfy the strict ordering
\begin{equation}
  0
  <
  \lambda_-
  <
  \lambda_+
  <
  \lambda_{\mathrm d}
  <
  1.
  \label{eq:ordering}
\end{equation}
\end{corollary}

\begin{proof}
The total multiplicity of the covariance eigenvalue with the negative sign is
\begin{equation}
  N+2\binom{N}{2}=N^2,\nonumber
\end{equation}
whereas the total multiplicity of the covariance eigenvalue with the positive
sign is
\begin{equation}
  2N+2\binom{N}{2}=N(N+1).\nonumber
\end{equation}
The displacement multiplicity is $2N$.

Since $\nu>1/2$ and $\nu_m\geq1/2$,
\begin{align}
  \lambda_+-\lambda_-
  &=
  \frac{1}{2(\nu+\nu_m)^2}
  >0,\nonumber\\
  \lambda_{\mathrm d}-\lambda_+
  &=
  \frac{\nu\nu_m-\frac14}
  {(\nu+\nu_m)^2}
  >0.\nonumber
\end{align}
Furthermore,
\begin{equation}
  \lambda_->0,
  \qquad
  1-\lambda_{\mathrm d}
  =
  \frac{\nu_m}{\nu+\nu_m}
  >0.\nonumber
\end{equation}
These relations establish Eq. \eqref{eq:ordering}.
\end{proof}

\begin{corollary}[Optimal seed within the aligned family]
\label{cor:matchedworstcase}
Among all aligned seeds of the form \eqref{eq:alignedSeedGeneral}, the robust
Fisher efficiency is maximized by the pure matched seed
\begin{equation}
  M_\mu^\star=\frac12\I_{2N},
  \qquad
  \sigma_m^\star=\frac12SS^T.
  \label{eq:optimalAlignedSeed}
\end{equation}
The corresponding optimal value is
\begin{equation}
  {
  \max_{\mu_1,\ldots,\mu_N\geq1/2}
  \lambda_{\min}(\FG)
  =
  \frac{
    \nu_{\min}-\frac12
  }{
    \nu_{\min}+\frac12
  }}.
  \label{eq:optimalWorstCase}
\end{equation}
\end{corollary}

\begin{proof}
Every eigenvalue listed in Theorem \ref{thm:Nmode} is a strictly decreasing
function of each seed eigenvalue $\mu_j$ that appears in its denominator.
Its numerator is independent of the seed, and is strictly positive because
$\nu_j>1/2$. Therefore, the simultaneous choice
\begin{equation}
  \mu_j=\frac12
  \qquad
  \text{for every }j\nonumber
\end{equation}
maximizes every eigenvalue individually and hence maximizes their minimum.

For this seed,
\begin{equation}
  \lambda_{jj,-}
  =
  f(\nu_j),
  \qquad
  f(x)
  :=
  \frac{x-\frac12}{x+\frac12}.
  \label{eq:fDefinition}
\end{equation}
The function $f$ is strictly increasing for $x>1/2$. Moreover, if
$\nu_j\leq\nu_k$, then
\begin{equation}
  \lambda_{jk,-}-f(\nu_j)
  =
  \frac{
    \nu_k-\nu_j
  }{
    2
    \left(\nu_j+\frac12\right)
    \left(\nu_k+\frac12\right)
  }
  \geq0.
  \label{eq:crossMinusComparison}
\end{equation}
Thus, every off-diagonal negative-sign eigenvalue is at least
$f(\nu_{\min})$.

The positive-sign covariance eigenvalues are strictly larger than their
negative-sign counterparts. In addition,
\begin{equation}
  \lambda_{\mathrm d,j}
  -
  \lambda_{jj,-}
  =
  \frac{
    \frac12
  }{
    \nu_j+\frac12
  }
  >0.\nonumber
\end{equation}
It follows that the smallest eigenvalue of the matched-seed operator is
attained in a diagonal negative-sign covariance direction associated with a
mode satisfying $\nu_j=\nu_{\min}$. Therefore,
\begin{equation}
  \lambda_{\min}(\FG)
  =
  f(\nu_{\min})
  =
  \frac{
    \nu_{\min}-\frac12
  }{
    \nu_{\min}+\frac12
  },\nonumber
\end{equation}
which proves Eq. \eqref{eq:optimalWorstCase}.
\end{proof}

\subsubsection{Thermal character of the least-efficient matched direction}
\label{rem:thermalWorstDirection}
For the matched seed, the minimum efficiency is attained in a diagonal
negative-sign covariance direction supported on a mode
$j_\star$ satisfying $\nu_{j_\star}=\nu_{\min}$. In the Williamson basis,
this direction is represented by
\begin{equation}
  \widetilde X_{j_\star}
  =
  0\oplus\cdots\oplus I_2\oplus\cdots\oplus0.\nonumber
\end{equation}
It changes the symplectic eigenvalue $\nu_{j_\star}$ while keeping the
symplectic transformation $S$ fixed and may therefore be interpreted as a
local thermal-occupation direction. In the original phase-space coordinates,
the corresponding perturbation is
\begin{equation}
  X_{j_\star}
  =
  S\widetilde X_{j_\star}S^T.\nonumber
\end{equation}

For a single thermal mode, $\sigma=\nu I_2$, measured with the vacuum
heterodyne seed $\sigma_m=I_2/2$, the thermal perturbation $X=I_2$ gives
\begin{equation}
  I_Q(X)=\frac{1}{\nu^2-\frac14},
  \qquad
  I_C(X)=\frac{1}{\left(\nu+\frac12\right)^2},
  \label{eq:thermalDirectionalFisher}
\end{equation}
and hence
\begin{equation}
  \frac{I_C(X)}{I_Q(X)}
  =
  \frac{\nu-\frac12}{\nu+\frac12}.\nonumber
\end{equation}
Thus, the multimode optimum is controlled by the heterodyne efficiency of
the thermal direction of the mode with the smallest symplectic eigenvalue.
This connects the limiting direction of the present robust optimization with
Gaussian thermometry \cite{Cenni2022}.

\subsubsection{Pure-state boundary and limitation of the unrestricted criterion}
The results above assume $\nu_j>1/2$ for every mode. If
$\nu_{\min}\to1/2^+$, the optimal aligned robust efficiency satisfies
\begin{equation}
  \frac{
    \nu_{\min}-\frac12
  }{
    \nu_{\min}+\frac12
  }
  \longrightarrow0.\nonumber
\end{equation}
This collapse is caused by the thermal direction of the mode with
$\nu_j=\nu_{\min}$. Equation \eqref{eq:thermalDirectionalFisher}, evaluated
at $\nu=\nu_{\min}$, shows that the directional QFI diverges as the
pure-state boundary is approached, whereas the CFI of the matched
finite-covariance general-dyne measurement remains finite. The vanishing
worst-direction ratio is therefore a genuine
limitation of taking the minimum over the full tangent space of faithful mixed
Gaussian states: it is dominated by a rank-changing thermal perturbation that
does not remain a two-sided tangent direction of the pure-state manifold.

At the boundary, $\mathcal D_\sigma$ becomes singular in precisely such
directions, and the faithful-state tangent space is no longer the appropriate
tangent space for a pure Gaussian model. The limit above consequently does
not imply that every physical tangent direction within the pure-state
manifold has zero classical Fisher information. Applications that protect
only a specified subspace, for example, fixed-symplectic-spectrum
perturbations, require the corresponding restricted or weighted efficiency
criterion. Pure reference states themselves require a restriction to their
physical tangent directions, or a separately defined limiting or
pseudoinverse treatment.

\subsection{Single-mode illustration}

For $N=1$, the covariance sector
$\Sym(2,\R)$ has dimension three, while the full Gaussian tangent space has
dimension five. Writing
\begin{equation}
  \nu_1=\nu,
  \qquad
  \mu_1=\nu_m,
  \qquad
  t=\nu+\nu_m,\nonumber
\end{equation}
Theorem \ref{thm:Nmode} gives
\begin{align}
  \lambda_-
  &=
  \frac{4\nu^2-1}{4t^2},
  &
  \operatorname{mult}(\lambda_-)
  &=1,\nonumber\\
  \lambda_+
  &=
  \frac{4\nu^2+1}{4t^2},
  &
  \operatorname{mult}(\lambda_+)
  &=2,\nonumber\\
  \lambda_{\mathrm d}
  &=
  \frac{\nu}{t},
  &
  \operatorname{mult}(\lambda_{\mathrm d})
  &=2.\nonumber
\end{align}

The one-dimensional eigenspace associated with $\lambda_-$ is the isotropic
covariance direction $X\propto\I_2.$ This direction changes the overall covariance scale and, for a thermal state, corresponds to an infinitesimal change in its thermal occupation.

The two-dimensional eigenspace associated with $\lambda_+$ is the traceless
symmetric covariance sector
\begin{equation}
  \operatorname{span}
  \left\{
    \begin{pmatrix}
      1&0\\
      0&-1
    \end{pmatrix},
    \begin{pmatrix}
      0&1\\
      1&0
    \end{pmatrix}
  \right\}.\nonumber
\end{equation}
These two directions describe infinitesimal changes in the magnitude and
orientation of the covariance ellipse. The remaining two-dimensional
eigenspace, associated with $\lambda_{\mathrm d}$, consists of displacement
perturbations.

As a simple numerical illustration, consider $\nu=1,
  \qquad
  \nu_m=\frac12.$ The value $\nu=1$ describes a faithful thermal state with mean occupation
number $ \bar n
  =
  \nu-\frac12
  =
  \frac12, $ while $\nu_m=1/2$ corresponds to a pure vacuum seed, namely, the smallest
isotropic seed covariance allowed by the uncertainty relation. The choice
$\nu=1$ is made only for illustration and is not required by the preceding
results. In this case,
\begin{equation}
  \lambda_-=\frac13,
  \qquad
  \lambda_+=\frac59,
  \qquad
  \lambda_{\mathrm d}=\frac23.
\end{equation}

\begin{theorem}[Globally optimal single-mode seed]
\label{thm:onemodeglobal}
Let
\begin{equation}
  \sigma=\nu SS^T,
  \qquad
  \nu>\frac12,
  \qquad
  S\in\Sp(2,\R).\nonumber
\end{equation}
Among all physical single-mode Gaussian seeds with finite covariance matrix,
the robust Fisher efficiency is uniquely maximized by
\begin{equation}
  \sigma_m^\star
  =
  \frac12SS^T.
  \label{eq:globalSeed1mode}
\end{equation}
The optimal value is
\begin{equation}
  \max_{\sigma_m\ \mathrm{physical}}
  \lambda_{\min}
  \!\left[
    \FG(\sigma,\sigma_m)
  \right]
  =
  \frac{\nu-\frac12}{\nu+\frac12}.
  \label{eq:globalOneModeValue}
\end{equation}
\end{theorem}

A detailed proof, including uniqueness, is given in
Appendix \ref{app:singleModeSchur}.

\subsubsection{Physical interpretation of the single-mode spectrum}
For every faithful single-mode state, $\nu>1/2$, and every physical isotropic
seed, $\nu_m\geq1/2$, the efficiencies satisfy
\begin{equation}
  \lambda_-<\lambda_+<\lambda_{\mathrm d}.\nonumber
\end{equation}
Thus, in the Williamson basis, displacement perturbations are measured most
efficiently, changes in the shape and orientation of the covariance ellipse
have intermediate efficiency, and the isotropic covariance direction,
associated with a change in thermal occupation, is the least efficiently
measured. These efficiencies refer to separate local one-parameter
perturbations and do not imply their simultaneous attainability in a
multiparameter estimation problem.

\subsection{Globally optimal multimode seed}

The following Schur-complement lemma is the second auxiliary result needed in
the multimode proof.

\begin{lemma}[Physicality of the effective single-mode covariance]
\label{lem:physicalConditionalCovariance}
Let $N\geq2$, and let $M$ be a physical $N$-mode Gaussian covariance matrix
partitioned as
\begin{equation}
  M
  =
  \begin{pmatrix}
    A&C\\
    C^T&B
  \end{pmatrix},\nonumber
\end{equation}
where $A\in\Sym(2,\R)$ corresponds to one mode. Let $V$ be a physical
covariance matrix of the remaining $N-1$ modes. Then
\begin{equation}
  M_{\mathrm{eff}}
  :=
  A-C(B+V)^{-1}C^T
  \label{eq:effectiveConditionalCovariance}
\end{equation}
is a physical single-mode covariance matrix:
\begin{equation}
  M_{\mathrm{eff}}+\frac{i}{2}J\succeq0.
  \label{eq:effectiveCovariancePhysicality}
\end{equation}
\end{lemma}

A proof based only on the uncertainty relation and the Schur-complement
criterion is given in Appendix \ref{app:singleModeSchur}.

\begin{theorem}[Globally optimal multimode Gaussian seed]
\label{thm:globalN}
Let $\rho$ be a faithful $N$-mode Gaussian state with covariance matrix
\begin{equation}
  \sigma
  =
  S
  \left(
    \bigoplus_{j=1}^{N}\nu_j\I_2
  \right)
  S^T,
  \qquad
  \nu_j>\frac12,\nonumber
\end{equation}
and define
\begin{equation}
  \nu_{\min}
  :=
  \min_{1\leq j\leq N}\nu_j.\nonumber
\end{equation}
Among all physical $N$-mode Gaussian measurement seeds with finite covariance
matrix,
\begin{equation}
  {
  \max_{\sigma_m\ \mathrm{physical}}
  \lambda_{\min}
  \!\left[
    \FG(\sigma,\sigma_m)
  \right]
  =
  \frac{
    \nu_{\min}-\frac12
  }{
    \nu_{\min}+\frac12
  }.}
  \label{eq:globalNvalue}
\end{equation}
This value is attained by the pure matched seed
\begin{equation}
  {
  \sigma_m^\star
  =
  \frac12SS^T.}
  \label{eq:globalSeedN}
\end{equation}
The matched seed is not necessarily the unique optimizer when the symplectic
spectrum is nondegenerate.
\end{theorem}

\begin{proof}
For $N=1$, the result is Theorem \ref{thm:onemodeglobal}. We therefore assume
$N\geq2$.

By Lemma \ref{lem:commonSymplecticCovariance}, applying the common symplectic
transformation $S^{-1}$ to the state and measurement seed preserves the
Fisher-efficiency spectrum. It is therefore sufficient to consider
\begin{equation}
  \sigma
  =
  D_\nu
  :=
  \bigoplus_{j=1}^{N}\nu_j\I_2.\nonumber
\end{equation}
Let
\begin{equation}
  M:=S^{-1}\sigma_mS^{-T}\nonumber
\end{equation}
be an arbitrary physical seed covariance matrix in this Williamson basis.

Choose an index $j_\star$ such that
\begin{equation}
  \nu_{j_\star}=\nu_{\min}.\nonumber
\end{equation}
After relabelling the modes if necessary, partition the state and seed with
respect to mode $j_\star$:
\begin{align}
  D_\nu
  &=
  \nu_{\min}\I_2
  \oplus
  D_{\setminus j_\star},
  \label{eq:stateMinimumModePartition}\\
  M
  &=
  \begin{pmatrix}
    A&C\\
    C^T&B
  \end{pmatrix},
  \label{eq:seedMinimumModePartition}
\end{align}
where
\begin{equation}
  D_{\setminus j_\star}
  :=
  \bigoplus_{k\neq j_\star}\nu_k\I_2.\nonumber
\end{equation}
The outcome covariance is
\begin{equation}
  \tau
  =
  D_\nu+M
  =
  \begin{pmatrix}
    \nu_{\min}\I_2+A&C\\
    C^T&D_{\setminus j_\star}+B
  \end{pmatrix}.
  \label{eq:tauMinimumModePartition}
\end{equation}

The lower-right block
$D_{\setminus j_\star}+B$ is positive definite. The block-inverse formula
therefore gives
\begin{equation}
  (\tau^{-1})_{j_\star j_\star}
  =
  \left(
    \nu_{\min}\I_2+M_{\mathrm{eff}}
  \right)^{-1},
  \label{eq:blockInverseEffectiveSeed}
\end{equation}
where
\begin{equation}
  M_{\mathrm{eff}}
  :=
  A
  -
  C
  \left(
    B+D_{\setminus j_\star}
  \right)^{-1}
  C^T.
  \label{eq:MeffGlobalProof}
\end{equation}
Since $D_{\setminus j_\star}$ is a physical covariance matrix,
Lemma \ref{lem:physicalConditionalCovariance} implies
\begin{equation}
  M_{\mathrm{eff}}+\frac{i}{2}J\succeq0.\nonumber
\end{equation}
Thus, $M_{\mathrm{eff}}$ is an admissible single-mode Gaussian seed
covariance matrix.

We now restrict the Gaussian tangent space to directions supported only on
mode $j_\star$. For a displacement direction, write
\begin{equation}
  \widetilde{\bm u}
  :=
  (\bm u,0),
  \qquad
  \bm u\in\R^2.\nonumber
\end{equation}
Equation \eqref{eq:blockInverseEffectiveSeed} gives
\begin{equation}
  \widetilde{\bm u}^{\,T}
  \tau^{-1}
  \widetilde{\bm v}
  =
  \bm u^T
  \left(
    \nu_{\min}\I_2+M_{\mathrm{eff}}
  \right)^{-1}
  \bm v.
  \label{eq:restrictedDisplacementCFI}
\end{equation}

For a covariance direction $X\in\Sym(2,\R)$, let
\begin{equation}
  \widetilde X
  :=
  \begin{pmatrix}
    X&0\\
    0&0
  \end{pmatrix}.\nonumber
\end{equation}
Write the inverse outcome covariance as
\begin{equation}
  \tau^{-1}
  =
  \begin{pmatrix}
    P&Q\\
    Q^T&R
  \end{pmatrix},
  \qquad
  P
  =
  \left(
    \nu_{\min}\I_2+M_{\mathrm{eff}}
  \right)^{-1}.
  \label{eq:tauInverseBlocks}
\end{equation}
Direct block multiplication gives
\begin{equation}
  \tau^{-1}\widetilde X
  =
  \begin{pmatrix}
    PX&0\\
    Q^TX&0
  \end{pmatrix}\nonumber
\end{equation}
and hence
\begin{equation}
  (\tau^{-1}\widetilde X)^2
  =
  \begin{pmatrix}
    PXPX&0\\
    Q^TXPX&0
  \end{pmatrix}.\nonumber
\end{equation}
Therefore,
\begin{equation}
  \Tr
  \!\left[
    (\tau^{-1}\widetilde X)^2
  \right]
  =
  \Tr(PXPX).
  \label{eq:restrictedCFIidentity}
\end{equation}
Thus, the restricted covariance contribution to the CFI is exactly
\begin{equation}
  \frac12
  \Tr\!\left[
    \left(
      \nu_{\min}\I_2+M_{\mathrm{eff}}
    \right)^{-1}
    X
    \left(
      \nu_{\min}\I_2+M_{\mathrm{eff}}
    \right)^{-1}
    X
  \right].\nonumber
\end{equation}

The QFI also restricts to the corresponding single-mode expression. Indeed,
because $D_\nu$ and $\Omega_N$ are direct sums, the superoperator
$\mathcal D_{D_\nu}$ preserves covariance perturbations supported only on
mode $j_\star$ and satisfies
\begin{equation}
  \mathcal D_{D_\nu}(\widetilde X)
  =
  \widetilde{
    \mathcal D_{\nu_{\min}\I_2}(X)
  }.
  \label{eq:restrictedDsuperoperator}
\end{equation} This follows directly from the block-diagonal forms of $D_\nu$ and $\Omega_N$:
when $X$ is supported only on mode $j_\star$, both terms in
$\mathcal D_{D_\nu}(X)$ remain supported on the same mode.

Since both superoperators are invertible,
\begin{equation}
  \mathcal D_{D_\nu}^{-1}(\widetilde X)
  =
  \widetilde{
    \mathcal D_{\nu_{\min}\I_2}^{-1}(X)
  }.\label{eq:restrictedInverseDsuperoperator}
\end{equation}
Consequently,
\begin{equation}
  \Tr\!\left[
    \widetilde X
    \mathcal D_{D_\nu}^{-1}(\widetilde X)
  \right]
  =
  \Tr\!\left[
    X
    \mathcal D_{\nu_{\min}\I_2}^{-1}(X)
  \right].
  \label{eq:restrictedQFIidentity}
\end{equation} Indeed, if
\begin{equation}
  \widetilde X=
  \begin{pmatrix}
    X&0\\
    0&0
  \end{pmatrix},\nonumber
\end{equation}
then the only nonzero diagonal block contributing to the trace is the block
containing $X$. Hence the full trace reduces to the corresponding
$2\times2$ trace.

Equations \eqref{eq:restrictedDisplacementCFI},
\eqref{eq:restrictedCFIidentity}, and
\eqref{eq:restrictedQFIidentity} show that the Fisher ratio on this restricted
subspace is exactly the Fisher ratio of the single-mode state
$\nu_{\min}\I_2$ measured with the physical seed $M_{\mathrm{eff}}$. Therefore,
\begin{equation}
  \inf_{\substack{
    \xi\neq0\\
    \xi\ \mathrm{supported\ on}\ j_\star
  }}
  \frac{\IC(\xi)}{\IQ(\xi)}
  =
  \lambda_{\min}
  \!\left[
    \FG
    \left(
      \nu_{\min}\I_2,
      M_{\mathrm{eff}}
    \right)
  \right].
  \label{eq:restrictedOneModeProblem}
\end{equation}
By Theorem \ref{thm:onemodeglobal},
\begin{equation}
  \lambda_{\min}
  \!\left[
    \FG
    \left(
      \nu_{\min}\I_2,
      M_{\mathrm{eff}}
    \right)
  \right]
  \leq
  \frac{
    \nu_{\min}-\frac12
  }{
    \nu_{\min}+\frac12
  }.
  \label{eq:oneModeEffectiveUpperBound}
\end{equation}

The minimum over the full tangent space cannot exceed the minimum over a
subspace. Hence,
\begin{align}
  \lambda_{\min}
  \!\left[
    \FG(D_\nu,M)
  \right]
  &=
  \inf_{\xi\neq0}
  \frac{\IC(\xi)}{\IQ(\xi)}\nonumber\\
  &\leq
  \inf_{\substack{
    \xi\neq0\\
    \xi\ \mathrm{supported\ on}\ j_\star
  }}
  \frac{\IC(\xi)}{\IQ(\xi)}\nonumber\\
  &\leq
  \frac{
    \nu_{\min}-\frac12
  }{
    \nu_{\min}+\frac12
  }.
  \label{eq:generalGlobalUpperBound}
\end{align}
This establishes the upper bound for every physical Gaussian seed, whether
pure or mixed, correlated or uncorrelated.

It remains to prove attainability. In the Williamson basis, choose
\begin{equation}
  M^\star=\frac12\I_{2N}.\nonumber
\end{equation}
This seed is aligned with $D_\nu$. By
Corollary \ref{cor:matchedworstcase},
\begin{equation}
  \lambda_{\min}
  \!\left[
    \FG
    \left(
      D_\nu,\frac12\I_{2N}
    \right)
  \right]
  =
  \frac{
    \nu_{\min}-\frac12
  }{
    \nu_{\min}+\frac12
  }.\nonumber
\end{equation}
Thus, the upper bound is attained. Transforming back to the original
phase-space coordinates gives
\begin{equation}
  \sigma_m^\star
  =
  SM^\star S^T
  =
  \frac12SS^T,\nonumber
\end{equation}
which proves Eqs. \eqref{eq:globalNvalue} and \eqref{eq:globalSeedN}.
\end{proof}

\subsubsection{Nonuniqueness of the multimode optimizer}
\label{rem:multimodeNonuniqueness}
The globally optimal seed need not be unique when the symplectic spectrum is
nondegenerate. For example, consider
\begin{equation}
  \sigma=I_2\oplus3I_2,
  \qquad
  \sigma_m(m)=\frac12I_2\oplus mI_2,
  \qquad m\geq\frac12.\nonumber
\end{equation}
Direct substitution into the spectrum of
Theorem \ref{thm:Nmode} gives
\begin{equation}
  \lambda_{\min}
  \!\left[
    \FG\bigl(\sigma,\sigma_m(m)\bigr)
  \right]
  =
  \min\left\{
    \frac13,\,
    \frac{35}{4(3+m)^2}
  \right\}.
  \label{eq:nonuniqueOptimizerFamily}
\end{equation}
The cross-mode eigenvalue satisfies
\begin{equation}
  \lambda_{12,-}
  =\frac{11}{6(3+m)}
  \geq
  \min\left\{
    \frac13,\frac{35}{4(3+m)^2}
  \right\},
  \qquad m\geq\frac12, \nonumber 
\end{equation}
and therefore does not change the minimum. The first value is the thermal-direction efficiency of the mode with $\nu_{\min}=1$, whereas the second is the corresponding efficiency of the
mode with symplectic eigenvalue $3$. Consequently,
$\lambda_{\min}
  \!\left[
    \FG\bigl(\sigma,\sigma_m(m)\bigr)
  \right]
  =\frac13,$ $\frac12\leq m\leq\frac{\sqrt{105}-6}{2}.$
Thus, the matched seed $I_4/2$ belongs to a continuous family of globally
optimal seeds. Theorem \ref{thm:globalN} therefore determines the optimal value
and provides a canonical optimizer, but the optimizer is not unique in
general.

\subsubsection{Scope of the global result}
In Theorem \ref{thm:globalN}, global optimality refers specifically to the class of covariant Gaussian general-dyne POVMs defined by a single physical seed with finite covariance matrix. Within this class, no assumption is made about the alignment, symmetry, purity, locality, separability, or correlation structure of the seed, and no bound is imposed on its energy, squeezing, entanglement, or circuit complexity. Randomized measurement schedules, adaptive protocols, singular homodyne limits, noncovariant Gaussian POVMs, and non-Gaussian measurements lie outside the optimization. The resulting optimal value depends only on the smallest symplectic eigenvalue $\nu_{\min}$, whereas the complete Fisher-efficiency spectrum generally retains a dependence on the full symplectic spectrum $(\nu_1,\ldots,\nu_N)$.

\subsection{Two-mode squeezed thermal states with arbitrary symplectic spectrum}
\label{sec:nondegenerateLocal}

We now remove the equal-symplectic-eigenvalue assumption from the
local-separable-collective comparison. Consider the two-mode Gaussian state
\begin{align}
  \sigma_{12}(r)
  &=S_{\mathrm{TMS}}(r)
  \left(\nu_1I_2\oplus\nu_2I_2\right)
  S_{\mathrm{TMS}}(r)^T,\nonumber\\
  &\nu_1,\nu_2>\frac12.
  \label{eq:generalThermalTMS}
\end{align}
where $\nu_1$ and $\nu_2$ are arbitrary and need not be equal.
The two-mode squeezing matrix is
\begin{equation}
  S_{\mathrm{TMS}}(r)
  =
  \begin{pmatrix}
    \cosh r\,I_2 & \sinh r\,Z\\
    \sinh r\,Z & \cosh r\,I_2
  \end{pmatrix},
  \qquad Z=\operatorname{diag}(1,-1).
\end{equation}
In standard form,
\begin{equation}
  \sigma_{12}(r)
  =
  \begin{pmatrix}
    a_rI_2&c_rZ\\
    c_rZ&b_rI_2
  \end{pmatrix},
  \qquad
  Z=\operatorname{diag}(1,-1),
  \label{eq:generalTMSstandard}
\end{equation}
with
\begin{align}
  a_r&=\nu_1\cosh^2r+\nu_2\sinh^2r,\nonumber\\
  b_r&=\nu_1\sinh^2r+\nu_2\cosh^2r,\nonumber\\
  c_r&=(\nu_1+\nu_2)\sinh r\cosh r.\nonumber
\end{align}

We first state the two auxiliary reductions used below. Their detailed proofs are given in Appendix \ref{app:productSeedAuxiliaries}.

\begin{lemma}[Dominance of the covariance sector]
\label{lem:covarianceControls}
Let $\sigma$ be the covariance matrix of a faithful Gaussian state and let
$\sigma_m$ be a finite physical Gaussian seed covariance. Then
\begin{equation}
  \lambda_{\min}
  \!\left[
    \FG(\sigma,\sigma_m)
  \right]
  =
  \lambda_{\min}
  \!\left[
    F_G^{(2)}(\sigma,\sigma_m)
  \right].
  \label{eq:covarianceControls}
\end{equation}
\end{lemma}

\begin{lemma}[Symmetry pinching of the seed covariance]
\label{lem:covariantPinching}
Let $U$ be an orthogonal symplectic matrix satisfying
\begin{equation}
  U\sigma U^T=\sigma,
  \qquad
  U^2=\pm I_{2N},\nonumber
\end{equation}
and define
\begin{equation}
  \mathcal E(A)
  :=
  \frac12\left(A+UAU^T\right).
  \label{eq:pinchingMap}
\end{equation} The condition $U^2=\pm I_{2N}$ ensures that $\mathcal E^2=\mathcal E$, so that $\mathcal E$ is a pinching map. This condition is not required for the efficiency inequality below. Then $\mathcal E(\sigma_m)$ is physical for every physical seed covariance
$\sigma_m$, and
\begin{equation}
  \lambda_{\min}
  \!\left[
    F_G^{(2)}(\sigma,\sigma_m)
  \right]
  \leq
  \lambda_{\min}
  \!\left[
    F_G^{(2)}(\sigma,\mathcal E(\sigma_m))
  \right].
  \label{eq:pinchingImproves}
\end{equation}
\end{lemma}

\begin{theorem}[Optimal product seed for arbitrary symplectic spectrum]
\label{thm:localEPRgeneral}
For the state \eqref{eq:generalThermalTMS}, the product heterodyne seed
\begin{equation}
  {
  \sigma_{m,\mathrm{het}}=\frac12I_4}
  \label{eq:heterodyneGeneral}
\end{equation}
maximizes the worst-direction Fisher efficiency over all physical product
Gaussian seeds, including mixed and anisotropic seeds with arbitrary local
orientations. Consequently,
\begin{equation}
  \eta_{\mathrm{loc}}^\star(\nu_1,\nu_2,r)
  =
  \lambda_{\min}\!\left[
    \FG\!\left(\sigma_{12}(r),\frac12I_4\right)
  \right].
  \label{eq:localGeneralOptimum}
\end{equation}

Operationally, Eq. \eqref{eq:heterodyneGeneral} represents two independent
heterodyne measurements, one on each signal mode. Each heterodyne detector
mixes its signal mode with a vacuum mode at a balanced beam splitter and
measures conjugate quadratures at the two outputs. The POVM is a product
measurement even though correlations in the input state generally produce
correlated measurement outcomes.
\end{theorem}

\begin{proof}
Consider
\begin{equation}
  U
  =
  R\!\left(\frac{\pi}{2}\right)
  \oplus
  R\!\left(-\frac{\pi}{2}\right),
  \qquad
  R(\theta)
  =
  \begin{pmatrix}
    \cos\theta&-\sin\theta\\
    \sin\theta&\cos\theta
  \end{pmatrix}.
  \label{eq:localPinchingSymmetry}
\end{equation}
This matrix is orthogonal and symplectic, satisfies $U^2=-I_4$, and leaves
the state invariant:
\begin{equation}
  U\sigma_{12}(r)U^T=\sigma_{12}(r).\nonumber
\end{equation}

Let
\begin{equation} \sigma_m=\gamma_A\oplus\gamma_B\nonumber\nonumber
\end{equation}
be an arbitrary physical product seed. Its image under
Eq. \eqref{eq:pinchingMap} is
\begin{equation}
  \mathcal E(\sigma_m)
  =
  \alpha I_2\oplus\beta I_2,
  \qquad
  \alpha=\frac12\Tr\gamma_A,
  \qquad
  \beta=\frac12\Tr\gamma_B.
  \label{eq:productPinching}
\end{equation}
Indeed, every real symmetric $2\times2$ matrix $\gamma$ satisfies
\begin{equation}
  \frac12\left(\gamma+J\gamma J^T\right)
  =
  \frac{\Tr\gamma}{2}I_2.\nonumber 
\end{equation}

The one-mode uncertainty relation implies
\begin{equation}
  \det\gamma_A\geq\frac14,
  \qquad
  \det\gamma_B\geq\frac14.\nonumber 
\end{equation}
By the arithmetic-geometric mean inequality \cite{HornJohnson2013} applied to the two positive
eigenvalues of each covariance matrix,
\begin{equation}
  \alpha
  =
  \frac12\Tr\gamma_A
  \geq
  \sqrt{\det\gamma_A}
  \geq\frac12,\nonumber 
\end{equation}
and similarly $\beta\geq1/2$. Hence
\begin{align}
  \mathcal E(\sigma_m)
  &=\frac12I_4+P,\nonumber\\
  P&=
  \left(\alpha-\frac12\right)I_2
  \oplus
  \left(\beta-\frac12\right)I_2
  \succeq0.
  \label{eq:pinchedSeedNoise}
\end{align}

Lemma \ref{lem:covariantPinching} and
Lemma \ref{lem:covarianceControls} give
\begin{equation}
  \lambda_{\min}\!\left[
    \FG(\sigma,\sigma_m)
  \right]
  \leq
  \lambda_{\min}\!\left[
    \FG(\sigma,\mathcal E(\sigma_m))
  \right].\nonumber 
\end{equation}
The covariance $P$ in Eq. \eqref{eq:pinchedSeedNoise} represents
parameter-independent Gaussian displacement noise added to the heterodyne
outcomes. Equivalently, the outcome distribution generated by
$\mathcal E(\sigma_m)=I_4/2+P$ is obtained from the heterodyne outcome
distribution by a fixed Gaussian convolution. Since this classical
post-processing is independent of the encoded parameter, the
Fisher-information data-processing inequality gives \cite{Ferrie2014}
\begin{equation}
  G_C\bigl(\mathcal E(\sigma_m)\bigr)
  \preceq
  G_C\left(\frac12I_4\right).
  \label{eq:pinchedDataProcessing}
\end{equation}
Because the QFI form is independent of the measurement seed, the
Rayleigh-Ritz characterization then implies
\begin{equation}
  \lambda_{\min}
  \!\left[
    \FG\bigl(\sigma,\mathcal E(\sigma_m)\bigr)
  \right]
  \leq
  \lambda_{\min}
  \!\left[
    \FG\left(\sigma,\frac12I_4\right)
  \right].
  \label{eq:pinchedHeterodyneBound}
\end{equation}
Since $I_4/2$ is itself a physical product seed, the upper bound is attained.
\end{proof}

We next derive a closed finite-dimensional expression for the optimal local
value. For compactness, let
\begin{equation}
  C:=\cosh(2r),
  \qquad
  S:=\sinh(2r),\nonumber 
\end{equation}
and define
\begin{equation}
  \Delta
  :=
  \nu_1\nu_2
  +\frac{\nu_1+\nu_2}{2}C
  +\frac14.
  \label{eq:DeltaGeneral}
\end{equation}
The coefficients entering the reduced matrix below are
\begin{equation}
  u:=\frac{\nu_2+C/2}{\Delta},
  \quad
  v:=\frac{\nu_1+C/2}{\Delta},
  \quad
  w:=\frac{S}{2\Delta}.
  \label{eq:uvwGeneral}
\end{equation}

\begin{proposition}[Closed expression for the optimal local efficiency]
\label{prop:localGeneralMatrix}
Under the assumptions of Theorem \ref{thm:localEPRgeneral}, the optimal local
worst-direction efficiency is
\begin{equation}
  {
  \eta_{\mathrm{loc}}^\star(\nu_1,\nu_2,r)
  =
  \lambda_{\min}(H_-),}
  \label{eq:localGeneralExact}
\end{equation}
where
\begin{widetext}
\begin{equation}
  H_-
  =
  \begin{pmatrix}
    u^2\left(\nu_1^2-\frac14\right)
    &
    w^2\left(\nu_1^2-\frac14\right)
    &
    \dfrac{uw}{2}\left(4\nu_1^2-1\right)
    \\[3mm]
    w^2\left(\nu_2^2-\frac14\right)
    &
    v^2\left(\nu_2^2-\frac14\right)
    &
    \dfrac{vw}{2}\left(4\nu_2^2-1\right)
    \\[3mm]
    \dfrac{uw}{4}\left(4\nu_1\nu_2+1\right)
    &
    \dfrac{vw}{4}\left(4\nu_1\nu_2+1\right)
    &
    \left(\nu_1\nu_2+\frac14\right)(uv+w^2)
  \end{pmatrix},
  \label{eq:HminusGeneral}
\end{equation}
with $u$, $v$, and $w$ defined in Eq. \eqref{eq:uvwGeneral}.

Although $H_-$ is not generally symmetric with respect to the Euclidean inner
product in the displayed coordinate basis, it is self-adjoint with respect to
the positive-definite metric
\begin{equation}
  G_-
  =
  \operatorname{diag}
  \left[
    \left(\nu_1^2-\frac14\right)^{-1},
    \left(\nu_2^2-\frac14\right)^{-1},
    2\left(\nu_1\nu_2+\frac14\right)^{-1}
  \right].
  \label{eq:GminusGeneral}
\end{equation}
\end{widetext}
In particular,
\begin{equation}
  G_-H_-=H_-^TG_-.
  \label{eq:HminusMetricSelfAdjoint}
\end{equation}
Consequently, $H_-$ is similar to the real symmetric matrix
\begin{equation}
  \widehat H_-
  :=
  G_-^{1/2}H_-G_-^{-1/2},
\label{eq:HminusSymmetricRepresentative}
\end{equation}
and therefore has only real eigenvalues.
\end{proposition}

The derivation of the matrix \(H_-\) in
Eq. \eqref{eq:HminusGeneral}, together with the proof of
Proposition \ref{prop:localGeneralMatrix}, is given in
Appendix \ref{app:HminusDerivation}.

The full two-mode covariance tangent sector has dimension ten. The preceding positivity and symmetry arguments show, however, that a least-efficient covariance direction can always be chosen in the three-dimensional invariant subspace $\mathcal V_-$. Consequently, the exact local benchmark for arbitrary
$\nu_1$, $\nu_2$, and $r$ is determined by the smallest eigenvalue of the
explicit matrix $H_-$.

Now, let $\mathfrak M_{\mathrm{prod}}$ and $\mathfrak M_{\mathrm{sep}}$
denote, respectively, the sets of finite physical two-mode Gaussian seed covariance matrices corresponding to product and separable
states with respect to the mode partition $1|2$. For the state family \eqref{eq:generalThermalTMS}, define
\begin{align}
  \eta_{\mathrm{coll}}^\star(\nu_1,\nu_2,r)
  &:=
  \sup_{\sigma_m\ \mathrm{physical}}
  \lambda_{\min}\!\left[
    \FG(\sigma_{12}(r),\sigma_m)
  \right],
  \\
  \eta_{\mathrm{sep}}^\star(\nu_1,\nu_2,r)
  &:=
  \sup_{\sigma_m\in\mathfrak M_{\mathrm{sep}}}
  \lambda_{\min}\!\left[
    \FG(\sigma_{12}(r),\sigma_m)
  \right],
\end{align}
Here, ``collective'' refers to a Gaussian seed, and hence to an effective POVM, that may jointly act on the two constituent modes in a single
experimental realization. It does not refer to a joint measurement across repeated copies of the state. The collective supremum runs over all physical two-mode Gaussian seed covariance matrices with finite entries.

\begin{theorem}[Separable seeds for arbitrary two-mode Gaussian states]
\label{thm:separableGeneral}
For every faithful two-mode Gaussian reference state,
\begin{equation}
  \sup_{\sigma_m\in\mathfrak M_{\mathrm{sep}}}
  \lambda_{\min}\!\left[
    \FG(\sigma,\sigma_m)
  \right]
  =
  \sup_{\sigma_m\in\mathfrak M_{\mathrm{prod}}}
  \lambda_{\min}\!\left[
    \FG(\sigma,\sigma_m)
  \right].\label{eq:generalSeparableProductEquality}
\end{equation}
For the state \eqref{eq:generalThermalTMS}, both suprema are attained by the
product heterodyne seed, and hence
\begin{equation}
  {
  \eta_{\mathrm{sep}}^\star(\nu_1,\nu_2,r)
  =
  \eta_{\mathrm{loc}}^\star(\nu_1,\nu_2,r)
  =
  \lambda_{\min}(H_-).}
  \label{eq:separableGeneralOptimum}
\end{equation}
\end{theorem}

\begin{proof}

By the Werner-Wolf characterization \cite{WernerWolf2001}, every separable
bipartite Gaussian seed covariance admits physical local covariance matrices
$\gamma_A$ and $\gamma_B$ such that
\begin{equation}
  \sigma_m
  \succeq
  \gamma_A\oplus\gamma_B.\nonumber 
\end{equation}
Thus
\begin{equation}
  \sigma_m
  =
  \gamma_A\oplus\gamma_B+P,
  \qquad
  P\succeq0.\nonumber 
\end{equation}
The covariance $P$ represents a
parameter-independent Gaussian post-processing of the outcomes obtained from
the product seed $\gamma_A\oplus\gamma_B$. The classical data-processing
inequality gives
\begin{equation}
  G_{\mathrm C}(\sigma_m)
  \preceq
  G_{\mathrm C}(\gamma_A\oplus\gamma_B).\nonumber 
\end{equation}
Consequently, for the symmetric representatives of the efficiency operators,
\begin{equation}
  \widehat{\mathbf F}_G(\sigma,\sigma_m)
  \preceq
  \widehat{\mathbf F}_G
  (\sigma,\gamma_A\oplus\gamma_B),\nonumber 
\end{equation}
and therefore
\begin{equation}
  \lambda_{\min}\!\left[
    \FG(\sigma,\sigma_m)
  \right]
  \leq
  \lambda_{\min}\!\left[
    \FG(\sigma,\gamma_A\oplus\gamma_B)
  \right].\nonumber 
\end{equation}
Hence no separable seed can outperform the best product seed. The reverse
inequality between the suprema follows from
$\mathfrak M_{\mathrm{prod}}\subseteq\mathfrak M_{\mathrm{sep}}$, proving
Eq. \eqref{eq:generalSeparableProductEquality}.

For the particular state \eqref{eq:generalThermalTMS},
Theorem \ref{thm:localEPRgeneral} shows that the product supremum is attained
by $I_4/2$, and Proposition \ref{prop:localGeneralMatrix} identifies its value
as $\lambda_{\min}(H_-)$.
\end{proof}

\begin{corollary}[Strict collective advantage for arbitrary symplectic spectrum]
\label{cor:collectiveAdvantageGeneral}
Let
\begin{equation}
  \nu_{\min}:=\min\{\nu_1,\nu_2\}.\nonumber 
\end{equation}
For every $r>0$,
\begin{equation}
  {
  \eta_{\mathrm{coll}}^\star
  =
  \frac{\nu_{\min}-\frac12}
       {\nu_{\min}+\frac12}
  >
  \lambda_{\min}(H_-)
  =
  \eta_{\mathrm{sep}}^\star
  =
  \eta_{\mathrm{loc}}^\star.}
  \label{eq:strictGeneralAdvantage}
\end{equation}
Consequently, every Gaussian general-dyne seed whose worst-direction
efficiency exceeds $\lambda_{\min}(H_-)$ must be entangled. For $r=0$, the
collective, product, and separable optima coincide.
\end{corollary}

\begin{proof}
Suppose first that
\begin{equation}
  \nu_1=\nu_{\min}.\nonumber 
\end{equation}
Because $G_-$ is diagonal and positive definite, the similarity
transformation in Eq. \eqref{eq:HminusSymmetricRepresentative} satisfies
\begin{equation}
  (\widehat H_-)_{ii}
  =
  (G_-)_{ii}^{1/2}(H_-)_{ii}(G_-)_{ii}^{-1/2}
  =
  (H_-)_{ii}.
  \label{eq:HminusDiagonalInvariance}
\end{equation}
Moreover, $H_-$ and $\widehat H_-$ have the same eigenvalues. Applying the
Rayleigh-Ritz principle to the real symmetric matrix $\widehat H_-$ and the
first Euclidean basis vector therefore gives (since $\lambda_{\min}(H_-) = \lambda_{\min}(\widehat H_-)$):
\begin{eqnarray}
  \lambda_{\min}(\widehat H_-)
  \leq
  \bm e_1^T\widehat H_-\bm e_1
  =
  (H_-)_{11} =
  u^2\left(\nu_1^2-\frac14\right). \nonumber \\ &&
  \label{eq:HminusDiagonalBound}
\end{eqnarray}
For $r>0$, $C=\cosh(2r)>1$. Direct calculation gives
\begin{align}
  \Delta
  -
  \left(\nu_2+\frac C2\right)
  \left(\nu_1+\frac12\right)
  &=
  \frac{C-1}{2}
  \left(\nu_2-\frac12\right)
  >0.
  \label{eq:strictUDifference}
\end{align}
Since
\begin{equation}
  u=\frac{\nu_2+C/2}{\Delta},\nonumber 
\end{equation}
Eq. \eqref{eq:strictUDifference} implies
\begin{equation}
  u<
  \frac1{\nu_1+\frac12}.\nonumber 
\end{equation}
Therefore,
\begin{align}
  \lambda_{\min}(H_-)
  &\leq
  u^2\left(\nu_1^2-\frac14\right)
  \\
  &<
  \frac{\nu_1^2-\frac14}
       {\left(\nu_1+\frac12\right)^2}
  =
  \frac{\nu_1-\frac12}
       {\nu_1+\frac12}.
  \label{eq:strictGeneralBoundNu1}
\end{align}

If instead $\nu_2=\nu_{\min}$, the same argument applied to $(H_-)_{22}$ and
$v$ yields
\begin{equation}
  \lambda_{\min}(H_-)
  <
  \frac{\nu_2-\frac12}
       {\nu_2+\frac12}.\nonumber 
\end{equation}
Theorem \ref{thm:globalN} identifies the right-hand side as the optimal
collective efficiency, while Theorem \ref{thm:separableGeneral} identifies
$\lambda_{\min}(H_-)$ as the common product and separable optimum.

For $r=0$, the state is
\begin{equation}
  \sigma_{12}(0)=\nu_1I_2\oplus\nu_2I_2,\nonumber 
\end{equation}
and the heterodyne seed $I_4/2$ is the globally matched seed. Hence
Theorem \ref{thm:globalN} gives equality between the collective, product, and
separable optima.
\end{proof}

Corollary \ref{cor:collectiveAdvantageGeneral} concerns entanglement of the
measurement seed within the class of covariant Gaussian general-dyne POVMs.
It does not identify separable seeds with LOCC measurement protocols, does not
assert that every entangled seed provides an advantage, and does not apply to
arbitrary non-Gaussian measurements. Moreover, the strict collective
advantage for $r>0$ does not require the measured state
\eqref{eq:generalThermalTMS} itself to be entangled.

\begin{figure}[h]
  \centering
  \includegraphics[width=\columnwidth]
  {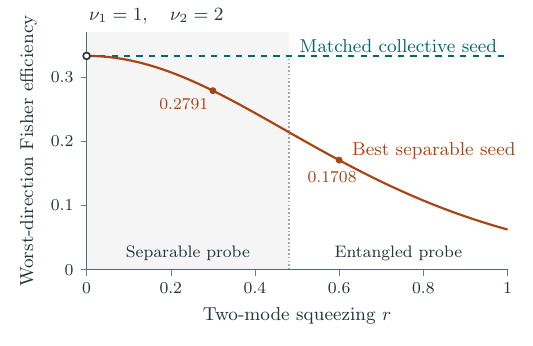}
  \caption{Worst-direction Fisher efficiency for a two-mode squeezed thermal state with symplectic eigenvalues $\nu_1=1$ and $\nu_2=2$. The dashed line is the matched collective optimum, $\eta_{\mathrm{coll}}^\star=1/3$, which is independent of $r$. The solid curve is $\lambda_{\min}(H_-)$, attained by product heterodyne and equal to the optimum over all separable Gaussian seeds. The shaded region denotes the range in which the probe is separable according to the Gaussian PPT criterion, $0\leq r\leq r_{\mathrm{sep}}$, with $r_{\mathrm{sep}}=\frac12\operatorname{arcosh}(3/2)\simeq0.4812$ \cite{Simon2000}. At $r=0.3$, the probe remains separable, while the optimal separable efficiency is approximately $0.2791$, below the collective value $1/3$.}
  \label{fig:collectiveAdvantageSeparableProbe}
\end{figure}

Figure \ref{fig:collectiveAdvantageSeparableProbe} illustrates this distinction for a nondegenerate two-mode squeezed thermal state. In particular, the collective advantage is already present in a range of squeezing parameters for which the measured state remains separable.

\section{Gaussian-optical implementations}
\label{sec:opticalImplementation}

The matched seed is not merely a formal solution of the optimization problem.
Every finite-dimensional symplectic transformation admits, in principle, a
Gaussian-unitary implementation. Therefore, whenever the required Gaussian
operations are experimentally available, the matched measurement can be
realized by applying the inverse symplectic transformation to the signal,
followed by standard heterodyne detection.

\begin{proposition}[Matched general-dyne measurement by Gaussian preprocessing]
\label{prop:opticalMatchedSeed}
Let $U_S$ be a Gaussian unitary implementing
$S\in\mathrm{Sp}(2N,\mathbb R)$ according to the convention
\begin{equation}
  U_S^\dagger\hat{\bm R}U_S
  =
  S\hat{\bm R}.
  \label{eq:GaussianUnitaryConvention}
\end{equation}
Let $H(\dd\bm z)$ be the standard $N$-mode heterodyne POVM, whose seed is the
multimode vacuum with covariance $I_{2N}/2$. Then the measurement procedure
\begin{equation}
  \rho
  \longmapsto
  U_S^\dagger\rho U_S
  \longmapsto
  H(\dd\bm z)
  \label{eq:preprocessedHeterodyne}
\end{equation}
is equivalent, after the invertible classical relabelling
\begin{equation}
  \bm r=S\bm z,
  \label{eq:classicalOutcomeRelabelling}
\end{equation}
to a covariant general-dyne POVM with seed covariance
\begin{equation}
  {
  \sigma_m^\star=\frac12SS^T.}
  \label{eq:opticalMatchedSeed}
\end{equation}
\end{proposition}

\begin{proof}
The standard heterodyne POVM is
\begin{equation}
  H(\dd\bm z)
  =
  \frac{\dd^{2N}\bm z}{(2\pi)^N}
  D(\bm z)\ket{0}\!\bra{0}D(\bm z)^\dagger,
  \label{eq:heterodynePOVMOptical}
\end{equation}
where $\ket{0}$ denotes the $N$-mode vacuum.

Measuring $H(\dd\bm z)$ after preprocessing the state by $U_S^\dagger$ gives
\begin{align}
  \Tr\!\left[
    U_S^\dagger\rho U_S H(\dd\bm z)
  \right]
  &=
  \Tr\!\left[
    \rho\,U_SH(\dd\bm z)U_S^\dagger
  \right].\nonumber 
\end{align}
Thus the effective POVM acting on the original state has effects
\begin{align}
  U_SH(\dd\bm z)U_S^\dagger
  &=\frac{\dd^{2N}\bm z}{(2\pi)^N}
  U_SD(\bm z)U_S^\dagger\nonumber\\
  &\quad\times U_S\ket{0}\!\bra{0}U_S^\dagger
  U_SD(\bm z)^\dagger U_S^\dagger.
  \label{eq:effectivePreprocessedEffect}
\end{align}
Under the convention \eqref{eq:GaussianUnitaryConvention}, the displacement
operators transform as
\begin{equation}
  U_SD(\bm z)U_S^\dagger
  =
  D(S\bm z).
  \label{eq:displacementCovariance}
\end{equation}
Moreover, the Gaussian state
\begin{equation}
  \rho_m^\star
  :=
  U_S\ket{0}\!\bra{0}U_S^\dagger\nonumber 
\end{equation}
has covariance matrix
\begin{equation}
  \sigma_m^\star
  =
  S\left(\frac12I_{2N}\right)S^T
  =
  \frac12SS^T.\nonumber 
\end{equation}
Since every symplectic matrix has determinant one,
\begin{equation}
  \dd^{2N}\bm r
  =
  \dd^{2N}\bm z
  \qquad
  \text{for }\bm r=S\bm z.\nonumber 
\end{equation}
Consequently, Eq. \eqref{eq:effectivePreprocessedEffect} becomes
\begin{equation}
  U_SH(\dd\bm z)U_S^\dagger
  =
  \frac{\dd^{2N}\bm r}{(2\pi)^N}
  D(\bm r)\rho_m^\star D(\bm r)^\dagger,\nonumber 
\end{equation}
which is precisely the covariant general-dyne POVM generated by the seed
$\rho_m^\star$.

As a consistency check, suppose that the input covariance is
\begin{equation}
  \sigma=SD_\nu S^T.\nonumber 
\end{equation}
The covariance of the raw heterodyne outcomes $\bm z$ is
\begin{equation}
  \tau_{\bm z}
  =
  S^{-1}\sigma S^{-T}
  +
  \frac12I_{2N}
  =
  D_\nu+\frac12I_{2N}.\nonumber 
\end{equation}
After the relabelling $\bm r=S\bm z$, the covariance becomes
\begin{equation}
  \tau_{\bm r}
  =
  S\tau_{\bm z}S^T
  =
  \sigma+\frac12SS^T
  =
  \sigma+\sigma_m^\star.\nonumber 
\end{equation}
Similarly, the mean transforms as
\begin{equation}
  S(S^{-1}\bm\varepsilon)=\bm\varepsilon.\nonumber 
\end{equation}
Thus the outcome statistics also reproduce the general-dyne relation
$\tau=\sigma+\sigma_m^\star$.
\end{proof}

The proposition shows that the matched seed need not be prepared as a
separate ancillary state. It can equivalently be implemented by undoing the
known symplectic transformation $S$ on the signal, performing local
heterodyne measurements at the output, and applying an invertible classical
relabeling of the outcomes. When the matched seed covariance $SS^T/2$ is not block diagonal with respect to the chosen mode partition, the effective POVM is collective, despite the
locality of the final heterodyne detectors.

\subsection{One mode: inverse squeezing followed by double homodyne detection}

Consider a single-mode squeezed thermal state without a phase-space rotation,
\begin{equation}
  S_1(r)
  =
  \operatorname{diag}(e^{-r},e^r),
  \qquad
  \sigma
  =
  \nu S_1(r)S_1(r)^T.
  \label{eq:singleModeSqueezedState}
\end{equation}
Its matched seed covariance is
\begin{equation}
  \sigma_{m,1}^\star
  =
  \frac12S_1(r)S_1(r)^T
  =
  \frac12
  \operatorname{diag}(e^{-2r},e^{2r}).
  \label{eq:singleModeMatchedOptical}
\end{equation}

\begin{figure}[t]
  \centering
  \includegraphics[
    width=\columnwidth
  ]{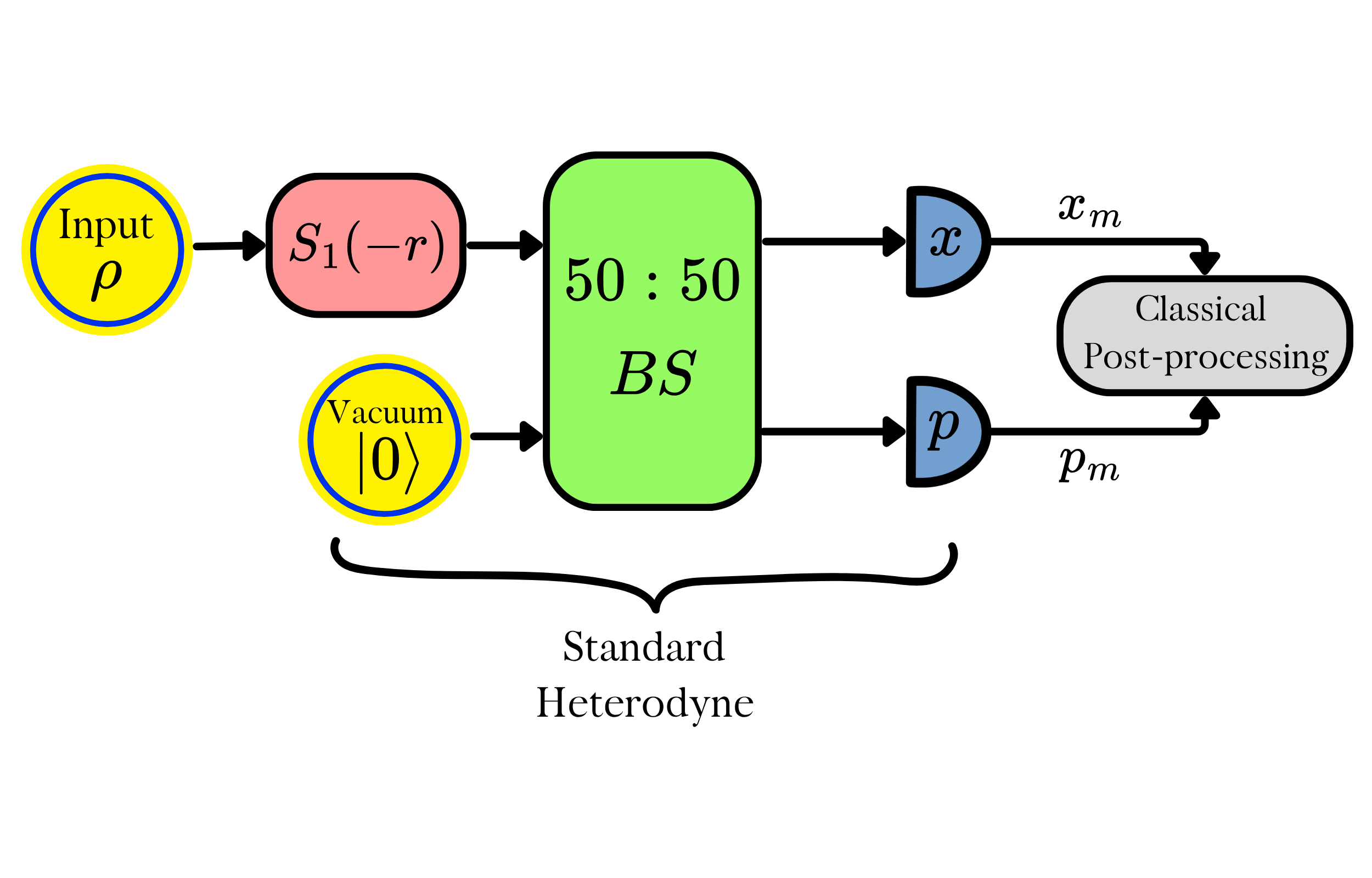}
  \caption{Single-mode implementation of the matched general-dyne
  measurement. The inverse squeezing transformation $S_1(-r)$ is first
  applied to the input state. The transformed mode is then mixed with a
  vacuum mode at a balanced beam splitter, and orthogonal quadratures are
  measured at the two outputs, realizing standard heterodyne detection. Let
  $\bm z=(x_m,p_m)^T$ denote the electronically calibrated heterodyne
  outcome, with the beam-splitter normalization factors absorbed into the
  calibration. The classical relabelling
  $\bm r=S_1(r)\bm z$ expresses the outcome in the original phase-space
  coordinates. The resulting effective seed covariance is
  $\sigma_{m,1}^\star=S_1(r)S_1(r)^T/2$.}
  \label{fig:singleModeOptical}
\end{figure}

Figure \ref{fig:singleModeOptical} shows a direct implementation. First, the
inverse transformation $S_1(r)^{-1}=S_1(-r)$ is applied to the signal. Standard heterodyne detection is then implemented
by mixing the transformed signal with a vacuum mode at a balanced beam
splitter and measuring orthogonal quadratures at the two outputs.

For a more general state
\begin{equation}
  \sigma
  =
  \nu R(\phi)S_1(r)S_1(r)^TR(\phi)^T,\nonumber 
\end{equation}
one may choose $S=R(\phi)S_1(r)$. The inverse preprocessing is then
\begin{equation}
  S^{-1}=S_1(-r)R(-\phi),\nonumber 
\end{equation}
corresponding to a phase shift by $-\phi$ followed by inverse squeezing.

\subsubsection{Passive implementation with an unbalanced beam splitter}

For one mode, the same pure general-dyne measurement can be implemented
without an active squeezer acting on the signal. The signal is mixed with a
vacuum mode at an unbalanced beam splitter, and orthogonal quadratures are
measured at the two outputs \cite{Genoni2014,Genoni2016}.

Let $T\in(0,1)$ denote the transmissivity from the signal input to the output
at which the $x$ quadrature is measured. With a fixed beam-splitter sign
convention, the measured output quadratures may be written as
\begin{align}
  \hat x_1'
  &=
  \sqrt{T}\,\hat x
  +
  \sqrt{1-T}\,\hat x_{\mathrm{vac}},
  \\
  \hat p_2'
  &=
  -\sqrt{1-T}\,\hat p
  +
  \sqrt{T}\,\hat p_{\mathrm{vac}}.
  \label{eq:unbalancedOutputQuadratures}
\end{align}
Define the calibrated outcomes
\begin{equation}
  x_m:=\frac{x_1'}{\sqrt T},
  \qquad
  p_m:=-\frac{p_2'}{\sqrt{1-T}}.
  \label{eq:unbalancedCalibratedOutcomes}
\end{equation}
They satisfy
\begin{align}
  x_m
  &=
  x+
  \sqrt{\frac{1-T}{T}}\,x_{\mathrm{vac}},
  \\
  p_m
  &=
  p-
  \sqrt{\frac{T}{1-T}}\,p_{\mathrm{vac}}.\nonumber 
\end{align}
Since the vacuum quadratures have variance $1/2$, the effective added-noise
covariance is
\begin{equation}
  \sigma_m(T)
  =
  \frac12
  \operatorname{diag}
  \left(
    \frac{1-T}{T},
    \frac{T}{1-T}
  \right).
  \label{eq:unbalancedSeedCovariance}
\end{equation}
Its determinant is
\begin{equation}
  \det\sigma_m(T)=\frac14,\nonumber 
\end{equation}
confirming that it represents a pure Gaussian seed.

Equating Eq. \eqref{eq:unbalancedSeedCovariance} with the matched seed
\eqref{eq:singleModeMatchedOptical} gives
\begin{equation}
  \frac{1-T}{T}=e^{-2r}.\nonumber 
\end{equation}
Therefore, the required transmissivity is
\begin{equation}
  {
  T=\frac{1}{1+e^{-2r}}.}
  \label{eq:unbalancedTransmissivity}
\end{equation}
If the two output ports or the measured quadratures are interchanged, the
equivalent convention is obtained by replacing $T$ with $1-T$.

{Note that as $r$ increases, the required transmissivity approaches unity, $ T\longrightarrow1,$ and the rescaling of the weak output quadrature involves a gain
proportional to $(1-T)^{-1/2}$. Consequently, losses, detector inefficiency,
and electronic noise are increasingly amplified in this regime. The passive
realization is therefore formally exact in the ideal model but becomes
progressively more sensitive to experimental imperfections at large
squeezing.}

This implementation converts the abstract matched-seed solution into a
directly testable prescription. After calibrating the outcomes according to
Eq. \eqref{eq:unbalancedCalibratedOutcomes}, the following predictions can be
tested:
\begin{enumerate}
  \item the measured outcome covariance is
  \begin{equation}
    \tau=\sigma+\sigma_{m,1}^\star;\nonumber 
  \end{equation}

  \item controlled infinitesimal displacement and covariance modulations can
  be used to reconstruct the classical Fisher form $G_{\mathrm C}$; combining
  it with $G_{\mathrm Q}$, computed from the calibrated reference covariance
  $\sigma$, yields the eigenvalues
  $\lambda_-$, $\lambda_+$, and $\lambda_{\mathrm d}$;

  \item the reconstructed worst-direction efficiency reaches
  \begin{equation}
    \lambda_{\min}(\FG)
    =
    \frac{\nu-\frac12}{\nu+\frac12}.\nonumber 
  \end{equation}
\end{enumerate}
Thus, in the single-mode case, the anisotropic matched measurement noise may
be realized either by active inverse squeezing followed by balanced
heterodyne detection or by an unbalanced double-homodyne scheme with
appropriately calibrated outcomes.

\subsection{Two modes: collective preprocessing followed by local heterodyne detection}

For the state \eqref{eq:generalThermalTMS},
Proposition \ref{prop:opticalMatchedSeed} prescribes the application of the
inverse two-mode squeezing transformation $S_{\mathrm{TMS}}(-r)$, followed by
independent heterodyne measurements on the two output modes. After the
classical outcome relabelling described in
Eq. \eqref{eq:classicalOutcomeRelabelling}, the effective seed covariance in
the original phase-space basis is
\begin{equation}
  {
  \sigma_{m,12}^\star
  =
  \frac12
  S_{\mathrm{TMS}}(r)S_{\mathrm{TMS}}(r)^T.}
  \label{eq:twoModeMatchedOptical}
\end{equation}

This is a pure two-mode Gaussian seed, since both of its symplectic eigenvalues
are equal to $1/2$. Its smallest partially transposed symplectic eigenvalue is
\begin{equation}
  \widetilde\nu_-^{(m)}
  =
  \frac12e^{-2r}.
  \label{eq:seedPToptical}
\end{equation}
Therefore,
\begin{equation}
  \widetilde\nu_-^{(m)}<\frac12
  \qquad
  \text{for every }r>0,\nonumber 
\end{equation}
and the matched measurement seed is entangled whenever the inverse
preprocessing is genuinely collective.

The inverse two-mode squeezing transformation may be implemented directly in
platforms providing a nondegenerate parametric amplifier. Reversing the
interaction sign, equivalently shifting the appropriate pump phase, implements
the inverse transformation. Alternatively, it admits a Bloch-Messiah
decomposition into balanced beam splitters and single-mode squeezers.

Define
\begin{equation}
  S_{\mathrm{sq}}(r)
  :=
  \operatorname{diag}(e^{-r},e^r)\nonumber 
\end{equation}
and
\begin{equation}
  B
  :=
  \frac1{\sqrt2}
  \begin{pmatrix}
    I_2&I_2\\
    -I_2&I_2
  \end{pmatrix}.
  \label{eq:balancedBeamSplitterMatrix}
\end{equation}
Then
\begin{equation}
  {
  S_{\mathrm{TMS}}(-r)
  =
  B
  \left[
    S_{\mathrm{sq}}(-r)
    \oplus
    S_{\mathrm{sq}}(r)
  \right]
  B^T.}
  \label{eq:TMSBlochMessiahOptical}
\end{equation}
Indeed,
\begin{align}
  &B
  \left[
    S_{\mathrm{sq}}(-r)
    \oplus
    S_{\mathrm{sq}}(r)
  \right]
  B^T
  \nonumber\\
  &\qquad=
  \begin{pmatrix}
    \cosh(r)I_2&-\sinh(r)Z\\
    -\sinh(r)Z&\cosh(r)I_2
  \end{pmatrix}
  =
  S_{\mathrm{TMS}}(-r).\nonumber 
\end{align}
The difference between the transformations $B$ and $B^T=B^{-1}$ can be
implemented by reversing the beam-splitter mixing angle or, equivalently, by
appropriate phase shifts. The phases of the homodyne local oscillators must be
chosen consistently with the quadratures selected in
Eq. \eqref{eq:TMSBlochMessiahOptical}.

\begin{figure}[t]
  \centering
  \includegraphics[
    width=\columnwidth
  ]{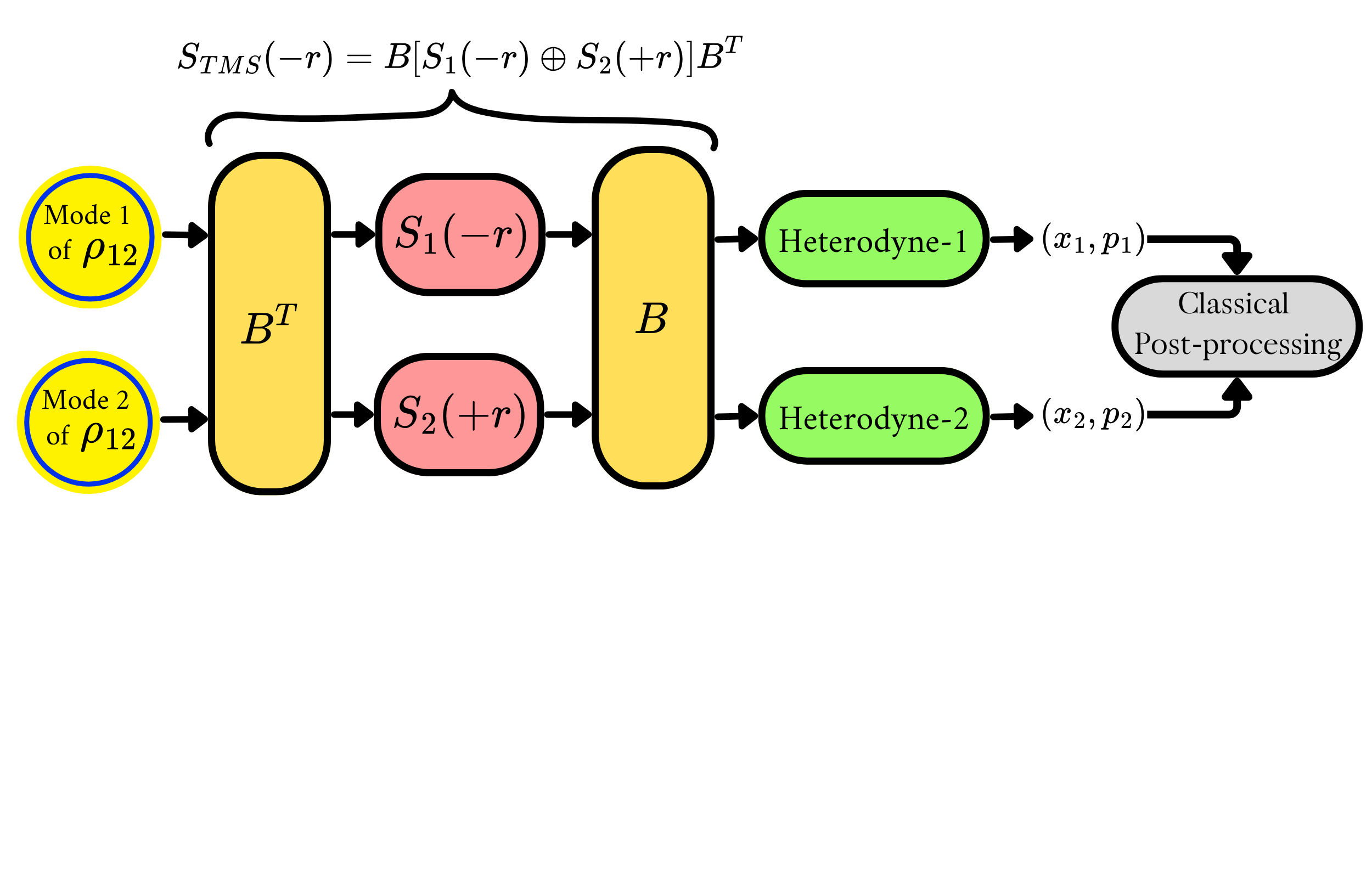}
  \vspace{-3cm}\caption{Two-mode optical implementation of the matched general-dyne
  measurement. The inverse two-mode squeezing transformation is implemented
  through the Bloch-Messiah decomposition
  $S_{\mathrm{TMS}}(-r)
  =B[S_{\mathrm{sq}}(-r)\oplus S_{\mathrm{sq}}(r)]B^T$,
  involving two balanced beam splitters and two local squeezers acting on
  orthogonal quadratures. Independent heterodyne measurements are then
  performed on the two output modes. Let
  $\bm z=(x_1,p_1,x_2,p_2)^T$ denote the vector of calibrated heterodyne
  outcomes. The classical relabelling
  $\bm r=S_{\mathrm{TMS}}(r)\bm z$ expresses these outcomes in the original
  phase-space coordinates, yielding the effective matched seed
  $\sigma_{m,12}^\star
  =S_{\mathrm{TMS}}(r)S_{\mathrm{TMS}}(r)^T/2$.
  Although the final heterodyne measurements are local, the preceding
  two-mode Gaussian transformation makes the effective POVM collective with
  respect to the original input modes.}
  \label{fig:twoModeOptical}
\end{figure}

The generation and characterization of optical EPR resources by interfering
single-mode squeezed states in orthogonal quadratures at a balanced beam
splitter has been demonstrated experimentally \cite{Fedorov2017}. That
experiment supports the availability of the principal optical components
appearing in Eq. \eqref{eq:TMSBlochMessiahOptical}, although the present
protocol uses them as an inverse transformation acting on an arbitrary
two-mode input rather than solely as a source of an EPR state. The complete implementation, including the subsequent local heterodyne
measurements and classical relabelling of the outcomes, is shown in
Fig. \ref{fig:twoModeOptical}.

The protocol provides a direct test of the central two-mode prediction. One
compares, under the same input conditions, the following two measurement
configurations:
\begin{enumerate}
  \item independent heterodyne detection applied directly to the two input
  modes, corresponding to the product seed $I_4/2$;

  \item the collective preprocessing $S_{\mathrm{TMS}}(-r)$ followed by the
  same two heterodyne modules, corresponding to the matched entangled seed
  \eqref{eq:twoModeMatchedOptical}.
\end{enumerate}
After the appropriate classical calibration of the outcomes, the collective
configuration must produce the covariance
\begin{equation}
  \tau_{\mathrm{coll}}
  =
  \sigma_{12}(r)+\sigma_{m,12}^\star.
  \label{eq:collectiveOutcomeCovariance}
\end{equation}
Controlled infinitesimal modulations of the displacement vector and covariance
matrix can then be used to reconstruct the classical Fisher form and,
consequently, the smallest eigenvalue of the Fisher-efficiency operator.

Varying $r$ tests the prediction that the ideal collective optimum remains
equal to
\begin{equation}
  \eta_{\mathrm{coll}}^\star
  =
  \frac{\nu_{\min}-\frac12}
       {\nu_{\min}+\frac12},
  \qquad
  \nu_{\min}=\min\{\nu_1,\nu_2\},
  \label{eq:collectiveOpticalPrediction}
\end{equation}
independently of the two-mode squeezing. Independently varying $\nu_1$ and
$\nu_2$ tests the extension to unequal symplectic eigenvalues.

\subsection{Proof-of-principle experimental proposal:
reconstruction of a \texorpdfstring{$3\times3$}{3x3} efficiency problem}
\label{sec:experimentalProposal}

The purpose of this protocol is to reconstruct experimentally the smallest
Fisher-efficiency eigenvalue for the product and collective measurement
configurations. This eigenvalue is not the direct reading of an individual
detector. It is inferred from measured covariances, locally estimated
derivatives, and the generalized eigenvalue problem
\begin{equation}
  G_{\mathrm C}\bm v
  =
  \lambda G_{\mathrm Q}\bm v.
  \label{eq:experimentalGeneralizedProblemIntro}
\end{equation}

The complete protocol consists of the following steps:
\begin{enumerate}
  \item prepare and independently calibrate a nominal two-mode Gaussian state
  with covariance $\sigma(\bm\theta_0)$;

  \item implement, in separate experimental runs, the product heterodyne
  measurement and the matched collective measurement;

  \item prepare small positive and negative variations of each parameter
  around the nominal point;

  \item record the four homodyne outcomes obtained in each experimental shot;

  \item estimate the outcome covariance matrices and the local covariance
  derivatives by centered finite differences;

  \item construct the restricted classical and quantum Fisher matrices
  $G_{\mathrm C}$ and $G_{\mathrm Q}$;

  \item solve the generalized eigenvalue problem and compare the uncertainty
  intervals of the reconstructed values
  $\lambda_{\min}^{\mathrm{prod}}$ and
  $\lambda_{\min}^{\mathrm{coll}}$.
\end{enumerate}

We consider the three-parameter family
\begin{align}
  \bm\theta&=(\nu_1,\nu_2,r),\nonumber\\
  \sigma_s(\bm\theta)
  &=S_{\mathrm{TMS}}(r)
  \left(\nu_1I_2\oplus\nu_2I_2\right)
  S_{\mathrm{TMS}}(r)^T.
  \label{eq:experimentalSignalFamily}
\end{align}
The reconstruction is performed locally around a nominal point
\begin{equation}
  \bm\theta_0
  =
  (\nu_{1,0},\nu_{2,0},r_0).\nonumber 
\end{equation}

The aim is not to reconstruct the complete
$14$-dimensional Gaussian Fisher-efficiency operator. Instead, we reconstruct its restriction to the three-dimensional tangent subspace generated by variations of $\nu_1$, $\nu_2$, and $r$. In the ideal covariance model considered below, this subspace contains a least-efficient direction for both measurement configurations.

\subsubsection{Two equivalent measurement configurations}

The product reference configuration uses two independent vacuum ancillas,
corresponding to the product heterodyne seed
\begin{equation}
  \sigma_m^{\mathrm{prod}}=\frac12I_4.
  \label{eq:experimentalProductSeed}
\end{equation}

For the Gaussian signal family considered here, the matched collective
outcome statistics can be implemented using either of the following two
equivalent schemes:
\begin{enumerate}
  \item apply $S_{\mathrm{TMS}}(-r_0)$ to the signal and then perform two local
  heterodyne measurements, as in Fig. \ref{fig:twoModeOptical};

  \item use the correlated-ancilla realization described below, in which the
  two signal modes are coupled locally to the two modes of a TMSV ancilla.
\end{enumerate}

In the second realization, the ancilla is prepared with covariance
\begin{equation}
  \sigma_a
  =
  \frac12
  S_{\mathrm{TMS}}(r_0)
  S_{\mathrm{TMS}}(r_0)^T
  =
  \sigma_m^\star.
  \label{eq:experimentalAncilla}
\end{equation}
This is the matched seed at the nominal value $r_0$.

\begin{figure}[t]
  \centering
  \includegraphics[
    width=\columnwidth
  ]{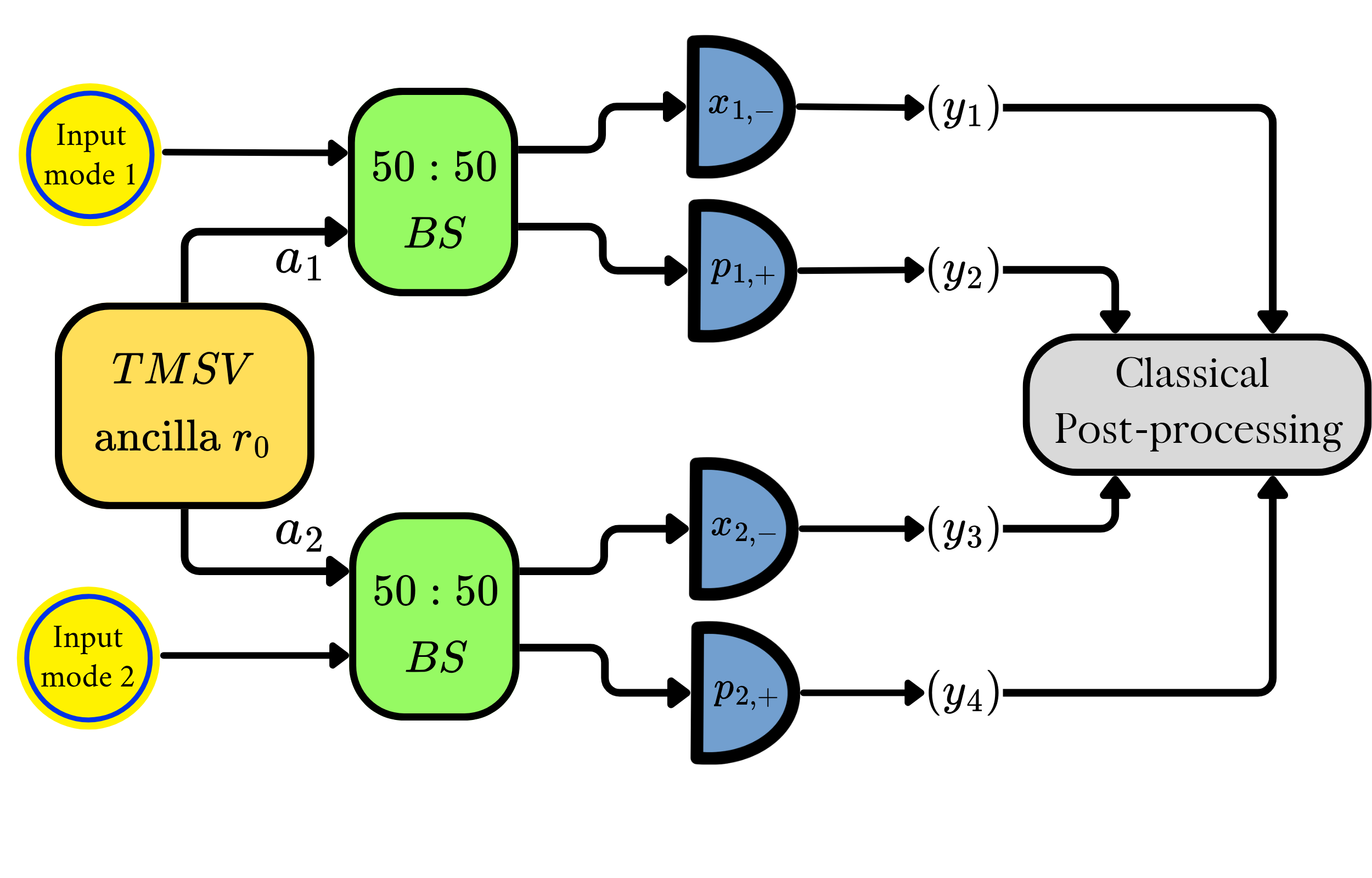}
  \caption{Correlated-ancilla realization of the matched collective
  measurement statistics. The two modes $a_1$ and $a_2$ of a TMSV ancilla
  with squeezing parameter $r_0$ are mixed separately with the two input
  signal modes at balanced beam splitters. Each balanced module measures the
  commuting quadratures
  $x_{j,-}=x_{s_j}-x_{a_j}$ and
  $p_{j,+}=p_{s_j}+p_{a_j}$.
  Beam-splitter normalization factors are absorbed into the electronic
  calibration, so that each experimental run produces
  $\bm y=(y_1,y_2,y_3,y_4)^T$ as defined in
  Eq. \eqref{eq:experimentalOutcomes}. At the nominal operating point, the
  ancilla covariance equals the matched seed covariance and the outcome
  covariance is
  $\operatorname{Cov}(\bm y)=\sigma_s+\sigma_m^\star$.
  The classical post-processing block denotes data collection and statistical
  reconstruction. Replacing the TMSV ancilla with two independent vacuum
  modes gives the product-heterodyne reference configuration.}
  \label{fig:experimentalProposal}
\end{figure}

Each signal mode $s_j$ is mixed with the corresponding ancilla mode $a_j$ at
a balanced beam splitter. One output is measured in the $x$ quadrature and
the other in the $p$ quadrature. After absorbing the factors of $\sqrt2$
introduced by the beam splitters into the electronic calibration, each
experimental shot produces the real outcome vector
\begin{equation}
  \bm y
  =
  \left(
    x_{s_1}-x_{a_1},
    \ p_{s_1}+p_{a_1},
    \ x_{s_2}-x_{a_2},
    \ p_{s_2}+p_{a_2}
  \right)^T.
  \label{eq:experimentalOutcomes}
\end{equation} The corresponding correlated-ancilla measurement circuit is shown in
Fig. \ref{fig:experimentalProposal}.

To verify the effective measurement covariance, define
\begin{equation}
  L
  :=
  \operatorname{diag}(-1,1,-1,1).
  \label{eq:experimentalLmatrix}
\end{equation}
Equation \eqref{eq:experimentalOutcomes} can then be written as
\begin{equation}
  \bm y=\hat{\bm R}_s+L\hat{\bm R}_a.\nonumber 
\end{equation}
Assuming that the signal and ancilla are independently prepared and centered,
their cross-covariance vanishes before the beam-splitter interactions.
Therefore,
\begin{equation}
  \tau
  :=
  \operatorname{Cov}(\bm y)
  =
  \sigma_s+L\sigma_aL^T.
  \label{eq:experimentalCovariance}
\end{equation}

For the TMSV covariance \eqref{eq:experimentalAncilla},
\begin{equation}
  L\sigma_aL^T=\sigma_a.
  \label{eq:TMSVLinvariance}
\end{equation}
Indeed, writing $L=(-Z)\oplus(-Z)$, the local blocks proportional to $I_2$
are invariant, while $(-Z)Z(-Z)=Z$
leaves the correlation blocks unchanged. Hence
\begin{equation}
  \tau
  =
  \sigma_s+\sigma_a
  =
  \sigma_s+\sigma_m^\star,
  \label{eq:experimentalMatchedCovariance}
\end{equation}
which coincides with the outcome covariance of the matched general-dyne
POVM. Both configurations also produce the same outcome mean. Since their
outcome distributions are Gaussian for every Gaussian input state in the
local model, equality of the first and second moments implies equality of
the complete outcome distributions. Consequently, the two configurations
have the same classical Fisher information on the Gaussian tangent space.

This establishes the statistical equivalence required for the present
Gaussian metrological protocol. By itself, the covariance calculation does
not assert equality of the two POVMs on arbitrary non-Gaussian input states.
A POVM-level equivalence requires fixing the Weyl, beam-splitter, and
phase-space-conjugation conventions explicitly; rigorous treatments of this
connection are given in Refs. \cite{KiukasLahti2008,Lahti2010}.

For each measurement configuration, the outcome covariance is estimated at
the nominal point $\bm\theta_0$ and at independently prepared neighboring
points $\bm\theta_0\pm\delta_a\bm e_a$, with
$a\in\{\nu_1,\nu_2,r\}$. Since the measurement seed is held fixed while the
signal parameters are varied, $\partial_a\tau=\partial_a\sigma_s$, and the
required covariance derivatives can be estimated by centered finite
differences. These estimates determine the restricted classical Fisher
matrix through Eq. \eqref{eq:CFI}. The corresponding quantum Fisher matrix is
not measured directly: it is calculated from the independently calibrated
reference covariance and Eq. \eqref{eq:QFI}. The experimentally reconstructed worst-direction efficiency is then the smallest generalized eigenvalue of Eq. \eqref{eq:experimentalGeneralizedProblemIntro}.

\subsubsection{Why the three-parameter reconstruction is sufficient}

At the nominal point, transform covariance perturbations to the Williamson
basis by
\begin{equation}
  X_a^{(0)}
  :=
  S_{\mathrm{TMS}}(-r_0)
  \left.
  \frac{\partial\sigma_s}{\partial\theta_a}
  \right|_{\bm\theta_0}
  S_{\mathrm{TMS}}(-r_0)^T.
  \label{eq:experimentalWilliamsonDerivatives}
\end{equation}
Direct differentiation of
Eq. \eqref{eq:experimentalSignalFamily} gives
\begin{align}
  X_{\nu_1}^{(0)}
  &=
  I_2\oplus0_2,\nonumber
  \\
  X_{\nu_2}^{(0)}
  &=
  0_2\oplus I_2,
  \label{eq:experimentalDerivativeBasis}\\
  X_r^{(0)}
  &=
  (\nu_{1,0}+\nu_{2,0})
  \begin{pmatrix}
    0_2&Z\\
    Z&0_2
  \end{pmatrix}.
  \nonumber
\end{align}
Because $\nu_{1,0}+\nu_{2,0}>0$, these three matrices are linearly
independent and span
\begin{equation}
  \mathcal V_-
  =
  \left\{
    \begin{pmatrix}
      xI_2&zZ\\
      zZ&yI_2
    \end{pmatrix}
    :
    x,y,z\in\mathbb R
  \right\}.
  \label{eq:experimentalReducedSector}
\end{equation}

This is precisely the invariant covariance sector appearing in Proposition \ref{prop:localGeneralMatrix}. For the ideal product heterodyne measurement, its smallest generalized eigenvalue is $\lambda_{\min}(H_-)$. For the ideal matched collective measurement, a
globally least-efficient thermal direction also belongs to this subspace. Therefore, under the state and measurement symmetries assumed above, the
$3\times3$ reconstruction accesses the complete worst-direction efficiency for both measurement configurations. Experimental imperfections that break these symmetries, such as asymmetric
losses, relative-phase errors, beam-splitter imbalance, or imperfect mode matching, can move the least-efficient direction outside this subspace. In that case, the smallest efficiency reconstructed within the three-dimensional subspace is an upper bound on the minimum over the complete Gaussian tangent space. Certifying the complete minimum would then require either an independent verification of the relevant symmetries or a reconstruction over a larger tangent subspace.

\subsection{Experimental signature and points of caution}

The relevant experimental signature is
\begin{equation}
  \widehat\lambda_{\min}^{\mathrm{coll}}
  >
  \widehat\lambda_{\min}^{\mathrm{prod}},
  \label{eq:experimentalSignature}
\end{equation}
with an uncertainty interval for their difference that excludes zero and with
both estimates consistent with an independently calibrated model of losses
and added noise.

The required number of samples $M$, the finite-difference steps $\delta_a$,
and the target covariance accuracy should not be specified without a
platform-dependent power analysis. The estimator of the smallest generalized
eigenvalue is nonlinear and may be biased at finite sample size. Its sampling
distribution and uncertainty interval should therefore be evaluated, for
example, by a parametric bootstrap or by a Monte Carlo simulation of the
complete estimation procedure.

Before claiming feasibility for a particular laboratory, the covariance model
must include at least:
\begin{enumerate}
  \item losses in the signal and ancilla preparation channels;
  \item finite purity and finite squeezing of the TMSV ancilla;
  \item homodyne detection inefficiency and electronic noise;
  \item beam-splitter imbalance and imperfect mode matching;
  \item relative-phase fluctuations;
  \item errors and drifts in the preparation of
  $\nu_1$, $\nu_2$, and $r$.
\end{enumerate}
If the first moments are not held fixed, their derivatives must also be
included in the Fisher matrices through the displacement contribution
\begin{equation}
  (\partial_a\bm\varepsilon)^T
  \tau^{-1}
  (\partial_b\bm\varepsilon).\nonumber 
\end{equation}
Furthermore, $G_{\mathrm Q}$ is not measured directly by this protocol. It is
calculated from the independently calibrated state covariance and its local
derivatives.

Accordingly, the proposed circuit is a conceptually feasible
proof-of-principle protocol. A quantitative claim of feasibility for a
specific experiment requires the measured parameters and noise model of that platform.

\section{Discussion}
\label{sec:discussion}

For a fixed symplectic spectrum, the simultaneous symplectic covariance of the
quantum and classical Fisher forms implies that the optimal collective
efficiency is independent of the particular symplectic transformation $S$.
Indeed, its value depends only on the smallest symplectic eigenvalue:
\begin{equation}
  \eta_{\mathrm{coll}}^\star
  =
  \frac{\nu_{\min}-\frac12}
       {\nu_{\min}+\frac12}.\nonumber
\end{equation}
By contrast, the canonical matched seed attaining this optimum is constructed
from $S$ and has covariance matrix
\begin{equation}
  \sigma_m^\star=\frac12SS^T.
  \label{eq:matchedSeedInterpretation}
\end{equation}
Thus, $\nu_{\min}$ determines the optimal achievable value, whereas $S$
determines the phase-space geometry, including the orientation, squeezing,
and intermode correlations, of this particular optimal measurement.

Theorem \ref{thm:separableGeneral} and
Corollary \ref{cor:collectiveAdvantageGeneral} show that the collective advantage is
not restricted to a comparison with product measurements. It persists
relative to every separable Gaussian seed, including correlated separable
seeds and states with unequal symplectic eigenvalues
$\nu_1\neq\nu_2$.

Within the class of covariant Gaussian general-dyne measurements, seed
entanglement is therefore necessary for exceeding the optimal separable
threshold. This is only a necessary condition: an arbitrary entangled seed
need not outperform the optimal product or separable seed. Moreover, this
statement concerns the measurement seed and does not require the measured
state itself to be entangled.

We refer to the measurement generated by
Eq. \eqref{eq:matchedSeedInterpretation} as the
\emph{collective covariant Gaussian measurement with matched seed}.
Proposition \ref{prop:opticalMatchedSeed} provides an optical implementation
through the inverse Gaussian transformation $U_S^\dagger$, followed by local
heterodyne detection on the transformed output modes.
Figures \ref{fig:singleModeOptical} and \ref{fig:twoModeOptical} illustrate
this construction for one and two modes.

Although the final heterodyne detectors act locally on the transformed modes,
the effective POVM on the original modes is a product measurement if and only
if the matched seed covariance $SS^T/2$ is block diagonal with respect to the
chosen mode partition. A nonlocal preprocessing transformation $S$ alone does
not imply a nonproduct POVM: for example, any passive symplectic transformation
satisfies $SS^T=I_{2N}$ and therefore yields the product heterodyne seed.
For the two-mode squeezing transformation considered here, the matched seed
is entangled for every $r>0$, and the effective POVM is consequently collective.

For finite squeezing, the matched measurement is a regular covariant
general-dyne POVM generated by a normalizable pure Gaussian seed. It should
not be identified automatically with an ideal continuous-variable Bell
measurement. The latter is obtained only in a singular infinite-squeezing
limit, which lies outside the finite-seed assumptions used in the present
theorems and requires a separate limiting treatment.

The results of this work are established under the following assumptions:
\begin{enumerate}
  \item the reference Gaussian state is faithful, with
  $\nu_j>1/2$;

  \item the measurement is a covariant Gaussian general-dyne POVM generated
  by a finite physical Gaussian seed covariance;

  \item the optimization is local around a known nominal reference state, so
  the measurement may depend on its calibrated covariance matrix;

  \item ``product'' and ``separable'' refer to the bipartition of the
  measurement seed covariance;

  \item no restriction is imposed on the energy, squeezing, entanglement, or
  circuit depth required to implement the seed.
\end{enumerate}
The results do not optimize over arbitrary non-Gaussian measurements,
adaptive strategies, or general LOCC protocols. They also do not imply that
every separable Gaussian seed admits an LOCC realization or that every
entangled seed provides a metrological advantage.

\section{Conclusion}
\label{sec:conclusion}

We have developed a spectral framework for the robust design of Gaussian
measurements. By representing the classical Fisher form relative to the
quantum Fisher form on the finite-dimensional Gaussian tangent space, the
Fisher-efficiency operator identifies how efficiently a fixed measurement
resolves each local one-parameter Gaussian perturbation of a reference state.
Its smallest eigenvalue provides a worst-direction figure of merit and turns
measurement design into a finite-dimensional spectral optimization problem,
without requiring a positive lower frame bound on the full
infinite-dimensional observable space.

Within this framework, we obtained the complete Fisher-efficiency spectrum for
arbitrary faithful multimode Gaussian states and aligned general-dyne seeds,
without assuming a degenerate symplectic spectrum. More importantly, we solved
the global optimization over all physical finite-covariance Gaussian seeds of
covariant general-dyne POVMs. A pure seed matched to the symplectic geometry of
the reference state is globally optimal for every number of modes. The optimal
worst-direction efficiency depends only on the smallest symplectic eigenvalue,
whereas the Williamson transformation determines how the optimal measurement
must be oriented, squeezed, and correlated in phase space. The least-efficient
direction can be chosen as an isotropic covariance perturbation of a mode with
the smallest symplectic eigenvalue, establishing a direct connection between
the global multimode result and single-mode thermal estimation.

For two-mode squeezed thermal states with arbitrary, possibly unequal,
symplectic eigenvalues, we established a sharp separation between product,
separable, and collective Gaussian measurements. Product heterodyne detection
is optimal among all product seeds, and allowing arbitrary correlations
compatible with Gaussian separability does not improve this benchmark. By
contrast, the matched collective seed achieves a strictly larger
worst-direction efficiency for every nonzero two-mode squeezing. Therefore,
within the covariant general-dyne class, the optimal separable efficiency lies
strictly below the collective optimum. Classical correlations in the
measurement seed cannot close this gap, and seed entanglement is a necessary,
although not sufficient, resource for surpassing the separable benchmark. This
conclusion concerns the measurement seed and does not require the measured
state itself to be entangled.

The optimal matched measurement also admits a direct Gaussian-optical
implementation. It can be realized by undoing the calibrated symplectic
transformation of the reference state and subsequently performing local
heterodyne detection. This establishes an operational connection between the
abstract spectral optimum and experimentally accessible Gaussian circuits, and the proposed two-mode protocol provides a proof-of-principle route for directly
comparing product and collective efficiencies.

The present results apply to faithful reference states and covariant
general-dyne measurements with finite Gaussian seeds, assuming that the nominal
reference covariance is known. Natural future extensions include optimization
under explicit constraints on energy, squeezing, seed entanglement, and circuit
depth; robustness against calibration errors, losses, and detector
imperfections; criteria restricted or weighted toward physically selected
tangent subspaces; and local-separable-collective comparisons for general
multimode partitions. These extensions build on, but are distinct from, the
unconstrained Gaussian measurement-design problem solved here.

\bigskip
\noindent\textit{Note added.}
During the finalization of this manuscript, we became aware of
Ref. \cite{Grzywa2026}, which independently develops a closely related comparison between the Fisher information recovered by Gaussian measurements and the SLD quantum Fisher information near Gaussian reference states. That work decomposes arbitrary perturbations into Hermite-Gauss orders and derives the corresponding efficiency matrices. Its first- and second-order sectors contain, respectively, the displacement and covariance directions considered here. In the present work, these directions are formulated intrinsically as the finite-dimensional tangent space of the manifold of faithful Gaussian states, with the classical and
quantum Fisher forms defined as bilinear forms and the Fisher-efficiency operator rigorously induced by the QFI inner product. In particular, both approaches yield the same single-mode thermal heterodyne efficiencies. Nevertheless, the two works address complementary questions. Reference \cite{Grzywa2026} focuses on the information recovered by homodyne and heterodyne detection across Gaussian and non-Gaussian perturbation orders, including randomized homodyne settings and pure Gaussian reference states. The present work focuses on robust minimax measurement design within the finite-covariance covariant general-dyne class, including the global optimization over physical Gaussian seeds and the exact product, separable, and collective thresholds for faithful two-mode squeezed thermal states.

\subsection*{Acknowledgments}
L.A.M.S. acknowledges FAPEMIG/MCI-Confap, grant APQ-06255-25.
G.A. acknowledges financial support from the Engineering and Physical Sciences Research Council  (EPSRC Grant No.~UKRI6536).

\appendix

\section{Relation to the state-regularized frame operator}

We now relate $\FG$ to the state-regularized frame formalism of
Innocenti et al. \cite{Innocenti2026}. In this subsection, $\rho_0$ denotes
the faithful reference state, while $\sigma$ continues to denote its
covariance matrix.

Let $\mathcal B(\mathcal H)_{\mathrm h}$ denote the real vector space of
bounded Hermitian operators on the bosonic Hilbert space $\mathcal H$. Define
\begin{equation}
  \langle A,B\rangle_{\rho_0}
  :=
  \frac12
  \Tr\!\left[
    \rho_0(AB+BA)
  \right]
  =
  \operatorname{Re}\Tr(\rho_0AB).
  \label{eq:rhoRegularizedInnerProduct}
\end{equation}
This form is real, bilinear, and symmetric, and satisfies
\begin{equation}
  \langle A,A\rangle_{\rho_0}
  =
  \Tr(\rho_0A^2)\geq0.\nonumber
\end{equation}
Since $\rho_0$ is faithful, equality holds only for $A=0$. Thus,
Eq. \eqref{eq:rhoRegularizedInnerProduct} defines an inner product on
$\mathcal B(\mathcal H)_{\mathrm h}$. Following
Ref. \cite{Innocenti2026}, we denote its Hilbert-space completion by
$L_{\mathrm h}^2(\rho_0)$.

Suppose that the continuous POVM admits a bounded operator-valued density
$\Pi(y)$ with respect to Lebesgue measure $\dd y$ on the outcome space
$\mathcal Y$, so that
\begin{equation}
  \int_{\mathcal Y}\Pi(y)\,\dd y=\I.\nonumber
\end{equation}
For the general-dyne measurements considered here,
$\mathcal Y=\R^{2N}$. Define
\begin{equation}
  p_0(y)
  :=
  \Tr[\rho_0\Pi(y)].\nonumber
\end{equation}
Because $\rho_0$ is faithful,
\begin{equation}
  p_0(y)>0\nonumber
\end{equation}
whenever $\Pi(y)$ is a nonzero positive operator.

For bounded Hermitian operators
$A\in\mathcal B(\mathcal H)_{\mathrm h}$, define
\begin{equation}
  (T_{\rho_0}^{(0)}A)(y)
  :=
  \frac{
    \Tr[\rho_0\{A,\Pi(y)\}]
  }{
    2\sqrt{p_0(y)}
  }.
  \label{eq:boundedAnalysisOperator}
\end{equation}
As proved in Lemma \ref{lem:boundedAnalysisExtension}, this map is contractive
with respect to the state-regularized norm and therefore extends uniquely by
continuity to the Hilbert-space completion \cite{Conway1990}:
\begin{equation}
  T_{\rho_0}:
  L_{\mathrm h}^2(\rho_0)
  \longrightarrow
  L^2(\mathcal Y,\dd y),
  \qquad
  \|T_{\rho_0}\|\leq1.
  \label{eq:extendedAnalysisOperator}
\end{equation}
The associated state-regularized frame operator is \cite{Innocenti2026}
\begin{equation}
  \mathfrak F_{\rho_0}
  :=
  T_{\rho_0}^{\dagger}T_{\rho_0}.
  \label{eq:fullFrameOperator}
\end{equation}
Consequently, $\mathfrak F_{\rho_0}$ is bounded, positive, and
self-adjoint on $L_{\mathrm h}^2(\rho_0)$, and satisfies
\begin{equation}
  0\preceq\mathfrak F_{\rho_0}\preceq\I.\nonumber
\end{equation}
Although suppressed from the notation, both $T_{\rho_0}$ and
$\mathfrak F_{\rho_0}$ depend on the chosen POVM and hence, in the present
setting, on its Gaussian seed.
Its quadratic form quantifies how well different
observable directions are resolved by the measurement, while its behavior
near the lower end of the spectrum determines the stability of observable
reconstruction \cite{Innocenti2026}.

For each $\xi\in\mathcal T_G$, let $L_\xi$ be the SLD determined by
\begin{equation}
  \dot\rho_\xi
  =
  \frac12\{\rho_0,L_\xi\}.
  \label{eq:SLDdirection}
\end{equation}
For a faithful Gaussian reference state, $L_\xi$ is at most quadratic in the
canonical operators and has a finite $\rho_0$-second moment
 \cite{Monras2013}. Hence, it defines an element of
$L_{\mathrm h}^2(\rho_0)$. The map $\xi\mapsto L_\xi$ is linear, and we
denote its image by
\begin{equation}
  \mathcal L_G
  :=
  \left\{
    L_\xi:\xi\in\mathcal T_G
  \right\}
  \subset L_{\mathrm h}^2(\rho_0).\nonumber
\end{equation}
Define
\begin{equation}
  \iota:
  \mathcal T_G
  \longrightarrow
  L_{\mathrm h}^2(\rho_0),
  \qquad
  \iota(\xi):=L_\xi.
  \label{eq:iotaDefinition}
\end{equation}

\begin{proposition}[Metrological compression of the full-space operator]
\label{prop:frameCompression}
For every $\xi,\zeta\in\mathcal T_G$,
\begin{align}
  \langle L_\xi,L_\zeta\rangle_{\rho_0}
  &=
  G_Q(\xi,\zeta),
  \label{eq:QFIasFrameNorm}\\
  \left\langle
    L_\xi,
    \mathfrak F_{\rho_0}L_\zeta
  \right\rangle_{\rho_0}
  &=
  G_C(\xi,\zeta).
  \label{eq:CFIasFrameNorm}
\end{align}
Consequently,
\begin{equation}
  {
  \FG
  =
  \iota^\dagger
  \mathfrak F_{\rho_0}
  \iota,}
  \label{eq:FGcompression}
\end{equation}
where
$\iota^\dagger:L_{\mathrm h}^2(\rho_0)\to\mathcal T_G$ is defined by
\begin{equation}
  G_Q(\xi,\iota^\dagger A)
  =
  \langle\iota(\xi),A\rangle_{\rho_0}\nonumber
\end{equation}
for every $\xi\in\mathcal T_G$ and
$A\in L_{\mathrm h}^2(\rho_0)$.
\end{proposition}

\begin{proof}
The SLD representation of the quantum Fisher metric gives
\begin{equation}
  G_Q(\xi,\zeta)
  =
  \frac12
  \Tr\!\left[
    \rho_0\{L_\xi,L_\zeta\}
  \right]
  =
  \langle L_\xi,L_\zeta\rangle_{\rho_0},
  \label{eq:SLDmetricProof}
\end{equation}
which proves Eq. \eqref{eq:QFIasFrameNorm}
 \cite{Monras2013,Saini2026}.

For the trace-norm differentiable Gaussian local models considered here,
the SLD $L_\xi$ belongs to $L_{\mathrm h}^2(\rho_0)$. By
Lemma \ref{lem:analysisOperatorOnSLD}, the continuous extension of the
analysis operator satisfies
\begin{equation}
  (T_{\rho_0}L_\xi)(y)
  =
  \frac{\partial_\xi p_0(y)}{\sqrt{p_0(y)}}
  \label{eq:analysisOnSLD}
\end{equation}
as an equality in $L^2(\mathcal Y,\dd y)$. Using
$\mathfrak F_{\rho_0}=T_{\rho_0}^{\dagger}T_{\rho_0}$, we obtain
\begin{align}
  \left\langle
    L_\xi,
    \mathfrak F_{\rho_0}L_\zeta
  \right\rangle_{\rho_0}
  &=
  \left\langle
    T_{\rho_0}L_\xi,
    T_{\rho_0}L_\zeta
  \right\rangle_{L^2(\mathcal Y,\dd y)}
  \nonumber\\
  &=
  \int_{\mathcal Y}
  \frac{
    \partial_\xi p_0(y)\,
    \partial_\zeta p_0(y)
  }{
    p_0(y)
  }
  \,\dd y\nonumber\\
  &=
  G_C(\xi,\zeta).\nonumber
\end{align}
This proves Eq. \eqref{eq:CFIasFrameNorm}.

Finally, for arbitrary $\xi,\zeta\in\mathcal T_G$,
\begin{align}
  G_Q\!\left(
    \xi,
    \iota^\dagger
    \mathfrak F_{\rho_0}
    \iota\zeta
  \right)
  &=
  \left\langle
    \iota\xi,
    \mathfrak F_{\rho_0}\iota\zeta
  \right\rangle_{\rho_0}\nonumber\\
  &=
  G_C(\xi,\zeta)\nonumber\\
  &=
  G_Q(\xi,\FG\zeta).\nonumber
\end{align}
Since $G_Q$ is nondegenerate, it follows that
\begin{equation}
  \FG\zeta
  =
  \iota^\dagger
  \mathfrak F_{\rho_0}
  \iota\zeta.\nonumber
\end{equation}
Because $\zeta$ is arbitrary, Eq. \eqref{eq:FGcompression} follows.
\end{proof}

\subsubsection{Consistency with the quantum Fisher upper bound}
\label{rem:compressionUpperBound}
Since
\begin{equation}
  0\preceq\mathfrak F_{\rho_0}\preceq\I\nonumber
\end{equation}
on $L_{\mathrm h}^2(\rho_0)$ and
$\iota:(\mathcal T_G,G_Q)\to L_{\mathrm h}^2(\rho_0)$ is an isometry, the
compression identity gives
\begin{equation}
  0\preceq
  \FG
  =
  \iota^\dagger\mathfrak F_{\rho_0}\iota
  \preceq\I\nonumber
\end{equation}
on $(\mathcal T_G,G_Q)$. Consequently,
\begin{equation}
  \lambda_{\max}(\FG)\leq1.\nonumber
\end{equation}
This independently reproduces the upper bound obtained earlier from the
monotonicity of Fisher information under quantum measurements.

Equation \eqref{eq:QFIasFrameNorm} shows that $\iota$ is an isometric
embedding. Since $\mathcal L_G$ is finite-dimensional, it is a closed subspace of
$L_{\mathrm h}^2(\rho_0)$ \cite{Conway1990}. Let
\begin{equation}
  P_G:=\iota\iota^\dagger\nonumber
\end{equation}
be the orthogonal projection onto $\mathcal L_G$. Under the isometric
identification $\mathcal T_G\simeq\mathcal L_G$,
Eq. \eqref{eq:FGcompression} is equivalent to
\begin{equation}
  \iota\FG\iota^{-1}
  =
  \left.
  P_G\mathfrak F_{\rho_0}
  \right|_{\mathcal L_G}.
  \label{eq:trueCompression}
\end{equation}
Thus, $\FG$ is the pullback to $\mathcal T_G$ of the compression of
$\mathfrak F_{\rho_0}$ to the Gaussian SLD subspace. Its eigenvalues are the
Ritz values of $\mathfrak F_{\rho_0}$ on
$\mathcal L_G$ \cite{HornJohnson2013}.

The subspace $\mathcal L_G$ need not be invariant under
$\mathfrak F_{\rho_0}$. Consequently, the Ritz values of the compression need
not belong to the point spectrum of the full operator. If $\mathcal L_G$ is
invariant, the compression coincides with the restriction of
$\mathfrak F_{\rho_0}$ to $\mathcal L_G$, and its eigenvalues are genuine
eigenvalues of the full operator.

\subsubsection{Vanishing full-space bound and positive Gaussian efficiency}
\label{rem:zeroFullFrameBound}
For a faithful single-mode Gaussian reference state and a Gaussian seed,
Innocenti et al. \cite{Innocenti2026} prove that the state-regularized
full-space operator has no strictly positive lower frame bound:
\begin{equation}
  \inf_{\substack{
    A\in L_{\mathrm h}^2(\rho_0)\\
    \|A\|_{\rho_0}=1
  }}
  \left\langle
    A,\mathfrak F_{\rho_0}A
  \right\rangle_{\rho_0}
  =
  0.
  \label{eq:zeroGlobalBound}
\end{equation}
Their proof reduces the problem to a thermal reference state and employs a
sequence of appropriately normalized Fock projectors with increasing photon
number. These test operators explore non-Gaussian directions in the
infinite-dimensional observable space.

By contrast, the Gaussian compression restricts the Rayleigh quotient to the
finite-dimensional subspace $\mathcal L_G$, whose dimension is
$2N^2+3N$. Equation \eqref{eq:GCpositive} and the compactness of the
$G_Q$-unit sphere imply
\begin{equation}
  \lambda_{\min}(\FG)
  =
  \min_{0\neq L\in\mathcal L_G}
  \frac{
    \left\langle
      L,\mathfrak F_{\rho_0}L
    \right\rangle_{\rho_0}
  }{
    \langle L,L\rangle_{\rho_0}
  }
  >
  0.\nonumber
\end{equation}
There is therefore no contradiction between a vanishing lower frame bound on
the complete observable space and a strictly positive minimum efficiency on
the Gaussian SLD subspace.

This construction does not establish stable tomography for arbitrary bosonic
observables. It establishes strictly positive Fisher efficiency only for the
finite-dimensional family of infinitesimal Gaussian perturbations considered
in this work.

\section{Domain of the state-regularized analysis operator}
\label{app:analysisOperatorDomain}

The pointwise expression for the state-regularized analysis operator is
initially defined on bounded Hermitian operators. This appendix establishes
the continuous extension required for the generally unbounded Gaussian SLDs.

\begin{lemma}[Bounded extension of the analysis operator]
\label{lem:boundedAnalysisExtension}
Let $\rho_0$ be faithful, and let $\Pi(y)$ be a bounded positive
operator-valued density satisfying
\begin{equation}
  \int_{\mathcal Y}\Pi(y)\,\dd y=\I,
  \qquad
  p_0(y):=\Tr[\rho_0\Pi(y)]>0\nonumber
\end{equation}
for almost every $y\in\mathcal Y$. For
$A\in\mathcal B(\mathcal H)_{\mathrm h}$, define
\begin{equation}
  (T_{\rho_0}^{(0)}A)(y)
  :=
  \frac{\Tr[\rho_0\{A,\Pi(y)\}]}
       {2\sqrt{p_0(y)}}.\nonumber
\end{equation}
Then
\begin{equation}
  \|T_{\rho_0}^{(0)}A\|_{L^2(\mathcal Y,\dd y)}^2
  \leq
  \Tr(\rho_0A^2).
  \label{eq:analysisContraction}
\end{equation}
Consequently, $T_{\rho_0}^{(0)}$ extends uniquely to a bounded operator
\begin{equation}
  T_{\rho_0}:
  L_{\mathrm h}^2(\rho_0)
  \longrightarrow
  L^2(\mathcal Y,\dd y),
  \qquad
  \|T_{\rho_0}\|\leq1.\nonumber
\end{equation}
\end{lemma}

\begin{proof}
For Hermitian $A$ and positive $\Pi(y)$,
\begin{equation}
  \frac12\Tr[\rho_0\{A,\Pi(y)\}]
  =
  \operatorname{Re}\Tr[\rho_0\Pi(y)A].\nonumber
\end{equation}
Define
\begin{equation}
  X_y:=\Pi(y)^{1/2}\rho_0^{1/2},
  \qquad
  Y_y:=\Pi(y)^{1/2}A\rho_0^{1/2}.\nonumber
\end{equation}
The Hilbert-Schmidt Cauchy-Schwarz inequality gives
\begin{align}
  \left|\Tr[\rho_0\Pi(y)A]\right|^2
  &=
  \left|\Tr(X_y^\dagger Y_y)\right|^2
  \nonumber\\
  &\leq
  \Tr(X_y^\dagger X_y)\Tr(Y_y^\dagger Y_y)
  \nonumber\\
  &=
  p_0(y)\Tr[\rho_0A\Pi(y)A].\nonumber
\end{align}
Therefore,
\begin{align}
  \|T_{\rho_0}^{(0)}A\|_{L^2}^2
  &\leq
  \int_{\mathcal Y}
  \Tr[\rho_0A\Pi(y)A]\,\dd y
  \nonumber\\
  &=
  \Tr\!\left[
    \rho_0A
    \left(\int_{\mathcal Y}\Pi(y)\,\dd y\right)
    A
  \right]
  \nonumber\\
  &=
  \Tr(\rho_0A^2).\nonumber
\end{align}
Thus $T_{\rho_0}^{(0)}$ is contractive on the dense subspace
$\mathcal B(\mathcal H)_{\mathrm h}$ and extends uniquely to its
completion $L_{\mathrm h}^2(\rho_0)$.
\end{proof}

\begin{lemma}[Action of the extended operator on Gaussian SLDs]
\label{lem:analysisOperatorOnSLD}
Assume that $\theta\mapsto\rho_\theta$ is trace-norm differentiable at
$\theta=0$, with faithful reference state $\rho_0$, and let $L_\xi$ be an
SLD satisfying
\begin{equation}
  \dot\rho_\xi
  =
  \frac12\{\rho_0,L_\xi\}.\nonumber
\end{equation}
Suppose that
\begin{equation}
  \Tr(\rho_0L_\xi^2)<\infty
  \qquad\text{and}\qquad
  \frac{\partial_\xi p_0}{\sqrt{p_0}}
  \in L^2(\mathcal Y,\dd y).\nonumber
\end{equation}
Then $L_\xi\in L_{\mathrm h}^2(\rho_0)$ and
\begin{equation}
  T_{\rho_0}L_\xi
  =
  \frac{\partial_\xi p_0}{\sqrt{p_0}}\nonumber
\end{equation}
as an equality in $L^2(\mathcal Y,\dd y)$.
\end{lemma}

\begin{proof}
Let
\begin{equation}
  L_\xi
  =
  \int_{\mathbb R}\lambda\,\dd E_\xi(\lambda)\nonumber
\end{equation}
be the spectral resolution of the self-adjoint SLD, and define its bounded
spectral truncations by
\begin{equation}
  L_{\xi,n}
  :=
  \int_{[-n,n]}\lambda\,\dd E_\xi(\lambda).\nonumber
\end{equation}
Introduce the finite positive measure
\begin{equation}
  \mu_{\rho_0,\xi}(B)
  :=
  \Tr[\rho_0E_\xi(B)]\nonumber
\end{equation}
on the Borel subsets $B\subseteq\mathbb R$. Since
$\Tr(\rho_0L_\xi^2)<\infty$, we have
\begin{align}
  \|L_{\xi,n}-L_\xi\|_{\rho_0}^2
  &=
  \Tr\!\left[
    \rho_0(L_{\xi,n}-L_\xi)^2
  \right]
  \\
  &=
  \int_{\{|\lambda|>n\}}
  \lambda^2\,\dd\mu_{\rho_0,\xi}(\lambda)
  \longrightarrow0\nonumber
\end{align}
by dominated convergence. Strictly speaking, $L_{\mathrm h}^2(\rho_0)$ is the abstract Hilbert-space
completion of $\mathcal B(\mathcal H)_{\mathrm h}$. In this completion, the
notation $L_\xi$ denotes the equivalence class represented by the Cauchy
sequence of bounded spectral truncations $(L_{\xi,n})_{n\geq1}$. This class
is canonically associated with the original unbounded SLD through its
spectral truncations.

The contractivity established in
Lemma \ref{lem:boundedAnalysisExtension} implies
\begin{equation}
  T_{\rho_0}^{(0)}L_{\xi,n}
  \longrightarrow
  T_{\rho_0}L_\xi
  \qquad\text{in }L^2(\mathcal Y,\dd y).\nonumber
\end{equation}

For every bounded real measurable function $f$, the weak operator integral
\begin{equation}
  B_f:=\int_{\mathcal Y}f(y)\Pi(y)\,\dd y\nonumber
\end{equation}
defines a bounded Hermitian operator. Passing to the limit in the corresponding
pairings gives
\begin{align}
  \int_{\mathcal Y}
  f(y)\sqrt{p_0(y)}
  (T_{\rho_0}L_\xi)(y)\,\dd y
  &=
  \frac12\Tr[\rho_0\{L_\xi,B_f\}]
  \nonumber\\
  &=
  \Tr(\dot\rho_\xi B_f)
  \nonumber\\
  &=
  \int_{\mathcal Y}f(y)\partial_\xi p_0(y)\,\dd y.\nonumber
\end{align}
Thus
\begin{equation}
  \int_{\mathcal Y}
  f(y)\sqrt{p_0(y)}
  \left[
    (T_{\rho_0}L_\xi)(y)
    -
    \frac{\partial_\xi p_0(y)}{\sqrt{p_0(y)}}
  \right]\dd y
  =0\nonumber
\end{equation}
for every bounded real measurable $f$. Since $p_0(y)>0$ almost everywhere,
the functions $f\sqrt{p_0}$ form a total subset of
$L^2(\mathcal Y,\dd y)$. The expression in square brackets therefore
vanishes in $L^2(\mathcal Y,\dd y)$.
\end{proof}

For a faithful Gaussian reference state, every Gaussian SLD is at most
quadratic in the canonical operators. Gaussian moments of all finite orders
exist, and hence
\begin{equation}
  \Tr(\rho_0L_\xi^2)
  =
  I_{\mathrm Q}(\xi)<\infty.\nonumber
\end{equation}
Moreover, the explicit Gaussian CFI formula ensures that
$\partial_\xi p_0/\sqrt{p_0}\in L^2$. Thus the preceding lemma applies to
all Gaussian tangent directions considered in this work.

\section{Common symplectic covariance}
\label{app:commonSymplecticCovariance}

This appendix proves Lemma \ref{lem:commonSymplecticCovariance}. In particular,
we show that a common symplectic transformation preserves both Fisher
bilinear forms and induces a similarity transformation between the
corresponding Fisher-efficiency operators.

\begin{proof}[Proof of Lemma \ref{lem:commonSymplecticCovariance}]
Set
\begin{equation}
  A:=S^{-1}.\nonumber 
\end{equation}
Since $S$ is symplectic, $A$ is also symplectic and satisfies
\begin{equation}
  A\Omega_NA^T=\Omega_N.
  \label{eq:Asymplectic}
\end{equation}
The transformed seed remains physical because
\begin{align}
  \sigma_{m,0}+\frac{i}{2}\Omega_N
  &=
  A\sigma_mA^T+\frac{i}{2}A\Omega_NA^T\\
  &=
  A
  \left(
    \sigma_m+\frac{i}{2}\Omega_N
  \right)
  A^T
  \succeq0.\nonumber 
\end{align}

Let
\begin{equation}
  \tau:=\sigma+\sigma_m.\nonumber 
\end{equation}
The transformed outcome covariance is
\begin{equation}
  \tau_0
  :=
  \sigma_0+\sigma_{m,0}
  =
  A\tau A^T,
  \label{eq:tauTransformation}
\end{equation}
and therefore
\begin{equation}
  \tau_0^{-1}
  =
  A^{-T}\tau^{-1}A^{-1}.
  \label{eq:tauInverseTransformation}
\end{equation}

The corresponding transformation of the Gaussian tangent space is
\begin{equation}
  \Phi_A:
  \mathcal T_G\longrightarrow\mathcal T_G,
  \qquad
  \Phi_A(\bm u,X)
  :=
  (A\bm u,AXA^T).
  \label{eq:tangentSymplecticMap}
\end{equation}
This map is invertible, with
\begin{equation}
  \Phi_A^{-1}(\bm u_0,X_0)
  =
  (A^{-1}\bm u_0,A^{-1}X_0A^{-T}).\nonumber 
\end{equation}

We first verify the invariance of the displacement contributions. From
Eq. \eqref{eq:tauInverseTransformation},
\begin{align}
  (A\bm u)^T\tau_0^{-1}(A\bm v)
  &=
  \bm u^T
  A^TA^{-T}
  \tau^{-1}
  A^{-1}A
  \bm v\nonumber \\
  &=
  \bm u^T\tau^{-1}\bm v.
  \label{eq:CFIdisplacementInvariant}
\end{align}
The same calculation with $\sigma$ in place of $\tau$ gives
\begin{equation}
  (A\bm u)^T\sigma_0^{-1}(A\bm v)
  =
  \bm u^T\sigma^{-1}\bm v.
  \label{eq:QFIdisplacementInvariant}
\end{equation}

For the covariance contribution to the classical Fisher form,
\begin{align}
  &\Tr\!\left[
    \tau_0^{-1}(AXA^T)
    \tau_0^{-1}(AYA^T)
  \right]
  \nonumber\\
  &\quad=
  \Tr\!\left[
    A^{-T}\tau^{-1}
    X\tau^{-1}Y
    A^T
  \right]
  \nonumber\\
  &\quad=
  \Tr\!\left[
    \tau^{-1}X\tau^{-1}Y
  \right],
  \label{eq:CFIcovarianceInvariant}
\end{align}
where cyclicity of the trace was used in the last step.

It remains to establish the transformation law for the covariance contribution
to the QFI. Equation \eqref{eq:Asymplectic} implies
\begin{equation}
  \Omega_NA^{-T}=A\Omega_N,
  \qquad
  A^{-1}\Omega_N=\Omega_NA^T.
  \label{eq:symplecticIdentities}
\end{equation}
Using $\sigma_0=A\sigma A^T$, we obtain, for every
$Z\in\Sym(2N,\R)$,
\begin{align}
  \mathcal D_{\sigma_0}
  \!\left(
    A^{-T}ZA^{-1}
  \right)
  &=2\sigma_0
  \left(A^{-T}ZA^{-1}\right)
  \sigma_0\nonumber\\
  &\quad+\frac12\Omega_N
  \left(A^{-T}ZA^{-1}\right)
  \Omega_N\nonumber \\
  &=
  A
  \left(
    2\sigma Z\sigma
    +
    \frac12\Omega_NZ\Omega_N
  \right)
  A^T\nonumber \\
  &=
  A\mathcal D_\sigma(Z)A^T.
  \label{eq:Dsigmacovariance}
\end{align}
Since both $\sigma$ and $\sigma_0$ correspond to faithful Gaussian states,
$\mathcal D_\sigma$ and $\mathcal D_{\sigma_0}$ are invertible on
$\Sym(2N,\R)$. Equation \eqref{eq:Dsigmacovariance} therefore implies
\begin{align}
  \mathcal D_{\sigma_0}^{-1}
  \!\left(AYA^T\right)
  &=A^{-T}\mathcal D_\sigma^{-1}(Y)\nonumber\\
  &\hspace{10mm}{}\times A^{-1}.
  \label{eq:DinverseCovariance}
\end{align}
Consequently,
\begin{align}
  \Tr\!\left[
    (AXA^T)
    \mathcal D_{\sigma_0}^{-1}
    \!\left(
      AYA^T
    \right)
  \right]
  &=
  \Tr\!\left[
    AX
    \mathcal D_\sigma^{-1}(Y)
    A^{-1}
  \right]\nonumber \\
  &=
  \Tr\!\left[
    X\mathcal D_\sigma^{-1}(Y)
  \right].
  \label{eq:QFIcovarianceInvariant}
\end{align}

Equations \eqref{eq:CFIdisplacementInvariant},
\eqref{eq:QFIdisplacementInvariant},
\eqref{eq:CFIcovarianceInvariant}, and
\eqref{eq:QFIcovarianceInvariant} establish
\begin{align}
  G_Q^{(0)}
  \!\left(
    \Phi_A\xi,\Phi_A\zeta
  \right)
  &=
  G_Q(\xi,\zeta),
  \label{eq:QFIsymplecticInvariance}\\
  G_C^{(0)}
  \!\left(
    \Phi_A\xi,\Phi_A\zeta
  \right)
  &=
  G_C(\xi,\zeta),
  \label{eq:CFIsymplecticInvariance}
\end{align}
where the superscript $(0)$ denotes the Fisher forms evaluated at
$(\sigma_0,\sigma_{m,0})$.

Using the defining relation of the Fisher-efficiency operator,
\begin{align}
  &G_Q^{(0)}
  \!\left(
    \Phi_A\xi,
    \FG(\sigma_0,\sigma_{m,0})
    \Phi_A\zeta
  \right)
  \nonumber\\
  &\quad=
  G_C^{(0)}
  \!\left(
    \Phi_A\xi,\Phi_A\zeta
  \right)\nonumber \\
  &\quad=
  G_C(\xi,\zeta)\nonumber \\
  &\quad=
  G_Q
  \!\left(
    \xi,
    \FG(\sigma,\sigma_m)\zeta
  \right)\nonumber \\
  &\quad=
  G_Q^{(0)}
  \!\left(
    \Phi_A\xi,
    \Phi_A\FG(\sigma,\sigma_m)\zeta
  \right).\nonumber 
\end{align}
Since $G_Q^{(0)}$ is nondegenerate and $\Phi_A$ is invertible,
\begin{equation}
  \FG(\sigma_0,\sigma_{m,0})\Phi_A
  =
  \Phi_A\FG(\sigma,\sigma_m).
  \label{eq:FGintertwining}
\end{equation}
Equivalently,
\begin{equation}
  \FG(\sigma_0,\sigma_{m,0})
  =
  \Phi_A
  \FG(\sigma,\sigma_m)
  \Phi_A^{-1}.
  \label{eq:FGsymplecticSimilarity}
\end{equation}
Thus, the two Fisher-efficiency operators are similar and have identical
spectra.

Finally, if $\sigma=SD_\nu S^T$ and $\sigma_m=SM_\mu S^T$, then
\begin{equation}
  S^{-1}\sigma S^{-T}=D_\nu,
  \qquad
  S^{-1}\sigma_mS^{-T}=M_\mu,\nonumber 
\end{equation}
which proves the last statement.
\end{proof}

\subsubsection{Physical interpretation}
Lemma \ref{lem:commonSymplecticCovariance} expresses a change of canonical
phase-space coordinates, rather than a change of the physical estimation
problem. When the state, measurement seed, and Gaussian tangent directions are
transformed simultaneously, both the QFI and CFI are preserved. Consequently,
the Fisher-efficiency spectrum is invariant; only the representation of its
eigendirections changes.

\subsubsection{Covariance of $\mathcal D_\sigma$}
The contragredient transformation in
Eq. \eqref{eq:Dsigmacovariance} is essential. For a symplectic matrix $A$, the
correct transformation law is
\begin{equation}
  \mathcal D_{A\sigma A^T}
  \!\left(
    A^{-T}ZA^{-1}
  \right)
  =
  A\mathcal D_\sigma(Z)A^T.\nonumber 
\end{equation}
In general,
\begin{equation}
  \mathcal D_{A\sigma A^T}
  \!\left(
    AZA^T
  \right)
  \neq
  A\mathcal D_\sigma(Z)A^T\nonumber 
\end{equation}
when $A$ is nonorthogonal. A direct substitution in the term
$2\sigma Z\sigma$ would introduce additional factors of $A^TA$. Therefore,
the QFI covariance contribution must be transformed using the dual
congruence appearing in Eq. \eqref{eq:Dsigmacovariance}, or equivalently the
inverse relation \eqref{eq:DinverseCovariance}.

\section{Auxiliary single-mode and Schur-complement results}
\label{app:singleModeSchur}

\subsection{Proof of the single-mode optimum}

The single-mode case admits a stronger result than
Corollary \ref{cor:matchedworstcase}. The symplectically matched seed is
optimal among all physical single-mode Gaussian seeds, rather than only among
those aligned with a previously fixed Williamson basis.

For a state with covariance matrix
\begin{equation}
  \sigma=\nu SS^T,
  \qquad
  \nu>\frac12,\nonumber 
\end{equation}
the \emph{matched seed} is defined as
\begin{equation}
  \sigma_m^{\mathrm{match}}
  :=
  \frac12SS^T.
  \label{eq:matchedDefinition}
\end{equation}
It shares the symplectic axes and squeezing encoded by $S$, but replaces the
state's symplectic eigenvalue $\nu$ by the minimum physical seed eigenvalue
$1/2$. Thus, it matches the geometry of the state without reproducing its
thermal noise.

\begin{proof}[Proof of Theorem \ref{thm:onemodeglobal}]
By Lemma \ref{lem:commonSymplecticCovariance}, we may transform the state and
seed by $S^{-1}$ without changing the Fisher-efficiency spectrum. It is
therefore sufficient to prove the result for
\begin{equation}
  \sigma=\nu\I_2.\nonumber 
\end{equation}

Every physical single-mode Gaussian seed covariance matrix can be written as
\begin{equation}
  \sigma_m
  =
  mR(\phi)
  \begin{pmatrix}
    e^{-2z}&0\\
    0&e^{2z}
  \end{pmatrix}
  R(\phi)^T,
  \qquad
  m\geq\frac12,
  \label{eq:generalOneModeSeed}
\end{equation}
where $R(\phi)\in\Sp(2,\R)\cap O(2)$ is a phase-space rotation and
$z\in\R$ is the seed squeezing parameter.

Because the thermal covariance is isotropic,
\begin{equation}
  R(\phi)(\nu I_2)R(\phi)^T=\nu I_2,\nonumber 
\end{equation}
and therefore the rotation angle of the seed does not affect the spectrum. Lemma \ref{lem:commonSymplecticCovariance} then shows that the Fisher-efficiency spectrum is independent of $\phi$.

We next show that it is sufficient to consider pure seeds. For fixed $z$ and
$\phi$, define
\begin{equation}
  \sigma_m^{\mathrm{pure}}
  :=
  \frac12
  R(\phi)
  \begin{pmatrix}
    e^{-2z}&0\\
    0&e^{2z}
  \end{pmatrix}
  R(\phi)^T.\nonumber 
\end{equation}
For $m>1/2$, write the seed covariance as
\begin{equation}
  \sigma_m=\sigma_m^{(0)}+C,
  \qquad C\succeq0,
\end{equation}
where $\sigma_m^{(0)}$ is the pure seed with the same squeezing and orientation.
Let $g_C$ denote the centered Gaussian probability measure with covariance
$C$. The outcome distribution associated with $\sigma_m$ is then the Gaussian
convolution
\begin{align}
  q_\theta(\bm y)
  &=\int_{\mathbb R^{2}}
  K_C(\bm y|\bm r)\,p_\theta(\bm r)\,\dd\bm r,\nonumber\\
  K_C(\bm y|\bm r)&:=g_C(\bm y-\bm r).\nonumber 
\end{align}
Here $K_C(\bm y|\bm r)$ is a classical Markov kernel: it is the conditional
probability density of the noisy outcome $\bm y$ given the original outcome
$\bm r$. Since $C$ is independent of the encoded parameter $\theta$, the
kernel is also independent of $\theta$.

Under the usual regularity assumptions allowing differentiation under the
integral, the score of the processed distribution satisfies
\begin{equation}
  \partial_\theta\log q_\theta(\bm y)
  =
  \mathbb E_\theta\!\left[
    \partial_\theta\log p_\theta(\bm R)
    \,\middle|\,\bm Y=\bm y
  \right].\nonumber 
\end{equation}
Conditional Jensen's inequality then gives
\begin{equation}
  I_C[q_\theta]
  \leq
  I_C[p_\theta].\nonumber 
\end{equation}
Therefore parameter-independent Gaussian convolution cannot increase classical
Fisher information \cite{Ferrie2014}.

Thus, adding the noise covariance $C$ cannot increase the CFI in any Gaussian
tangent direction. Since the QFI is unchanged, replacing a mixed seed by the
corresponding pure seed cannot decrease
$\lambda_{\min}(\FG)$. It is therefore sufficient to optimize over pure
seeds.

After removing the irrelevant rotation, write the pure seed as
\begin{align}
  \sigma_m&=\begin{pmatrix}a&0\\0&b\end{pmatrix},&
  a&:=\frac12e^{-2z},\nonumber\\
  b&:=\frac12e^{2z},&ab&=\frac14.
  \label{eq:pureDiagonalSeed}
\end{align}
The outcome covariance is
\begin{align}
  \tau&=\begin{pmatrix}t_1&0\\0&t_2\end{pmatrix},\nonumber\\
  t_1&:=\nu+a,&t_2&:=\nu+b.\nonumber 
\end{align}

Consider the invariant two-dimensional subspace of diagonal covariance
perturbations,
\begin{equation}
  X=\diag(x,y).\nonumber 
\end{equation}
On this subspace,
\begin{equation}
  \mathcal N_\tau(X)
  =
  \tau^{-1}X\tau^{-1}
  =
  \diag
  \left(
    \frac{x}{t_1^2},
    \frac{y}{t_2^2}
  \right).\nonumber 
\end{equation}
For a diagonal matrix $\diag(\alpha,\beta)$,
\begin{equation}
  J
  \diag(\alpha,\beta)
  J
  =
  -\diag(\beta,\alpha).\nonumber 
\end{equation}
Using
\begin{equation}
  F_G^{(2)}
  =
  \frac12
  \mathcal D_{\nu \I_2}\circ\mathcal N_\tau,\nonumber 
\end{equation}
we obtain
\begin{equation}
  F_G^{(2)}(X)
  =
  \diag
  \left(
    \frac{\nu^2x}{t_1^2}
    -
    \frac{y}{4t_2^2},
    \frac{\nu^2y}{t_2^2}
    -
    \frac{x}{4t_1^2}
  \right).\nonumber 
\end{equation}
Hence, in the coordinate vector $(x,y)^T$, the restriction of
$F_G^{(2)}$ to the diagonal covariance subspace is represented by
\begin{equation}
  M(z)
  =
  \begin{pmatrix}
    \dfrac{\nu^2}{t_1^2}
    &
    -\dfrac{1}{4t_2^2}
    \\[6pt]
    -\dfrac{1}{4t_1^2}
    &
    \dfrac{\nu^2}{t_2^2}
  \end{pmatrix}.
  \label{eq:MoneMode}
\end{equation}
The matrix $M(z)$ need not be symmetric in the Euclidean inner product.
Nevertheless, it represents the restriction of a QFI-self-adjoint operator
to an invariant subspace. Since the diagonal covariance subspace is invariant under $F_G^{(2)}$, the
restriction of $F_G^{(2)}$ to this subspace is self-adjoint in the restricted
QFI inner product. Its eigenvalues are therefore real
 \cite{HornJohnson2013}. Denote them by
\begin{equation}
  \mu_-(z)\leq\mu_+(z).\nonumber 
\end{equation}
Since these are eigenvalues of an invariant restriction, they also belong to
the spectrum of the complete Fisher-efficiency operator. In particular,
\begin{equation}
  \lambda_{\min}(\FG)
  \leq
  \mu_-(z).
  \label{eq:fullMinimumVsDiagonal}
\end{equation} Indeed, the infimum of the Rayleigh quotient over the full tangent space
cannot exceed its infimum over any subspace. Equivalently, because the
subspace is invariant, the eigenvalues of the restricted operator belong to
the spectrum of the full operator. Set
\begin{equation}
  c:=\cosh(2z)\geq1.\nonumber
\end{equation}
Using
\begin{equation}
  a+b=c,
  \qquad
  ab=\frac14,\nonumber
\end{equation}
we find
\begin{align}
  p
  &:=
  t_1t_2
  =
  \nu^2+\nu c+\frac14,\nonumber\\
  s
  &:=
  t_1^2+t_2^2
  =
  2\nu^2+2\nu c+c^2-\frac12.\nonumber
\end{align}
The trace and determinant of $M(z)$ are
\begin{align}
  \Tr M(z)
  &=
  \frac{\nu^2s}{p^2},
  \label{eq:Mtrace}\\
  \det M(z)
  &=
  \frac{\nu^4-\frac1{16}}{p^2}.
  \label{eq:Mdeterminant}
\end{align}

Define the efficiency attained by the unsqueezed vacuum seed as
\begin{equation}
  L_0
  :=
  \frac{\nu-\frac12}{\nu+\frac12}.
  \label{eq:L0definition}
\end{equation}
The characteristic polynomial of $M(z)$ is
\begin{equation}
  q_z(\lambda)
  :=
  \lambda^2
  -
  \Tr[M(z)]\lambda
  +
  \det M(z).\nonumber
\end{equation}
Direct algebraic simplification gives
\begin{equation}
  q_z(L_0)
  =
  -
  \frac{
    8\nu(c-1)(2\nu-1)
    \left(
      4c\nu+8\nu^2+2\nu+1
    \right)
  }{
    (2\nu+1)^2
    \left(
      4c\nu+4\nu^2+1
    \right)^2
  }
  \leq0.
  \label{eq:qfactorization}
\end{equation}
All denominator factors are strictly positive, while
\(\nu>1/2\) and \(c\geq1\). Equality holds if and only if \(c=1\), equivalently
\(z=0\).

Because $M(z)$ has the real eigenvalues
$\mu_-(z)\leq\mu_+(z)$, its monic characteristic polynomial is nonpositive
precisely between these roots. Therefore,
\begin{equation}
  q_z(L_0)\leq0
  \quad\Longrightarrow\quad
  \mu_-(z)\leq L_0.\nonumber
\end{equation}
Together with Eq. \eqref{eq:fullMinimumVsDiagonal}, this gives
\begin{equation}
  \lambda_{\min}(\FG)
  \leq
  L_0\nonumber
\end{equation}
for every pure single-mode Gaussian seed. For $z\neq0$, the inequality is
strict because \(q_z(L_0)<0\).

For every pure seed with $z\neq0$, the inequality is strict:
\begin{equation}
  \lambda_{\min}(\FG)<L_0.\nonumber
\end{equation}
For a mixed seed with $z\neq0$, classical data processing gives an efficiency
no larger than that of the corresponding pure seed, and hence it is also
strictly smaller than $L_0$.

It remains to consider isotropic mixed seeds, for which $z=0$ and
$\sigma_m=mI_2$. Their smallest eigenvalue is
\begin{equation}
  \lambda_-(m)
  =
  \frac{4\nu^2-1}{4(\nu+m)^2}.\nonumber
\end{equation}
This function is strictly decreasing for $m\geq1/2$, and
\begin{equation}
  \lambda_-\!\left(\frac12\right)
  =
  L_0.\nonumber
\end{equation}
Therefore, equality is possible only for
\begin{equation}
  m=\frac12,
  \qquad
  z=0.\nonumber
\end{equation}
The rotation angle is then irrelevant because the covariance matrix is
$I_2/2$. Thus the vacuum covariance is the unique optimizer in the
Williamson basis. Applying the inverse common symplectic transformation
gives the unique optimal covariance
\begin{equation}
  \sigma_m^\star=\frac12SS^T\nonumber
\end{equation}
in the original phase-space coordinates.
\end{proof}

\subsubsection{Infinite-squeezing limit}
\label{rem:homodyneWorstLimit}
For a pure single-mode seed with squeezing parameter $z$, the smallest
efficiency satisfies
\begin{equation}
  \mu_-(z)
  \sim
  \frac{
    4\left(\nu^4-\frac1{16}\right)
  }{
    \nu^2
  }
  e^{-4|z|}
  \qquad
  \text{as }|z|\to\infty.\nonumber
\end{equation}
Thus,
\begin{equation}
  \lambda_{\min}(\FG)\longrightarrow0\nonumber
\end{equation}
in the ideal-homodyne limit. Excluding the singular infinite-squeezing seed
from the finite-covariance measurement class therefore does not remove a
better optimizer of the robust Fisher-efficiency criterion; this singular
limit is maximally unfavorable to the least-efficient Gaussian direction. This limit also illustrates the conservative character of the
worst-direction criterion. In the same limit the displacement
efficiency along the infinitely squeezed quadrature satisfies
$\lambda_{d} = \nu/(\nu + \tfrac{1}{2}e^{-2z}) \to 1$, so ideal
homodyne detection saturates the quantum Cram\'er-Rao bound for
that single tangent direction while being ranked as the least
favourable seed by $\lambda_{\min}(\mathcal{F}_G)$.

\subsubsection{Physical interpretation}
The optimization above imposes no constraint on the energy or squeezing of
the measurement seed. In the Williamson basis of the reference state, the
optimal seed is the vacuum and requires no squeezing. In the original
laboratory basis, however, the matched seed is the Gaussian state with
covariance $SS^T/2$ and may require the squeezing contained in $S$. Thus, the
optimal measurement follows the symplectic geometry of the state's covariance
ellipse.

\subsection{Proof of physicality of the effective covariance}

\begin{proof}[Proof of Lemma \ref{lem:physicalConditionalCovariance}]
Write
\begin{equation}
  \Omega_N
  =
  J\oplus\Omega_{N-1}.\nonumber
\end{equation}
The physicality of $M$ gives
\begin{equation}
  M+\frac{i}{2}\Omega_N
  =
  \begin{pmatrix}
    A+\dfrac{i}{2}J&C\\[2mm]
    C^T&B+\dfrac{i}{2}\Omega_{N-1}
  \end{pmatrix}
  \succeq0.
  \label{eq:MphysicalBlock}
\end{equation}
Since $V$ is physical,
\begin{equation}
  V+\frac{i}{2}\Omega_{N-1}\succeq0.\nonumber
\end{equation}
Taking the complex conjugate of this Hermitian matrix inequality yields
\begin{equation}
  V-\frac{i}{2}\Omega_{N-1}\succeq0.
  \label{eq:VconjugatePhysicality}
\end{equation}
Adding
\begin{equation}
  0_2\oplus
  \left(
    V-\frac{i}{2}\Omega_{N-1}
  \right)
  \succeq0\nonumber
\end{equation}
to the matrix in Eq. \eqref{eq:MphysicalBlock} gives
\begin{equation}
  \begin{pmatrix}
    A+\dfrac{i}{2}J&C\\[2mm]
    C^T&B+V
  \end{pmatrix}
  \succeq0.
  \label{eq:complexBlockPositive}
\end{equation}

Every finite-dimensional physical covariance matrix is real, symmetric, and
positive definite. Hence, $B+V\succ0$ and is invertible. Applying the
Schur-complement criterion to Eq. \eqref{eq:complexBlockPositive} gives
\begin{equation}
  A+\frac{i}{2}J
  -
  C(B+V)^{-1}C^T
  \succeq0.\nonumber
\end{equation}
By Eq. \eqref{eq:effectiveConditionalCovariance}, this is precisely
\begin{equation}
  M_{\mathrm{eff}}+\frac{i}{2}J\succeq0.\nonumber
\end{equation}
Moreover, $M_{\mathrm{eff}}$ is real and symmetric. It is therefore a physical
single-mode covariance matrix.
\end{proof}

Every real symmetric matrix satisfying
\begin{equation}
  M_{\mathrm{eff}}+\frac{i}{2}J\succeq0\nonumber
\end{equation}
is the covariance matrix of a centered one-mode Gaussian state. Hence
$M_{\mathrm{eff}}$ is a legitimate one-mode measurement-seed covariance.
It is not, in general, the marginal covariance $A$ of the selected mode:
rather, it is an effective covariance obtained after eliminating the remaining
blocks through the Schur complement. No conditional-state interpretation is
required in the proof.

\section{Auxiliary results for product-seed optimization}
\label{app:productSeedAuxiliaries}

\subsection{Proof of covariance-sector dominance}

\begin{proof}[Proof of Lemma \ref{lem:covarianceControls}]
Let
\begin{equation}
  \tau=\sigma+\sigma_m\nonumber
\end{equation}
and define
\begin{equation}
  k
  :=
  \max_{\bm v\neq0}
  \frac{\bm v^T\tau\bm v}
       {\bm v^T\sigma\bm v}.
  \label{eq:kDefinition}
\end{equation}
Since every finite physical Gaussian covariance matrix is positive definite,
$\sigma_m\succ0$, and hence $k>1$.

The minimum efficiency in the displacement sector is
\begin{equation}
  \eta_1
  :=
  \lambda_{\min}\!\left[F_G^{(1)}\right]
  =
  \min_{\bm u\neq0}
  \frac{\bm u^T\tau^{-1}\bm u}
       {\bm u^T\sigma^{-1}\bm u}
  =
  \frac1k.
  \label{eq:displacementMinimumK}
\end{equation}
The last equality follows by inverting the corresponding positive-definite
generalized eigenvalue problem.

Let
\begin{equation}
  \eta_2
  :=
  \lambda_{\min}\!\left[F_G^{(2)}\right].\nonumber
\end{equation}
Inverting the covariance-sector generalized eigenvalue problem gives
\begin{equation}
  \frac1{\eta_2}
  =
  \max_{\substack{Y=Y^T\\Y\neq0}}
  \frac{
    2\Tr(Y\tau Y\tau)
  }{
    \Tr[Y\mathcal D_\sigma(Y)]
  },
  \label{eq:reciprocalCovarianceRQ}
\end{equation}
where
\begin{equation}
  \mathcal D_\sigma(Y)
  =
  2\sigma Y\sigma+\frac12\Omega_NY\Omega_N.\nonumber
\end{equation}

Choose a real vector $\bm v$ attaining the maximum in
Eq. \eqref{eq:kDefinition} and set
\begin{equation}
  Y=\bm v\bm v^T.\nonumber
\end{equation}
Since $\Omega_N$ is antisymmetric,
\begin{equation}
  \bm v^T\Omega_N\bm v=0.\nonumber
\end{equation}
Consequently,
\begin{align}
  2\Tr(Y\tau Y\tau)
  &=
  2(\bm v^T\tau\bm v)^2,\nonumber\\
  \Tr[Y\mathcal D_\sigma(Y)]
  &=
  2(\bm v^T\sigma\bm v)^2.\nonumber
\end{align}
Equation \eqref{eq:reciprocalCovarianceRQ} therefore implies
\begin{equation}
  \frac1{\eta_2}\geq k^2,
  \qquad\text{or equivalently}\qquad
  \eta_2\leq\frac1{k^2}=\eta_1^2.\nonumber
\end{equation}
Because $k>1$, we have
\begin{equation}
  \eta_2\leq\eta_1^2<\eta_1.\nonumber
\end{equation}
Thus the minimum eigenvalue of the complete Gaussian Fisher-efficiency
operator necessarily belongs to the covariance sector.
\end{proof}

The next lemma permits the reciprocal problem to be extended temporarily from
real symmetric matrices to the complex matrix space. This is useful because
it allows us to employ the positive cone of Hermitian matrices.

\begin{lemma}[Positive extension of the reciprocal problem]
\label{lem:positiveExtension}
Let $\sigma$ be faithful. Then $\mathcal D_\sigma$ is a positive-definite
self-adjoint operator on the complex matrix space equipped with the
Hilbert-Schmidt inner product. Moreover,
$\mathcal D_\sigma^{-1}$ maps positive semidefinite Hermitian matrices to
positive semidefinite Hermitian matrices.

The largest eigenvalue of the reciprocal problem
\eqref{eq:reciprocalCovarianceRQ} admits a nonzero real symmetric positive
semidefinite eigenmatrix. Consequently, extending the optimization from
$\Sym(2N,\mathbb R)$ to the full complex matrix space does not change its
maximum value.
\end{lemma}

\begin{proof}
Define
\begin{equation}
  A_\sigma
  :=
  \sigma^{-1/2}\Omega_N\sigma^{-1/2},
  \qquad
  Z:=\sigma^{1/2}Y\sigma^{1/2}.\nonumber
\end{equation}
The matrix $A_\sigma$ is real and antisymmetric. Moreover,
\begin{equation}
  \sigma^{-1/2}A_\sigma\sigma^{1/2}
  =
  \sigma^{-1}\Omega_N,\nonumber
\end{equation}
so $A_\sigma$ is similar to $\sigma^{-1}\Omega_N$. The eigenvalues of
$\sigma^{-1}\Omega_N$ are $\pm i/\nu_j$, where $\{\nu_j\}$ are the
symplectic eigenvalues of $\sigma$. Since a real antisymmetric matrix is
normal, its singular values are the absolute values of its eigenvalues.
Therefore, the singular values of $A_\sigma$ are $1/\nu_j$, each repeated
twice. Since the state is faithful,
\begin{equation}
  \|A_\sigma\|_2
  =
  \frac1{\nu_{\min}}
  <2.\nonumber
\end{equation}
For every complex matrix $Y$,
\begin{align}
  \Tr\!\left[
    Y^\dagger\mathcal D_\sigma(Y)
  \right]
  &=
  2\|Z\|_F^2
  +
  \frac12
  \Tr\!\left[
    Z^\dagger A_\sigma ZA_\sigma
  \right] \nonumber
  \\
  &\geq
  \left(
    2-\frac12\|A_\sigma\|_2^2
  \right)
  \|Z\|_F^2
  >0
  \label{eq:Dpositive}
\end{align}
whenever $Y\neq0$. Thus $\mathcal D_\sigma$ is positive definite. Its
self-adjointness follows directly from cyclicity of the trace.

We next prove that its inverse preserves positivity. The equation
\begin{equation}
  \mathcal D_\sigma(Y)=X\nonumber
\end{equation}
can be rewritten as
\begin{equation}
  Y-KYK^T
  =
  \frac12\sigma^{-1}X\sigma^{-1},
  \qquad
  K:=\frac12\sigma^{-1}\Omega_N.
  \label{eq:DinverseSteinEquation}
\end{equation}
Indeed, Eq. \eqref{eq:DinverseSteinEquation} follows by multiplying the
original equation on both sides by $\sigma^{-1}$ and dividing by two.

The matrix $K$ is similar to $A_\sigma/2$, because
\begin{equation}
  \sigma^{1/2}K\sigma^{-1/2}
  =
  \frac12A_\sigma.\nonumber
\end{equation}
Hence,
\begin{equation}
  \rho(K)
  =
  \frac{1}{2\nu_{\min}}
  <1,\nonumber
\end{equation}
where
\begin{equation}
  \rho(K)
  :=
  \max\left\{
    |\lambda|:
    \lambda\in\operatorname{spec}(K)
  \right\}\nonumber
\end{equation}
denotes the spectral radius of $K$.
The corresponding convergent Neumann series gives
\begin{equation}
  \mathcal D_\sigma^{-1}(X)
  =
  \frac12
  \sum_{n=0}^{\infty}
  K^n\sigma^{-1}X\sigma^{-1}(K^T)^n.
  \label{eq:DinverseSeries}
\end{equation}
If $X\succeq0$, every term in this series is positive semidefinite.
Therefore,
\begin{equation}
  X\succeq0
  \quad\Longrightarrow\quad
  \mathcal D_\sigma^{-1}(X)\succeq0.\nonumber
\end{equation}

Now define the positive self-adjoint map
\begin{equation}
  \mathcal C_\tau(Y):=2\tau Y\tau\nonumber
\end{equation}
and consider
\begin{equation}
  \mathcal T
  :=
  \mathcal D_\sigma^{-1}\circ\mathcal C_\tau.\nonumber
\end{equation}
Both factors preserve the cone of positive semidefinite Hermitian matrices,
and therefore so does $\mathcal T$. The real vector space of Hermitian matrices is finite-dimensional, and its
cone of positive semidefinite matrices is a proper cone: it is closed,
convex, pointed, and has nonempty interior. Since $\mathcal T$ preserves this cone, the finite-dimensional Perron-Frobenius theorem for positive linear maps implies that the spectral radius of $\mathcal T$ is an eigenvalue with a nonzero positive semidefinite Hermitian eigenmatrix
$Y_\star$ \cite{EvansHoeghKrohn1978}.

Furthermore, $\mathcal T$ is similar to the positive self-adjoint operator
\begin{equation}
  \mathcal D_\sigma^{-1/2}
  \circ\mathcal C_\tau
  \circ\mathcal D_\sigma^{-1/2}.
  \label{eq:reciprocalSymmetricOperator}
\end{equation}
Its spectral radius is therefore its largest eigenvalue and coincides with the
maximum reciprocal generalized eigenvalue in
Eq. \eqref{eq:reciprocalCovarianceRQ}.

All coefficients of $\mathcal T$ are real. Hence the real part of $Y_\star$
is also an eigenmatrix with the same eigenvalue. Since $Y_\star\succeq0$ and
$Y_\star\neq0$, its real part is a nonzero real symmetric positive
semidefinite matrix. Thus the maximum reciprocal eigenvalue is already
attained in $\Sym(2N,\mathbb R)$.
\end{proof}

\subsubsection{Role of the positive extension}
The lemma does not assert that every covariance tangent direction is positive
semidefinite. Rather, it shows that an eigenmatrix associated with the largest
eigenvalue of the reciprocal covariance problem can be chosen inside the
positive-semidefinite cone. This prevents the subsequent symmetry averages
from vanishing. After the symmetry reduction, the corresponding
least-efficiency eigenmatrix of the direct problem is recovered through the
map $\mathcal D_\sigma$, as shown explicitly in
Proposition \ref{prop:localGeneralMatrix}.

\subsection{Proof of the pinching lemma}

\begin{proof}[Proof of Lemma \ref{lem:covariantPinching}]
Because $U$ is symplectic, $U\sigma_mU^T$ is physical whenever $\sigma_m$ is
physical. The set of physical covariance matrices is convex, so
$\mathcal E(\sigma_m)$ is also physical.

On the full vectorized matrix space, the two operators entering the reciprocal
Rayleigh quotient are represented by
\begin{align}
  \mathbf C_\tau
  &=
  2\tau\otimes\tau,\nonumber\\
  \mathbf D_\sigma
  &=
  2\sigma\otimes\sigma
  -
  \frac12\Omega_N\otimes\Omega_N.
\label{eq:vectorizedReciprocalOperators}
\end{align}
The minus sign in the second expression follows from
$\Omega_N^T=-\Omega_N$ and the identity
\begin{equation}
  \operatorname{vec}(AYB)
  =
  (B^T\otimes A)\operatorname{vec}(Y).\nonumber
\end{equation}

By Lemma \ref{lem:positiveExtension}, the largest reciprocal eigenvalue is
the smallest scalar $\kappa$ such that
\begin{equation}
  2\tau\otimes\tau
  \preceq
  \kappa\mathbf D_\sigma.
  \label{eq:tensorOrder}
\end{equation}
Extend $\mathcal E$ complex-linearly to $M_d(\mathbb C)$, where $d=2N$.
Since $U$ is real and orthogonal, $U^T=U^\dagger$. Under the canonical
identification
\begin{equation}
  M_{d^2}(\mathbb C)
  \cong
  M_d(\mathbb C)\otimes M_d(\mathbb C),\nonumber
\end{equation}
the tensor-product map acts as
\begin{equation}
  (\mathcal E\otimes\mathcal E)(X)
  =
  \frac14
  \sum_{V,W\in\{I_d,U\}}
  (V\otimes W)X(V\otimes W)^\dagger.
  \label{eq:EtensorEpositive}
\end{equation}
If $X\succeq0$, every term on the right-hand side is positive semidefinite.
Consequently,
\begin{equation}
  X\succeq0
  \quad\Longrightarrow\quad
  (\mathcal E\otimes\mathcal E)(X)\succeq0,\nonumber
\end{equation}
so $\mathcal E\otimes\mathcal E$ preserves the positive-semidefinite order.

Since $U\sigma U^T=\sigma$ and $U$ is symplectic,
\begin{equation}
  \mathcal E(\sigma)=\sigma,
  \qquad
  \mathcal E(\Omega_N)=\Omega_N.\nonumber
\end{equation}
Therefore,
\begin{equation}
  (\mathcal E\otimes\mathcal E)(\mathbf D_\sigma)
  =
  \mathbf D_\sigma.\nonumber
\end{equation}
Applying $\mathcal E\otimes\mathcal E$ to
Eq. \eqref{eq:tensorOrder} gives
\begin{equation}
  2\mathcal E(\tau)\otimes\mathcal E(\tau)
  \preceq
  \kappa\mathbf D_\sigma.\nonumber
\end{equation}
Moreover,
\begin{equation}
  \mathcal E(\tau)
  =
  \sigma+\mathcal E(\sigma_m).\nonumber
\end{equation}
Thus the largest reciprocal covariance-sector eigenvalue cannot increase
under the pinching map. Taking reciprocals proves
Eq. \eqref{eq:pinchingImproves}.
\end{proof}

\subsubsection{Tensor-product action after vectorization}
Under vectorization, the matrices $\mathbf C_\tau$ and
$\mathbf D_\sigma$ act on
$\mathbb C^d\otimes\mathbb C^d$, with $d=2N$. Accordingly,
$\mathcal E\otimes\mathcal E$ acts on the matrix representatives of these
superoperators. This tensor-product action must not be confused with applying
$\mathcal E$ directly to a linear map on $M_d(\mathbb C)$.

\section{Derivation of the matrix $H_-$}
\label{app:HminusDerivation}

\begin{proof}[Proof of Proposition \ref{prop:localGeneralMatrix}]
By Theorem \ref{thm:localEPRgeneral}, the optimal product seed is the
heterodyne seed
\begin{equation}
  \sigma_{m,\mathrm{het}}=\frac12I_4.\nonumber
\end{equation}
It remains to determine the smallest eigenvalue of the Gaussian
Fisher-efficiency operator associated with this seed.

Apply the common symplectic transformation $S_{\mathrm{TMS}}(-r)$ to the
reference state, the measurement seed, and the Gaussian tangent directions.
By Lemma \ref{lem:commonSymplecticCovariance}, this transformation does not
change the Fisher-efficiency spectrum. The state covariance becomes
\begin{equation}
  D_\nu
  =
  \nu_1I_2\oplus\nu_2I_2,\nonumber
\end{equation}
whereas the heterodyne seed becomes
\begin{align}
  M_{\mathrm h}
  &=\frac12\begin{pmatrix}CI_2&-SZ\\-SZ&CI_2\end{pmatrix},\nonumber\\
  C&=\cosh(2r),&S&=\sinh(2r).
  \label{eq:heterodyneWilliamsonGeneral}
\end{align}
with $Z=\operatorname{diag}(1,-1).$
Thus the outcome covariance in the Williamson basis is
\begin{equation}
  \tau_0
  :=
  D_\nu+M_{\mathrm h}
  =
  \begin{pmatrix}
    \left(\nu_1+\dfrac C2\right)I_2
    &
    -\dfrac S2Z
    \\[2mm]
    -\dfrac S2Z
    &
    \left(\nu_2+\dfrac C2\right)I_2
  \end{pmatrix}.
  \label{eq:tauWilliamsonGeneral}
\end{equation}

Since $Z^2=I_2$, direct block inversion gives
\begin{equation}
  K
  :=
  \tau_0^{-1}
  =
  \begin{pmatrix}
    uI_2&wZ\\
    wZ&vI_2
  \end{pmatrix},
  \label{eq:inverseTauGeneral}
\end{equation}
where
\begin{align}
  u&=\frac{\nu_2+C/2}{\Delta}, \nonumber \\
  v&=\frac{\nu_1+C/2}{\Delta},\nonumber\\
  w&=\frac{S}{2\Delta},\notag\\
  \Delta
  &=
  \left(\nu_1+\frac C2\right)
  \left(\nu_2+\frac C2\right)
  -
  \frac{S^2}{4}
  \notag\\
  &=
  \nu_1\nu_2
  +
  \frac{\nu_1+\nu_2}{2}C
  +
  \frac14.\nonumber
\end{align}
In the last equality we used $C^2-S^2=1$. The positive sign of $w$ follows
from the negative off-diagonal block of $\tau_0$.

We next show that the smallest covariance-sector eigenvalue can be found in a
three-dimensional invariant subspace. Consider the reciprocal generalized
eigenvalue problem
\begin{equation}
  2\tau_0Y\tau_0
  =
  \kappa\,\mathcal D_{D_\nu}(Y).
  \label{eq:reciprocalProblemHminus}
\end{equation}
Its largest eigenvalue satisfies
\begin{equation}
  \kappa_{\max}
  =
  \frac{1}{
    \lambda_{\min}\!\left[
      F_G^{(2)}(D_\nu,M_{\mathrm h})
    \right]
  }.
  \label{eq:reciprocalDirectRelation}
\end{equation}
By Lemma \ref{lem:positiveExtension}, Eq. \eqref{eq:reciprocalProblemHminus}
admits a nonzero real symmetric positive semidefinite eigenmatrix
$Y_\star\succeq0$ associated with $\kappa_{\max}$.

Define the orthogonal symplectic matrices
\begin{equation}
  U_\theta
  :=
  R(\theta)\oplus R(-\theta).\nonumber
\end{equation}
Direct multiplication gives
\begin{equation}
  U_\theta D_\nu U_\theta^T=D_\nu,
  \qquad
  U_\theta M_{\mathrm h}U_\theta^T=M_{\mathrm h},\nonumber
\end{equation}
where
\begin{equation}
  R(\theta)I_2R(\theta)^T=I_2,
  \qquad
  R(\theta)ZR(\theta)=Z.\nonumber
\end{equation}
Therefore, both sides of Eq. \eqref{eq:reciprocalProblemHminus} are covariant
under
\begin{equation}
  Y\longmapsto U_\theta YU_\theta^T.\nonumber
\end{equation}
Consequently,
\begin{equation}
  \overline Y_\star
  :=
  \frac{1}{2\pi}
  \int_0^{2\pi}
  U_\theta Y_\star U_\theta^T\,\dd\theta
  \label{eq:rotationAverageY}
\end{equation}
is another eigenmatrix associated with $\kappa_{\max}$. Moreover,
$\overline Y_\star\succeq0$, and it is nonzero because conjugation preserves
the trace:
\begin{equation}
  \Tr(\overline Y_\star)=\Tr(Y_\star)>0.\nonumber
\end{equation}

Solving the rotational invariance condition
\begin{equation}
  U_\theta\overline Y_\star U_\theta^T
  =
  \overline Y_\star
  \qquad
  \text{for every }\theta\nonumber
\end{equation}
shows that $\overline Y_\star$ has the form
\begin{equation}
  \overline Y_\star
  =
  \begin{pmatrix}
    xI_2&zZ+sW\\
    zZ+sW&yI_2
  \end{pmatrix},
  \qquad
  W=
  \begin{pmatrix}
    0&1\\
    1&0
  \end{pmatrix}.
  \label{eq:rotationInvariantForm}
\end{equation}

Now define
\begin{equation}
  Q:=Z\oplus Z.\nonumber
\end{equation}
Since
\begin{equation}
  ZJZ=-J,\nonumber
\end{equation}
one has
\begin{equation}
  Q\Omega_2Q^T=-\Omega_2,\nonumber
\end{equation}
so $Q$ is antisymplectic. Nevertheless,
\begin{equation}
  QD_\nu Q^T=D_\nu,
  \qquad
  QM_{\mathrm h}Q^T=M_{\mathrm h}.\nonumber
\end{equation}
Furthermore, the sign change of $\Omega_2$ cancels in
$\mathcal D_{D_\nu}$ because $\Omega_2$ occurs twice. Hence
Eq. \eqref{eq:reciprocalProblemHminus} is invariant under
\begin{equation}
  Y\longmapsto QYQ^T.\nonumber
\end{equation}
The additional average
\begin{equation}
  Y_-
  :=
  \frac12
  \left(
    \overline Y_\star
    +
    Q\overline Y_\star Q^T
  \right)
  \label{eq:QaverageY}
\end{equation}
is therefore a nonzero positive semidefinite eigenmatrix associated with
$\kappa_{\max}$. Since
\begin{equation}
  ZZZ=Z,
  \qquad
  ZWZ=-W,\nonumber
\end{equation}
the $sW$ component is removed, and
\begin{equation}
  Y_-
  \in
  \mathcal V_-,\nonumber
\end{equation}
where
\begin{equation}
  \mathcal V_-
  :=
  \left\{
    X(x,y,z)
    =
    \begin{pmatrix}
      xI_2&zZ\\
      zZ&yI_2
    \end{pmatrix}
    :
    x,y,z\in\R
  \right\}.
  \label{eq:VminusDefinition}
\end{equation}

The preceding positivity argument concerns the reciprocal eigenmatrix $Y_-$.
To return to the direct Fisher-efficiency problem, define
\begin{equation}
  X_-:=\mathcal D_{D_\nu}(Y_-).\nonumber
\end{equation}
Since $\mathcal D_{D_\nu}$ is invertible, $X_-\neq0$. Moreover,
$\mathcal D_{D_\nu}$ preserves $\mathcal V_-$, so
\begin{equation}
  X_-\in\mathcal V_-.\nonumber
\end{equation}
From Eq. \eqref{eq:reciprocalProblemHminus},
\begin{equation}
  2\tau_0Y_-\tau_0
  =
  \kappa_{\max}X_-.\nonumber
\end{equation}
Multiplying from the left and right by $\tau_0^{-1}=K$ gives
\begin{equation}
  KX_-K
  =
  \frac{2}{\kappa_{\max}}Y_-.\nonumber
\end{equation}
Therefore,
\begin{align}
  F_G^{(2)}(X_-)
  &=
  \frac12
  \mathcal D_{D_\nu}(KX_-K)
  \nonumber\\
  &=
  \frac{1}{\kappa_{\max}}
  \mathcal D_{D_\nu}(Y_-)
  \nonumber\\
  &=
  \frac{1}{\kappa_{\max}}X_-.\nonumber
\end{align}
Thus a direct eigenmatrix associated with the smallest covariance-sector
efficiency can indeed be chosen in $\mathcal V_-$. Consequently, it is
sufficient to compute the restriction of $F_G^{(2)}$ to this subspace.

Let
\begin{equation}
  X(x,y,z)
  =
  \begin{pmatrix}
    xI_2&zZ\\
    zZ&yI_2
  \end{pmatrix}
  \in\mathcal V_-.\nonumber
\end{equation}
Using
\begin{equation}
  JI_2J=-I_2,
  \qquad
  JZJ=Z,\nonumber
\end{equation}
we obtain
\begin{equation}
  \mathcal D_{D_\nu}[X(x,y,z)]
  =
  \begin{pmatrix}
    2a_1xI_2&2bzZ\\
    2bzZ&2a_2yI_2
  \end{pmatrix},
  \label{eq:DactionVminus}
\end{equation}
where
\begin{equation}
  a_1:=\nu_1^2-\frac14,
  \qquad
  a_2:=\nu_2^2-\frac14,
  \qquad
  b:=\nu_1\nu_2+\frac14.
  \label{eq:a1a2bDefinitions}
\end{equation}

Direct block multiplication using Eq. \eqref{eq:inverseTauGeneral} gives
\begin{equation}
  KX(x,y,z)K
  =
  X(A,B,C_z),\nonumber
\end{equation}
where
\begin{align}
  A
  &=
  u^2x+w^2y+2uwz,
  \label{eq:KXKA}\\
  B
  &=
  w^2x+v^2y+2vwz,
  \label{eq:KXKB}\\
  C_z
  &=
  uwx+vwy+(uv+w^2)z.
  \label{eq:KXKC}
\end{align}
Since
\begin{equation}
  F_G^{(2)}(X)
  =
  \frac12
  \mathcal D_{D_\nu}(KXK),\nonumber
\end{equation}
Eqs. \eqref{eq:DactionVminus}-\eqref{eq:KXKC} yield
\begin{align}
  x'
  &=
  a_1A
  =
  a_1u^2x
  +
  a_1w^2y
  +
  2a_1uwz,
  \label{eq:xprimeHminus}\\
  y'
  &=
  a_2B
  =
  a_2w^2x
  +
  a_2v^2y
  +
  2a_2vwz,
  \label{eq:yprimeHminus}\\
  z'
  &=
  bC_z
  =
  buwx
  +
  bvwy
  +
  b(uv+w^2)z.
  \label{eq:zprimeHminus}
\end{align}
Therefore,
\begin{equation}
  F_G^{(2)}[X(x,y,z)]
  =
  X(x',y',z'),\nonumber
\end{equation}
with
\begin{equation}
  \begin{pmatrix}
    x'\\y'\\z'
  \end{pmatrix}
  =
  \begin{pmatrix}
    a_1u^2&a_1w^2&2a_1uw\\
    a_2w^2&a_2v^2&2a_2vw\\
    buw&bvw&b(uv+w^2)
  \end{pmatrix}
  \begin{pmatrix}
    x\\y\\z
  \end{pmatrix}.
  \label{eq:HminusIntermediate}
\end{equation}
Substituting the definitions
\begin{equation}
  2a_j
  =
  \frac{4\nu_j^2-1}{2},
  \qquad
  b
  =
  \frac{4\nu_1\nu_2+1}{4},\nonumber
\end{equation}
into Eq. \eqref{eq:HminusIntermediate} gives precisely the matrix
$H_-$ displayed in Eq. \eqref{eq:HminusGeneral}.

It remains to identify the inner product in which this matrix is
self-adjoint. From Eq. \eqref{eq:DactionVminus},
\begin{equation}
  \mathcal D_{D_\nu}^{-1}[X(x,y,z)]
  =
  \begin{pmatrix}
    \dfrac{x}{2a_1}I_2
    &
    \dfrac{z}{2b}Z
    \\[3mm]
    \dfrac{z}{2b}Z
    &
    \dfrac{y}{2a_2}I_2
  \end{pmatrix}.\nonumber
\end{equation}
Consequently,
\begin{align}
  I_{\mathrm Q}[X(x,y,z)]
  &=
  \Tr\!\left[
    X(x,y,z)
    \mathcal D_{D_\nu}^{-1}[X(x,y,z)]
  \right]
  \\
  &=
  \frac{x^2}{a_1}
  +
  \frac{y^2}{a_2}
  +
  \frac{2z^2}{b}.
  \label{eq:QFIonVminus}
\end{align}
Thus the coordinate matrix of the restricted QFI inner product is
\begin{equation}
  G_-
  =
  \operatorname{diag}
  \left(
    \frac1{a_1},
    \frac1{a_2},
    \frac2b
  \right),\nonumber
\end{equation}
which is Eq. \eqref{eq:GminusGeneral}. Since $\nu_1,\nu_2>1/2$, one has
$a_1>0$, $a_2>0$, and $b>0$, so $G_-$ is positive definite.

Multiplying $H_-$ from the left by $G_-$ gives
\begin{equation}
  G_-H_-
  =
  \begin{pmatrix}
    u^2&w^2&2uw\\
    w^2&v^2&2vw\\
    2uw&2vw&2(uv+w^2)
  \end{pmatrix},
  \label{eq:GminusHminusExplicit}
\end{equation}
which is real and symmetric. Therefore,
\begin{equation}
  G_-H_-=H_-^TG_-.\nonumber
\end{equation}
It follows that
\begin{equation}
  \widehat H_-
  =
  G_-^{1/2}H_-G_-^{-1/2}\nonumber
\end{equation}
is a real symmetric matrix similar to $H_-$. Hence all eigenvalues of $H_-$
are real.

We have shown that the smallest covariance-sector eigenvalue is attained in
$\mathcal V_-$ and equals $\lambda_{\min}(H_-)$. Finally,
Lemma \ref{lem:covarianceControls} implies that the minimum eigenvalue of the
complete Gaussian Fisher-efficiency operator belongs to the covariance
sector. Combining this fact with Theorem \ref{thm:localEPRgeneral} yields
\begin{equation}
  \eta_{\mathrm{loc}}^\star(\nu_1,\nu_2,r)
  =
  \lambda_{\min}(H_-),\nonumber
\end{equation}
which proves Eq. \eqref{eq:localGeneralExact}.
\end{proof}

\section{Faithful-state assumption and the pure-state boundary}
\label{app:pureStateBoundary}

All Fisher-efficiency results in this work are formulated for faithful Gaussian
reference states,
\begin{equation}
  \nu_j>\frac12
  \qquad
  \text{for every }j.
  \label{eq:faithfulAppendix}
\end{equation}
Under this assumption, the superoperator
\begin{equation}
  \mathcal D_\sigma(Y)
  =
  2\sigma Y\sigma+\frac12\Omega_NY\Omega_N\nonumber
\end{equation}
is positive definite and invertible on
$\Sym(2N,\mathbb R)$. Consequently, the covariance contribution to the QFI,
\begin{equation}
  I_{\mathrm Q}^{(2)}(X)
  =
  \Tr\!\left[
    X\mathcal D_\sigma^{-1}(X)
  \right],\nonumber
\end{equation}
is well defined for every real symmetric covariance perturbation $X$.

This statement does not extend without qualification to the pure-state
boundary. A pure Gaussian covariance matrix satisfies
\begin{equation}
  \sigma\Omega_N\sigma
  =
  \frac14\Omega_N.
  \label{eq:pureCovarianceCondition}
\end{equation}
At such a covariance, $\mathcal D_\sigma$ is singular. For example, for the
multimode vacuum,
\begin{equation}
  \sigma=\frac12I_{2N},\nonumber
\end{equation}
one has
\begin{equation}
  \mathcal D_\sigma(I_{2N})
  =
  \frac12I_{2N}
  +
  \frac12\Omega_N^2
  =
  0.
  \label{eq:DvacuumKernel}
\end{equation}
Therefore, the ordinary inverse
$\mathcal D_\sigma^{-1}$ does not exist on the complete symmetric-matrix
space at the pure boundary.

Moreover, not every $X\in\Sym(2N,\mathbb R)$ is tangent to the manifold of
pure Gaussian states. Differentiating
Eq. \eqref{eq:pureCovarianceCondition} along a pure-state family gives the
constraint
\begin{equation}
  X\Omega_N\sigma
  +
  \sigma\Omega_NX
  =
  0. \label{eq:pureGaussianTangentConstraint}
\end{equation}
Thus the covariance tangent space of the pure Gaussian manifold is a proper
subspace of $\Sym(2N,\mathbb R)$. Directions that violate
Eq. \eqref{eq:pureGaussianTangentConstraint} generally move the state away
from the pure-state stratum and may require a rank-changing statistical model.

There are three possible treatments of this boundary:
\begin{enumerate}
  \item \emph{Faithful-state limit.} Perform the calculation for
  $\nu_j>1/2$ and then take a specified limit
  $\nu_j\to1/2^+$. The limit must be evaluated direction by direction and
  need not remain finite.

  \item \emph{Moore-Penrose inverse.} Replace
  $\mathcal D_\sigma^{-1}$ with its Moore-Penrose pseudoinverse after
  restricting the perturbations to the appropriate support. This procedure
  requires verifying that the chosen tangent directions and statistical
  model satisfy the corresponding support and regularity conditions.

  \item \emph{Pure-manifold formulation.} Work directly on the tangent space
  defined by Eq. \eqref{eq:pureGaussianTangentConstraint} and use the QFI
  appropriate to pure Gaussian families.
\end{enumerate}

The first approach may be used when a particular pure-state limit is required,
but none of the theorems in the main text relies on such a limiting argument.
Accordingly, the results proved for faithful Gaussian states should not be
interpreted as automatically established at the pure-state boundary. Notice,
however, that the measurement seed itself may be pure: faithfulness is required
for the reference state entering the QFI metric, not for the Gaussian seed
defining the general-dyne POVM.

\bibliography{frame}

\end{document}